\documentclass[onecolumn]{aa}  

\usepackage{graphicx}
\usepackage{txfonts}
\usepackage{amsmath}	% Advanced maths commands
\usepackage{amssymb}	% Extra maths symbols
\usepackage{xspace}
\usepackage{lscape}
\usepackage{color}
\usepackage{longtable}
\usepackage{array}
\usepackage{etoc}
\usepackage{inputenc}
\newcolumntype{L}[1]{>{\raggedright\let\newline\\\arraybackslash\hspace{0pt}}m{#1}}
\newcolumntype{C}[1]{>{\centering\let\newline\\\arraybackslash\hspace{0pt}}m{#1}}
\newcolumntype{R}[1]{>{\raggedleft\let\newline\\\arraybackslash\hspace{0pt}}m{#1}}

\newcommand{\OII}{$\left[\mathrm{O\textrm{\textsc{ii}}}\right]$\xspace}

\begin{document} 

\title{Stellar population properties for 2 million galaxies from SDSS DR14 and DEEP2 DR4 from full spectral fitting}
\titlerunning{Stellar populations in SDSS and DEEP2}
\author{
Johan Comparat\inst{1} 
\and
Claudia Maraston\inst{2} 
\and
Daniel Goddard\inst{2}
\and
Violeta Gonzalez-Perez\inst{2,} \inst{10}
\and
Jianhui Lian\inst{2}
\and
Sofia Meneses-Goytia\inst{2,} \inst{11}
\and
Daniel Thomas\inst{2} 
\and
Joel R. Brownstein\inst{3}  
\and
Rita Tojeiro\inst{4} 
\and
Alexis Finoguenov\inst{1,9} 
\and
Andrea Merloni\inst{1} 
\and
Francisco Prada\inst{5}
\and
Mara Salvato\inst{1} 
\and
Guangtun B. Zhu\inst{6}
\and
Hu Zou\inst{7}
\and
Jonathan Brinkmann\inst{8}
}
\institute{
Max-Planck-Institut f\"{u}r extraterrestrische Physik (MPE), Giessenbachstrasse 1, D-85748 Garching bei München, Germany\\
\email{comparat@mpe.mpg.de}
\and
Institute of Cosmology and Gravitation, University of Portsmouth, Portsmouth, PO1 3FX, UK\\
\and
Department of Physics and Astronomy, University of Utah, 115 S. 1400 E., Salt Lake City, UT 84112, USA \\
\and
School of Physics and Astronomy, North Haugh, St. Andrews KY16 9SS, UK \\
\and
Instituto de Astrof\'{\i}sica de Andaluc\'{\i}a (CSIC), Glorieta de la Astronom\'{\i}a, E-18080 Granada, Spain \\
\and
Center for Astrophysical Sciences, Department of Physics and Astronomy, Johns Hopkins University, 3400 North Charles Street, Baltimore, MD 21218, USA\\ 
\and 
Key Laboratory of Optical Astronomy, National Astronomical Observatories, Chinese Academy of Sciences, Beijing 100012, China\\
\and 
Apache Point Observatory, P.O. Box 59, Sunspot, NM 88349\\
\and
Department of Physics, University of Helsinki, Gustaf H\"allstr\"omin katu 2a, FI-00014 Helsinki, Finland\\
\and
Energy Lancaster, Lancaster University, Lancaster LA14YB, UK\\
\and
Department of Physics, University of Surrey, Guildford GU2 7XH, UK\\
}

\date{Received Nov 16, 2017}

\abstract
{}
{We determine the stellar population properties - age, metallicity, dust reddening, stellar mass and the star formation history - for all spectra classified as galaxies that were published by the Sloan Digital Sky Survey (SDSS data release 14) and by the DEEP2 (data release 4) galaxy surveys.}
{We perform full spectral fitting on individual spectra, making use of high spectral resolution stellar population models. Calculations are carried out for several choices of the model input, including three stellar initial mass functions and three input stellar libraries to the models. We study the accuracy of parameter derivation, in particular the stellar mass, as a function of the signal-to-noise of the galaxy spectra. We find that at low redshift, a signal to noise ratio per pixel around 20 (5) allows a statistical accuracy on $\log_{10}(M^{*}/M_{\odot})$ of 0.2 (0.4) dex, for the Chabrier IMF.}
{%We obtain the galaxy stellar mass function probed by SDSS, eBOSS and DEEP2 for galaxies with $0.2<z<0.8$. 
For the first time, we study DEEP2 galaxies selected by their \OII luminosity in the redshift range $0.83<z<1.03$, finding that they are consistent with a flat number density in stellar mass in the range $10^9<M/M_{\odot}<10^{11.5}$. 
We find the resulting stellar mass function based on SDSS or eBOSS in agreement with previous studies
\citep[e.g.][]{Maraston2013}.  
We publish all catalogs of properties as well as model spectra of the continuum for these galaxies as a value added catalog of the fourteenth data release of the SDSS. This catalog is about twice as large as its predecessors (DR12) and will aid a variety of studies on galaxy evolution and cosmology.}
{}

\keywords{galaxy evolution - stellar population model - galaxy surveys }
\maketitle

%%%%%%%%%%%%%%%%%%%%%%%%%%%%%%%%%%%%%%%%%%%%%%%%%%
%%%%%%%%%%%%%%%%% BODY OF PAPER %%%%%%%%%%%%%%%%%%
\clearpage
\setcounter{secnumdepth}{2}
\setcounter{tocdepth}{2}
\tableofcontents
\clearpage
\section{Introduction}
\label{sec:introduction}

In the current paradigm of galaxy evolution, structures and galaxies form hierarchically: larger halos are formed by the coalescence of smaller progenitors. 
From a macroscopic or thermodynamical point of view, galaxies are typically described as systems composed by  tightly interacting sub-systems: a dark matter halo, a central black hole, stars, cold gas, hot gas and dust. 
The visible component of galaxies is approximated as a tri-phased system made of stars, inter-stellar medium and circum-galactic medium. A galactic system is governed by various processes, such as star formation rate, supernovae rate, the active galactic nuclei which - when present - is held to regulate gas dynamics, such as winds, accretion and outflows \citep{mo2010book}. The composition of stellar populations in galaxies is thus a key aspect of galaxy formation and evolution. 

The classical method to infer the stellar properties of galaxies (e.g. stellar ages, chemical composition, dust, the star formation history and the stellar mass) is to fit stellar population model spectra to the observed spectral energy distribution. Several input physics and parameters enter this approach and determine the resulting galaxy properties, namely: the stellar population model and its input physics (i.e. stellar evolution and atmosphere models); the wavelength range spanned by observations and models; the fitting method to compare models and data (e.g. statistics, priors, etc.). 
In this study we use the \citet{Maraston_2011} (M11 hereafter) stellar population models together with the \textsc{firefly} fitting routine \citep[hereafter FF][]{Wilkinson_2015,Goddard2017MNRAS.465..688G,Goddard2017MNRAS.466.4731G,firefly2017MNRAS}.
These code and models have been shown to be able to accurately reconstruct a galaxy star formation history from spectra with signal to noise ratios (SNR) of about 5 per pixel \citep[see, ][and Sec. \ref{subsec:firefly:performances}]{firefly2017MNRAS}. 
FF was used in the following recent stellar population studies using SDSS-IV MANGA integral field spectroscopy data:  \citet{Wilkinson_2015,Goddard2017MNRAS.465..688G,Goddard2017MNRAS.466.4731G,2018MNRAS.474.1143L,2018MNRAS.476.3883L,2018MNRAS.477.3954P}.

We perform model fitting to the optical spectra measured by the Sloan Digital Sky Survey (SDSS DR14 \citealt{SDSS_DR14}) and the DEEP2 survey (DR4 \citealt{Newman_2013}). 
We chose these two medium resolution optical spectroscopic surveys because they sample the observed magnitude vs. redshift plane in a complementary manner. 
Stellar population model catalogs are available for the SDSS DR12. These consist of stellar properties obtained from broad-band SED fitting \citep{Maraston2013} and emission-line properties by \citet{Thomas2013a}. In this work we extend those approaches by employing full spectral fitting and we extend the calculations to a substantially larger data set. 

The paper is organised as follows. The adopted set of observed spectra are presented in Section \ref{sec:DATA}. 
The \textsc{firefly} fitting routine and the stellar population models are described in Section \ref{sec:SPS}. 
We present the results obtained for the four main fitted parameters - namely stellar ages, chemical composition, dust reddening and stellar mass - in Section \ref{sec:results:params} and in Section \ref{sec:results}, we discuss the global results obtained for each set of observed spectra. 
Finally in Section \ref{sec:application:elg}, we calculate for the first time the stellar mass function of \OII emitters in DEEP2 and discuss the scientific implications of our findings.

Throughout the work we assume a standard flat $\Lambda$CDM \citep{Planck_2014} cosmology. 
The software is available at the official \textsc{Firefly} page\footnote{\url{http://www.icg.port.ac.uk/firefly/}} and through GitHub\footnote{\url{https://github.com/FireflySpectra/}}.
The results are available through the \textsc{firefly} data repository
\footnote{\url{https://firefly.mpe.mpg.de/v1_1_0/}}.

%%%%%%%%%%%%%%%%%%%%%%%%%%%%%%%
%%%%%%%%%%%%%%%%%%%%%%%%%%%%%%%
% DATA
%%%%%%%%%%%%%%%%%%%%%%%%%%%%%%%
%%%%%%%%%%%%%%%%%%%%%%%%%%%%%%%
% \clearpage
\section{Spectroscopic data}
\label{sec:DATA}

In this analysis, we consider galaxy spectra taken from both the SDSS and DEEP2 spectroscopic surveys. 
The redshift range spanned by the data is $0<z<1.7$. As we shall see most derived stellar masses are in the range $10^6 M_\odot$ to $10^{12.5} M_\odot$. 
These two surveys cover the parameter space of galaxy evolution in a complementary fashion. 
SDSS covers the most luminous galaxies over a wide area (order of 10,000 deg$^2$) and DEEP2 samples fainter galaxies (by about 2 magnitudes, over a much smaller area (order of 2 deg$^2$). More details are provided in the next subsections.

\subsection{SDSS}
We consider galaxy spectra obtained with either the SDSS or BOSS spectrograph \citep{2006AJ....131.2332G,Smee2013} as in the fourteenth data release \citep{dawson_2016,blanton_2017,SDSS_DR14}. 
The SDSS (BOSS) spectrographs cover 3800-9200\text{\AA} ($3650-10,400$\text{\AA}) at a resolution $1500$ at $3800$\text{\AA} and $2500$ at $9000$\text{\AA} with 3 (2) arc seconds diameter fibers.
Due to the variety of target selection algorithms successively applied to target galaxies within SDSS, the magnitude limit assumes different values. In the $i$-band, the various magnitude limits lie mostly within the range 17 to 22.5. 
After the twelth data release of SDSS, the program using the BOSS spectrograph was extended into the extended-BOSS (eBOSS) program. In the following we use eBOSS to designate spectroscopic data acquired with the BOSS spectrograph and released in the DR14.

For the stellar population fitting, we consider objects classified as galaxies following criteria used in previous SDSS galaxy products\footnote{\url{http://www.sdss.org/dr12/spectro/galaxy/}}.
We consider objects for which a definite positive redshift was derived using galaxy templates 
(CLASS=="GALAXY", $Z>Z_{ERR}>0$, $ZWARNING==0$) in the current redshift pipeline \citep[][version v5\_10\_0]{2012AJ....144..144B}. 
For the data obtained with the BOSS spectrograph, we consider the "NOQSO" version of these quantities. 
We finally retrieve about 2.7 million optical galaxy spectra, of which $948,259$ were observed with the SDSS spectrograph setup and with $1,759,362$ with the BOSS spectrograph setup. 
The SDSS (BOSS) spectrograph allows 640 (1000) fibers of 3 (2) arc-seconds diameter per plate to be plugged and covers the wavelength range 3,800-9,200 (3,600-10,400) Angstroms with two arms. 
% The BOSS spectrograph allows 1000 fibers of 2 arc-seconds diameter per plate to be plugged and covers the wavelength range 3,600-10,400 Angstroms with two arms.
%, see Table \ref{table:single:spectra}. 

\subsubsection*{Data source}

The full set of observed spectra we processed occupies about 0.8T of disk space.This data is available via the SDSS server,
\begin{itemize}
\item BOSS spectrograph data\\ \url{https://dr14.sdss.org/sas/dr14/eboss/spectro/redux/v5_10_0/spectra/PLATE/spec-PLATE-MJD-FIBERID.fits}
\item SDSS spectrograph data\\ \url{https://dr14.sdss.org/sas/dr14/sdss/spectro/redux/26/spectra/PLATE/spec-PLATE-MJD-FIBERID.fits}
\end{itemize}
The spectra are described here \url{https://data.sdss.org/datamodel/files/BOSS_SPECTRO_REDUX/RUN2D/spectra/PLATE4/spec.html}. 

\subsection{DEEP2}
DEEP2 is a deep pencil beam survey that acquired spectra for galaxies brighter than $R<24.1$ to study the evolution of galaxies. The survey is split in four fields that cover 2.7 deg$^2$  \citep{Newman_2013}. 
The DEIMOS spectrograph at Keck was used, which covers approximately the wavelength range $6500-9300$\text{\AA} at a resolution $\sim$6000 \citep{Faber2003}. It accommodates of the order 120 slits per mask. 
Although DEEP2 is a major galaxy evolution survey and stellar masses for galaxies observed by DEEP2 are mentioned in several publications, there does not seems to be a publicly available catalogue of stellar mass and other galaxy properties \citep{kassin2007,covington2010,mostek2013,2017ApJ...838...87C}. With our work we fill this gap.

In our analysis we consider galaxy spectra classified with a flag $Z\_FLG\geq2$ and whose redshift lies in the range $0.7<z<1.2$. This redshift range allows the sampling of the 4000\text{\AA} break which is needed for a robust recovery of stellar ages. Out of the $50,319$ entries in the DEEP2 DR4 catalog, our redshift cut selects $\sim~22,873$ unique objects. We further sort the data according to the detection of emission line in the spectrum. All data is detailed in Table \ref{table:single:spectra:deep2}. Finally, we use flux-calibrated spectra which were published by \citet{Comparat2016LFs}. 

The spectra used in this analysis were obtained via the DEEP2 server, here: \url{http://deep.ps.uci.edu/DR4/spectra.html}. 
The subset of processed flux-calibrated spectra are available here \url{https://firefly.mpe.mpg.de/v1_1_0/DEEP2/spectra}. 

%%%%%%%%%%%%%%%%%%%%%%%%%%%%%%%
%%%%%%%%%%%%%%%%%%%%%%%%%%%%%%%%
% model
%%%%%%%%%%%%%%%%%%%%%%%%%%%%%%%
%%%%%%%%%%%%%%%%%%%%%%%%%%%%%%%%
% \clearpage
\section{Obtaining the stellar population properties of galaxies}
\label{sec:SPS}

We adopt the fitting code \textsc{Firefly} \citep{firefly2017MNRAS} described in Sec. \ref{subsec:firefly} and
the stellar population models of \citet{Maraston_2011}, described in Sec. \ref{subsec:SPMmodels}.
We run the fitting procedure with different options for the Initial Mass Function (IMF) and input stellar library (see \ref{subsec:parameters}).
The details of the various runs are described in Sec. \ref{subsec:processing}. 
The \textsc{firefly} data-model generated by our analysis is described in Sec. \ref{subsec:public:data}.
We discuss the resulting stellar age, stellar metallicity and stellar mass distributions in Sec. \ref{sec:results:params}.

\subsection{The Firefly fitting routine}
\label{subsec:firefly}

\textsc{Firefly}
\footnote{\url{http://www.icg.port.ac.uk/firefly}} 
is a chi-squared minimization fitting code that for a given input observed Spectral Energy Distribution (SED), 
compares combinations of single-burst stellar population models (SSP), 
following an iterative best-fitting process controlled by the Bayesian Information Criterion (BIC) until convergence is achieved. 
An important feature of this code is that no priors - other than the assumed models and model grid - are applied, rather all solutions within a statistical cut are retained with their weight. The weight of each component can be arbitrary and no regularisation is imposed afterwards (as instead in pPXF, \citealt{Cappellari2004}). 
Attenuation by dust is accounted for in a novel way, which is fully described in \citet{firefly2017MNRAS} and summarised below. 
As dust attenuation has the effect at distorting the intrinsic continuum shape, the attenuation can be deduced after the intrinsic continuum shape is recovered. To this end we first rectify the continuum shape of both observations and models by multiplying them by a function, which is referred to as High-Pass Filter (HPF). 
This removes the large-scale modes of the spectra, i.e. those with a width larger than $\sim~100$\AA.  
We then find the stellar population model best-fitting the rectified data. 
The best-fit model essentially is the one that best matches the spectral absorption features. 
The comparison between the original (not rectified) observed spectrum and best-fitting model finally gives the attenuation. The flux difference between the two is used to obtain an attenuation array (i.e. flux ratios per wavelength). 
The returned attenuation array is then matched to known analytical approximations to return an E(B-V) value. 
Note that this procedure allows for removal of large scale modes of the spectrum associated with dust but also poor flux calibration \citep[see][]{Wilkinson_2015}.	
Finally, \textsc{Firefly} provides both light- and mass-weighted stellar population properties (age and metallicity), E(B-V) values and the stellar mass for the best fitting model and its single-burst components.  
Errors on these properties are obtained by the likelihood of solutions within the statistical cut. 

In summary, the fitting routine follows these steps: i) resolution of models and data are matched (by, usually, downgrading the models); ii) emission lines are masked; iii) dust attenuation is determined as described before; iv) the best fitting stellar population model is obtained as a linear combination of single-burst models; v) $\chi^2$ are converted into probabilities and average properties and errors (both mass weighted and light weighted) are calculated.
For full details we refer the readers to \citet{firefly2017MNRAS}. 
The code and the models used to create this dataset are public via the SDSS server:
\begin{itemize}
\item Fitting code. The official website of the \textsc{Firefly} team is \url{http://www.icg.port.ac.uk/firefly} links to the up-to-date version of the \textsc{firefly} software. 
\item Stellar population models: \url{https://svn.sdss.org/public/data/sdss/stellarpopmodels/tags/v1_0_2/} 
\end{itemize}
Note that we output the actual (present) stellar mass, and its fraction locked in stellar remnants (white dwarfs, neutron stars and black holes) or lost via stellar evolution (returned fraction). 

\subsubsection{Performances of Firefly and reliability of derived galaxy properties.}
\label{subsec:firefly:performances}
\citet{firefly2017MNRAS} have throughly investigated the performances of \textsc{Firefly} to recover stellar population properties such as age, metallicity, star formation history and stellar mass, as a function of several variables, in particular the SNR, the amount of reddening, the assumed star formation history, the wavelength range spanned by the data, using both mock galaxies with known input properties as well as real Milky Way star clusters with ages and metallicity determined independently (i.e. via CMD fitting for ages and resolved spectroscopy for metallicity, see \citealt{firefly2017MNRAS} for full details). Here we summarise those findings and refer to the paper for full details and plots.

Firefly is able to recover the stellar population properties (age, metallicity, star formation history and stellar mass) down to a SNR$\sim5$, for moderately dusty systems (E(B-V)<0.75) (see Figures 8-16 in \citealt{firefly2017MNRAS}). 
At SNR$\sim20$, the recovery of the star formation history is remarkably good independently of reddening, unless the star formation is very extended ($\sim$10 Gyr). Even at lower SNR down to SNR$\sim0.5$, stellar masses are well recovered (e.g. Figure~14) in some cases due to compensating errors in the derived age and metallicity. Spectral stellar masses are in agreement (within errors) with estimates based on SED fitting on broad band magnitudes. Indeed, \citet{firefly2017MNRAS} compared the \textsc{firefly}-SDSS determinations to the results from \citet{CidFernandes2005MNRAS358363C,Tojeiro2009ApJS1851T} (section 6.2, Figs. 24, 25) for galaxies from the SDSS-II.

Not surprisingly, the age-metallicity-dust degeneracy is more severe in very dusty systems (e.g. Figure~13) or where the age/metallicity degeneracy peaks (e.g. when the red giant Branch becomes an important contributor, hence around 1 Gyr). These degeneracies make individual stellar ages and stellar metallicities less well determined (e.g. Figure~12). 

Regarding the effect of the wavelength range spanned by the fitted data,
\citet{firefly2017MNRAS} demonstrated via mock experiments as well as fitting of star clusters with known ages from CMD fitting (section 5.2, Fig. 19) the well-known fact that the 4000\text{\AA}-break spectral region is needed in order to reliably constrain stellar ages. While SDSS+eBOSS spectra all contain this region, this is not the case for the entire DEEP2 database, in particular for DEEP2 galaxy spectra only covering the wavelength range 6500-9300\text{\AA}. This implies that accurate ages for DEEP2 galaxies can only be recovered if the redshift is in the window 3800(1+z)$>$6500 and 4200(1+z)$<$9300 i.e. $0.7<z<1.2$.

In both cases, a qualitative agreement was found and residual discrepancies were attributed to the priors set in those other codes.

Overall we conclude that Firefly is an excellent tool to try and recover galaxy properties in regimes of low S/N.

\subsection{\citet{Maraston_2011} models}
\label{subsec:SPMmodels}

We use the \citet{Maraston_2011} stellar population models, which are offered for a variety of input stellar libraries at the same energetics. In particular, we shall use M11 models including the following three main libraries:
\begin{itemize}
\item STELIB, covers 3200–9300\text{\AA} with a 3.4\text{\AA} sampling at 5500\text{\AA}, i.e. at a resolution $R=1617$, with 85 eigenvectors \citep{leborgne2003},  
\item MILES, covers 3500–7430\text{\AA} with a 2.54\text{\AA} sampling at 5500\text{\AA}, i.e. at a resolution $R=2165$, with 156 eigenvectors \citep{MILES_2006,MILES_2011,2011AA...531A.109B},
\item ELODIE, covers 3900–6800\text{\AA} with a 0.55\text{\AA} sampling at 5500\text{\AA}, i.e. at a resolution $R=10000$, with 132 eigenvectors \citep{Prugniel2007}.
\end{itemize}
The wavelength range spanned by the models define the redshift range where the model fitting is possible given the wavelength ranges spanned by the data. Recall that the instruments on SDSS, BOSS and DEIMOS cover the ranges $3800-9200$\text{\AA}, $3650-10,400$\text{\AA}, and $6500-9300$\text{\AA} at R$\sim$2000, 2000, 6000, respectively. The mismatch between the wavelength coverage of DEEP2 and the models explain the lack of fits at low redshift. It is part of an ongoing effort to complete the low-redshift extension of DEEP2 with near-IR extended models, which we shall release in the future.

The reason for performing such a large variety of spectral modelling without choosing one single model rendition is manifold. 
First of all, it is currently unclear which one of these Milky Way-based empirical libraries is more correct, where correctness means that when implemented in population models, the models are able to recover age and metallicity of known stellar systems. 
\citet{Maraston_2011} noted that M11-STELIB models were recovering most accurately the turnoff age of the solar-metallicity Milky Way star cluster W3. 
On the other hand, STELIB has a limited coverage in sub-solar metallicity, implying that STELIB-based models cannot be  fit to metal-poor Milky Way globular clusters, for which MILES-based models were giving a good fit \citep{Maraston_2011,firefly2017MNRAS}.
Moreover, galaxy evolution studies in the literature are based on models with different input libraries, hence a proper comparison of studies based on the latest SDSS data releases with those early works requires the availability of a variety of models. 
Indeed, while early SDSS-based galaxy evolution studies were based on STELIB-type models 
(by \citealt{BruzualCharlot2003}, as e.g. in \citealt{2003MNRAS.346.1055K} or \citealt{Tremonti_2004}), 
the data analysis of SDSS-IV/MaNGA is mostly based on MILES-type models (by M11, \citealt{2016MNRAS.463.3409V} or \citealt{2014ApJ...780...33C}). 
In addition, we explore the ELODIE-based models which have not been used much in the literature, but allows a higher spectral resolution.

\subsection{Input parameters}
\label{subsec:parameters}

We perform spectral model fitting for each input M11 model described above (namely, M11-ELODIE, M11-MILES and M11-STELIB), and for  
three choices of the stellar initial mass function (IMF), namely Salpeter, \citet{Salpeter_1955},  
Chabrier, \citet{Chabrier2003} and Kroupa, \citet{Kroupa2001}, for each of these models.
Hence, in total we provide up to nine modelling of the continuum for each galaxy (depending on whether the parameter 'stellar mass' results to be constrained). 

The models span ages in the range $10^{6}<Age [yr]< 2 \times 10^{10}$ and metallicities in the range $-3<\log_{10}(Z/Z_\odot)<3$, with each M11 model spanning a different age/metallicity grid \citep[cf.][Table1]{firefly2017MNRAS}. Recalling, the total number of ages available for each model flavour are: 85 for M11-STELIB, 132 for M11-ELODIE and 156 for M11-MILES. 

The E(B-V) parameter boundaries are 0-0.7. 

\begin{figure*}
\begin{center}
\caption{\label{fig:firefly:output:sdss}
Example of a fit for a galaxy spectrum randomly taken from a SDSS plate, namely plate=0266, mjd=51602, fiberid=4. This galaxy lies at redshift 0.127. It has a S/N around the 4000A break of 10.75. 
The {\bf top panel} shows the observed spectrum (grey line) and its modelling (coloured lines). 
Each model fit is characterized by the number of SSPs given in the caption. Gaps in the spectrum correspond to regions masked due to the presence of emission lines. 
The {\bf second panel} shows the $\chi^2$ distribution per pixel for each model compared to a normal distribution (labeled N(0,1), dashed line).
The {\bf third and fourth rows} of panels show the derived galaxy parameters and their $1\sigma$ uncertainties. 
On the {\bf third row} from left to right: mass-weighted age vs. mass-weighted metallicity (Panel 1), mass-weighted age vs. stellar mass (Panel 2) and SSP weights vs. mass-weighted age of the individual SSP components (Panel 3). 
On the {\bf fourth row} from left to right: Panel 1 shows mass vs. mass-weighted metallicity, Panel 2 shows mass vs. reddening and Panel 3 shows the SSP weights vs. mass-weighted metallicities of the individual SSP components. 
Results consistently point towards a super-solar mass-weighted metallicity of 10$^{0.2}Z_\odot$, an old mass-weighted age of 10$^{10}$yr, an E(B-V) between 0.15 and 0.2. and a stellar mass about 10$^{10.75}$ -- 10$^{11}M_\odot$. 
STELIB-based models give the most massive and older solution than MILES and ELODIE.
The decomposition in SSPs shows the solution is constituted of two old bursts, one with solar metallicity and one with higher metallicity in comparable proportions.}
\includegraphics[width=14cm]{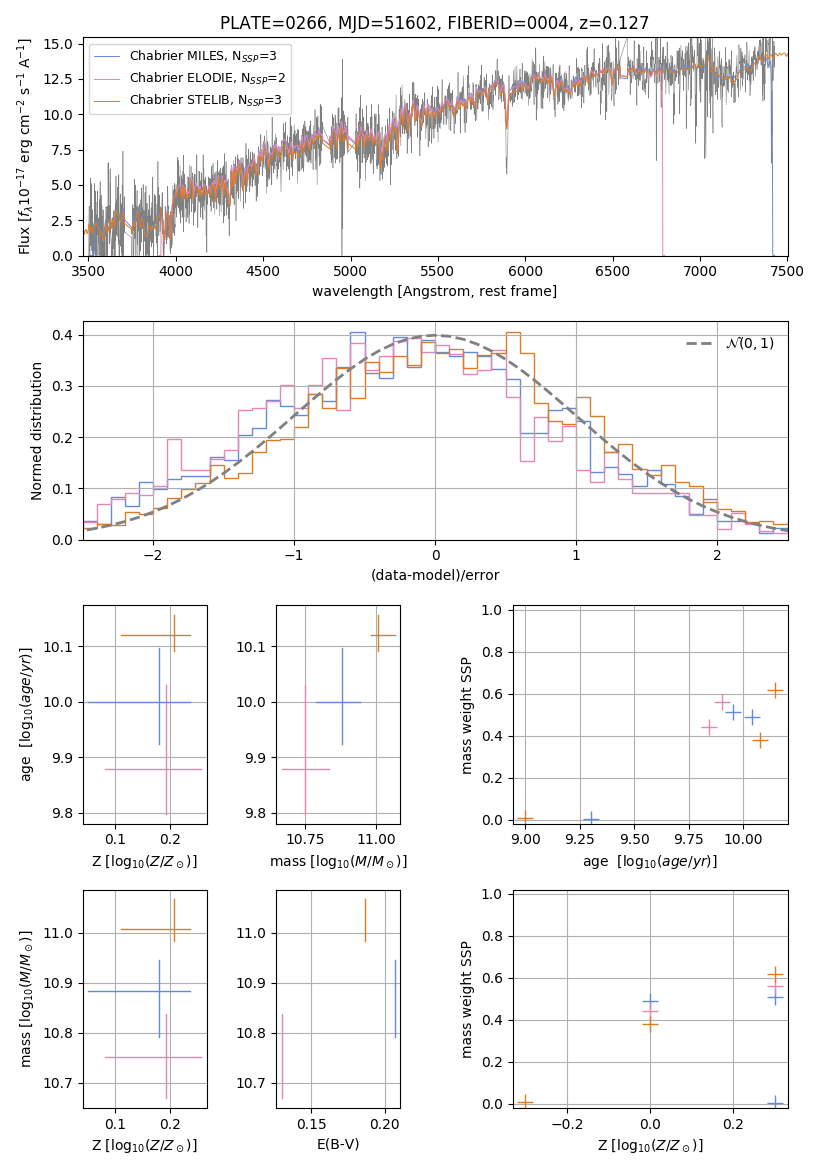}
\end{center}
\end{figure*}

\begin{figure*}
\begin{center}
\caption{\label{fig:firefly:output:boss}
Same as Fig. \ref{fig:firefly:output:sdss} for a galaxy spectrum randomly taken from the first BOSS plate, namely plate=3586, mjd=55181, fiberid=3, featuring a galaxy at redshift 0.588. 
It has a S/N around the 4000A break of 3.1.
The results point overall towards a metallicity around solar, an age larger than a few billion years, a larger reddening and a stellar mass around 10$^{11}M_\odot$. 
The fit is dominated by a single old SSP with solar metallicity. 
The uncertainties on the derived parameters are larger than in the SDSS case of Figure~1.}
\includegraphics[width=14cm]{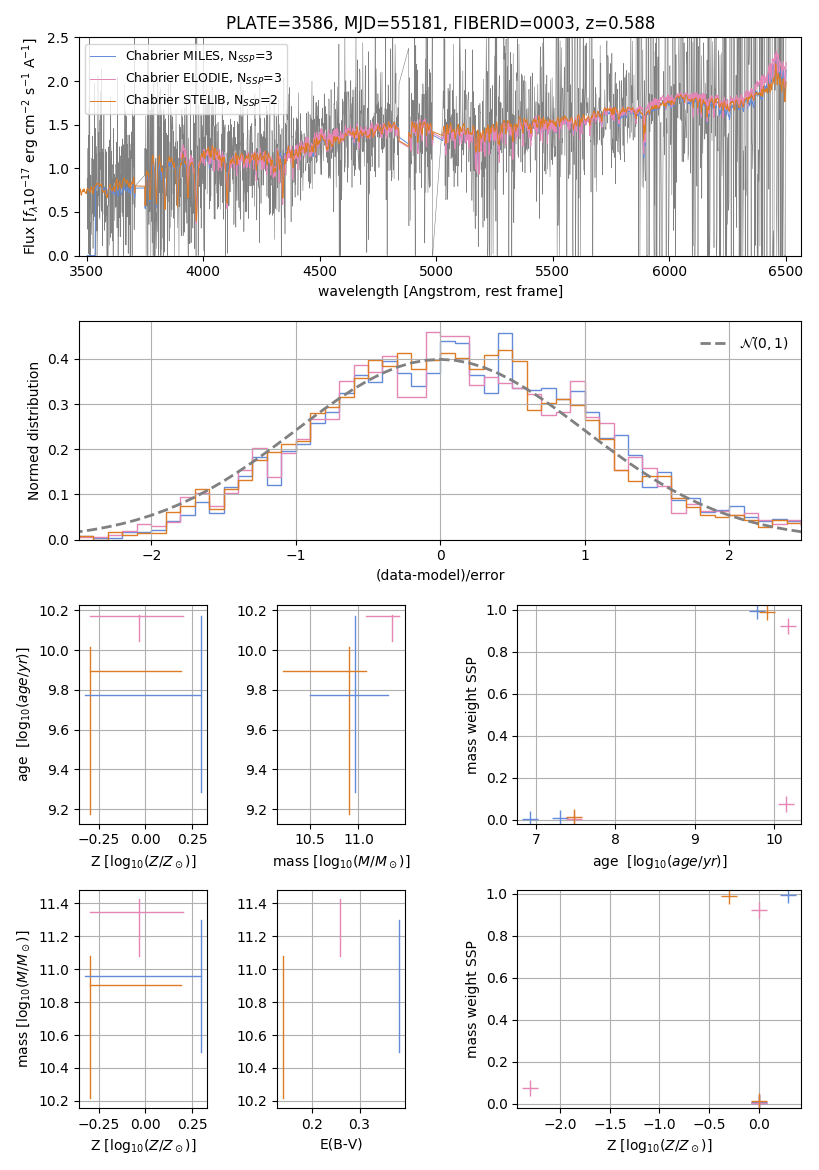}
\end{center}
\end{figure*}

\begin{figure*}
\begin{center}
\caption{\label{fig:firefly:output:deep2}
Same as Fig. \ref{fig:firefly:output:sdss} for the DEEP2 spectrum at mask=1103 and objno=11013914. 
It features a galaxy at redshift 0.784.
It has a median S/N of 5.4.
The results point overall towards a metallicity value of 10$^{-1}Z_\odot$, an age of 10$^{9.9}$yr, E(B-V) between 0.2 and 0.6 and a mass around 10$^{11}M_\odot$. 
The decomposition in SSPs with the ELODIE library suggest a combination of a young and an old SSPs both with sub solar metallicities. }
\includegraphics[width=14cm]{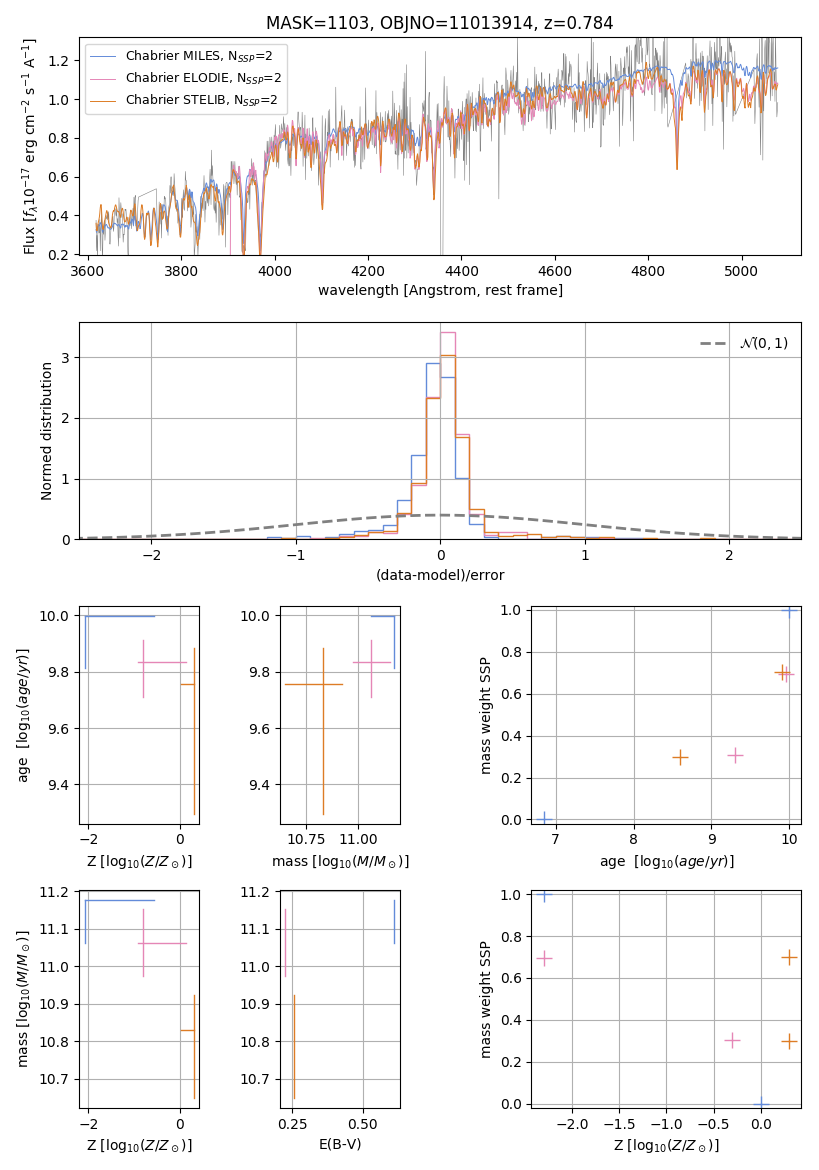}
\end{center}
\end{figure*}

Examples of spectral fitting results for a SDSS, an eBOSS and a DEEP2 galaxy spectrum are given in Figures \ref{fig:firefly:output:sdss}, \ref{fig:firefly:output:boss}, \ref{fig:firefly:output:deep2}, where we show the observed spectrum, the fitted models, the distribution of residuals and the main parameters derived. As the assumed IMF has little influence on the spectral modelling, mostly resulting in well-known offsets in the derived stellar mass \citep[e.g.][]{2012MNRAS.422.3285P}, for clarity we show the results for one IMF and the three model libraries.

\clearpage
\section{Resulting galaxy parameters}
\label{sec:results:params}

In this section we describe the results of the fitting procedure, present statistics on the four main considered parameters.
%: stellar age and metallicity (Sec. \ref{subsec:res:age:metal}), dust and stellar mass (Sec. \ref{subsec:res:dust:mass}).
\localtableofcontents

\subsection{Ages and metallicities}
\label{subsec:res:age:metal}
\begin{figure}
\begin{center}
\caption{\label{fig:distributions:AZ} 
2-D age/metallicity histograms for the SDSS (top row), eBOSS (middle row) and DEEP2 (bottom row) data in the 3 library setups, ELODIE, MILES and STELIB (rows from left to right) for the Chabrier IMF. 
These show how each library samples the age-metallicity plane. 
The difference between surveys come from the different redshift range, but also from targeting different type of galaxies.
Note that the axis of the colour maps are different for DEEP2.}
\includegraphics[height=5.5cm]{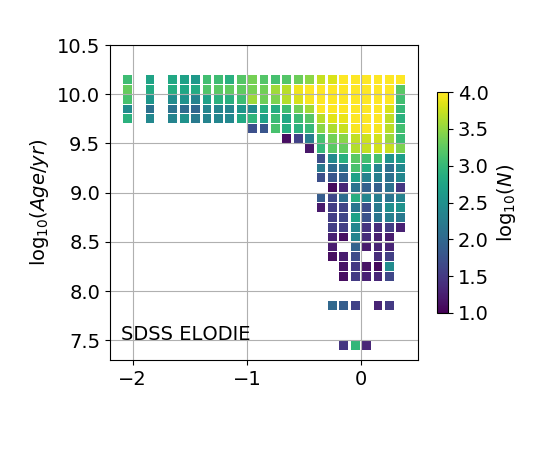}
\hspace*{-1.78cm}
\includegraphics[height=5.5cm]{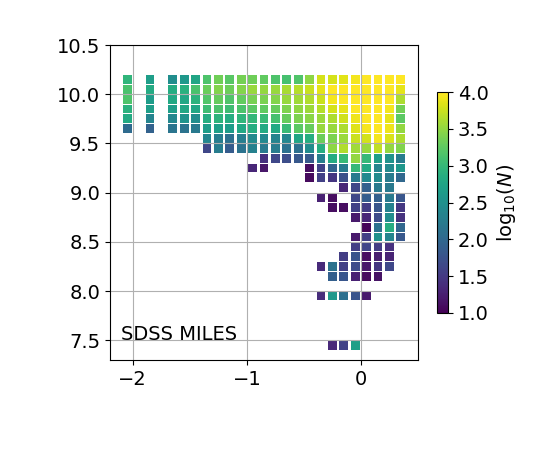}
\hspace*{-1.78cm}
\includegraphics[height=5.5cm]{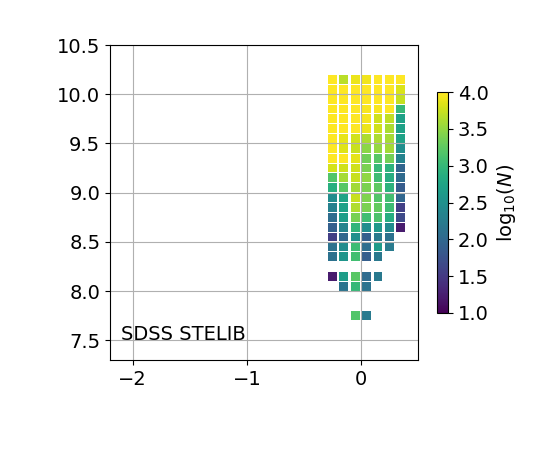} \\
\vspace*{-1.1cm}
\includegraphics[height=5.5cm]{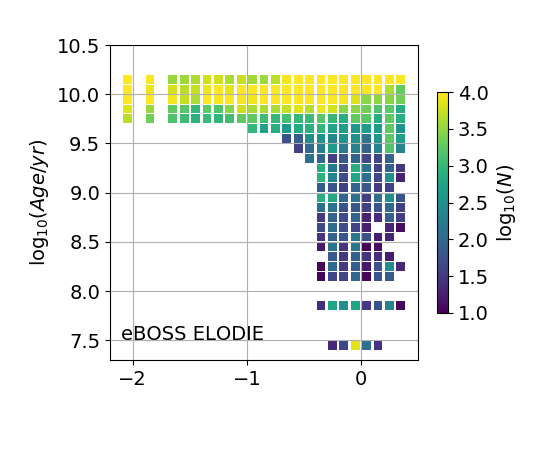}
\hspace*{-1.78cm}
\includegraphics[height=5.5cm]{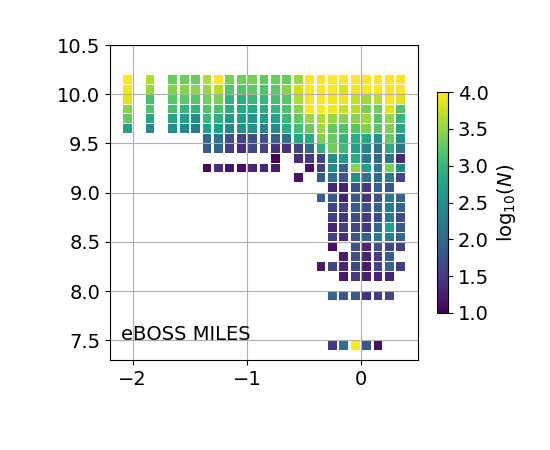}
\hspace*{-1.78cm}
\includegraphics[height=5.5cm]{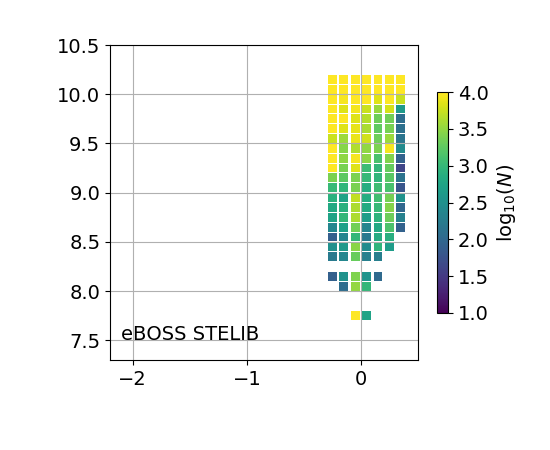}\\
\vspace*{-1.1cm}
\includegraphics[height=5.5cm]{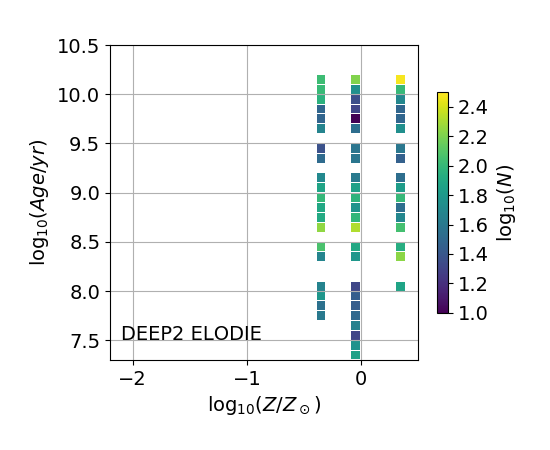}
\hspace*{-1.78cm}
\includegraphics[height=5.5cm]{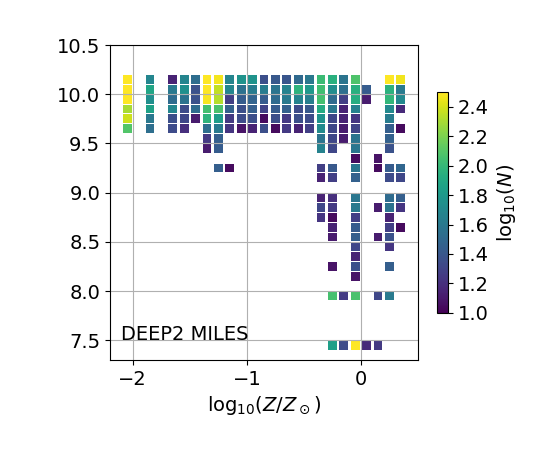}
\hspace*{-1.78cm}
\includegraphics[height=5.5cm]{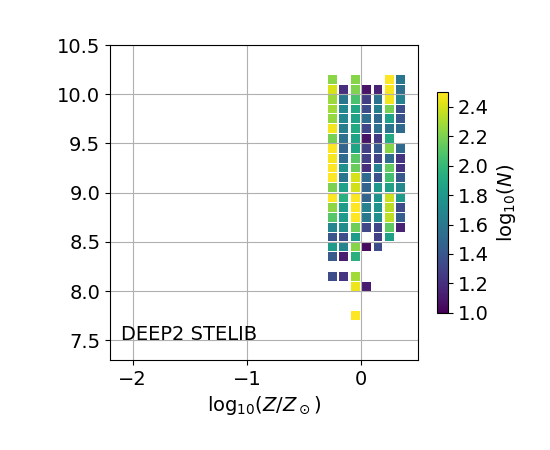}
\end{center}
\end{figure}

We show in Fig. \ref{fig:distributions:AZ} how each dataset occupies the input parameter space of ages and metallicities. SDSS galaxies mostly occupy the space of old and metal rich populations at low redshift ($z\sim0.1$). eBOSS due to its higher mean redshift ($z\sim0.5$) samples slightly younger ages and slightly lower metallicities. DEEP2 occupies the complementary space of lower ages and metallicities at higher redshifts ($z\sim0.8$). In the DEEP2 case, it seems that a certain number of fits occupy the lowest metallicity edge of the parameter space, meaning these values may be not very well constrained. The young ages found for a fraction of galaxies using STELIB models might just be an artefact of the models 

Regarding the input stellar library, the age/metallicity distributions obtained using models based on the MILES and the ELODIE libraries are overall similar, as already discussed on a smaller sample by \citet{firefly2017MNRAS}. For the eBOSS sample (second row) lying at higher redshift and based on a different target selection, ELODIE-type models give a distribution of metallicities stretching towards sub-solar values while MILES-based models remain more concentrated at solar metallicities. The STELIB library grants a smaller coverage in metallicity hence model results are confined within half-solar and twice-solar in chemical composition. Because of the smaller range of input metallicity and due to the age metallicity degeneracy, the range in ages found using STELIB-based models is larger and extend to younger ages with respect to the other two models. These results highlight how much galaxy evolution findings depend on the assumed model frame used to interpret galaxy spectra.
  
Quantitative comparisons between the obtained ages and metallicities are shown in Figures \ref{fig:distributions:MwA} and \ref{fig:distributions:MwZ}, described in the next paragraphs.

\subsubsection{Stellar age}
\label{subsec:res:age}

\begin{figure}
\begin{center}
\caption{\label{fig:distributions:MwA} 
Normed distribution (area is equal to 1) of the difference in age weighted by its $1\sigma$ uncertainty obtained when assuming different libraries and fixed IMF to Chabrier for the DEEP2 (left), the SDSS (middle) and eBOSS (right) samples. 
The difference between ages obtained with the MILES and the ELODIE is shown with a blue solid line. 
The difference between ages obtained with the STELIB and the ELODIE is shown with a pink solid line. 
This Figure quantifies the differences seen qualitatively between libraries in Fig. \ref{fig:distributions:AZ}. 
The Gaussian distribution (orange dashed line) shows the width of the distribution expected if the ages derived agree within errors. 
MILES and ELODIE ages are in good agreement but STELIB and ELODIE ages disagree.}  
\includegraphics[height=5.6cm]{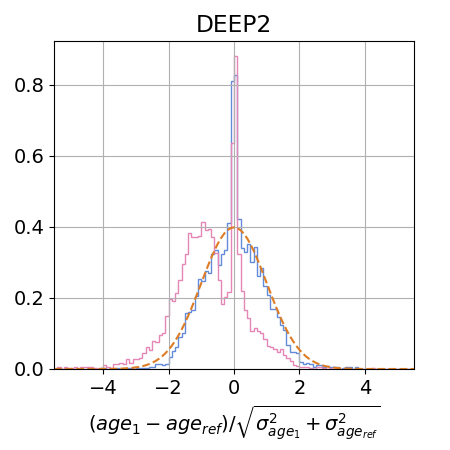}
\hspace*{-0.5cm}
\includegraphics[height=5.6cm]{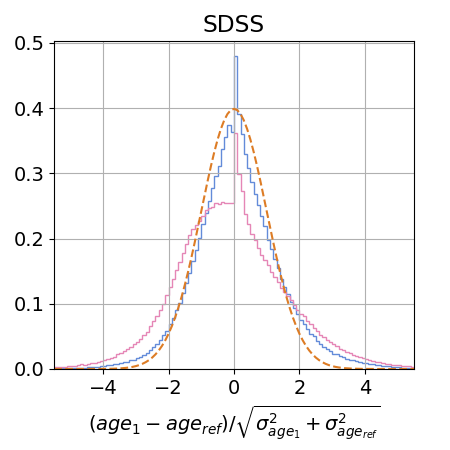}
\hspace*{-0.5cm}
\includegraphics[height=5.6cm]{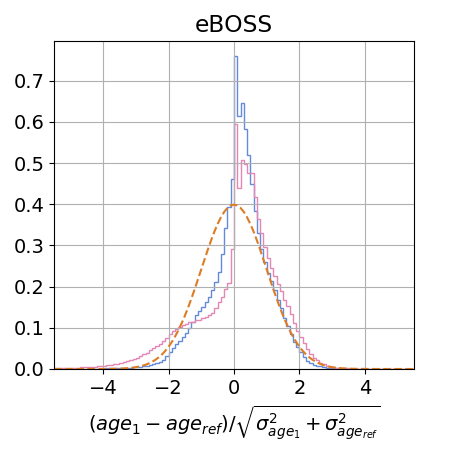}
\end{center}
\end{figure}

We find that - at fixed stellar library - variations in the IMF assumed in the models give ages that remain  always consistent within $1\sigma$. On the contrary, at fixed IMF, variations in the input stellar library produce the largest difference among models. While the ages obtained with MILES and ELODIE-type models (at a fixed Chabrier IMF) are consistent within $1\sigma$ with each other in both the SDSS and eBOSS samples, those
obtained with STELIB are discrepant from the other two sets of models due to the small coverage in metallicity of the library. 
Fig. \ref{fig:distributions:MwA} shows the distribution of age difference normed by their $1\sigma$ uncertainties when varying the input libraries, ELODIE, MILES and STELIB (at fixed Chabrier IMF). 
Absolute values of age for the sample galaxies are presented in Section \ref{sec:results}. 

\subsubsection{Stellar metallicity}
\label{subsec:res:metal}

\begin{figure}
\begin{center}
\caption{\label{fig:distributions:MwZ} 
Same as Fig. \ref{fig:distributions:MwA} for the stellar metallicity.
The difference between metallicities obtained with the MILES and the ELODIE is shown with a blue solid line. 
The difference between metallicities obtained with the STELIB and the ELODIE is shown with a pink solid line. 
The Gaussian distribution (orange dashed line) shows the width of the distribution expected if the ages derived agree within errors. 
}  
\includegraphics[height=5.6cm]{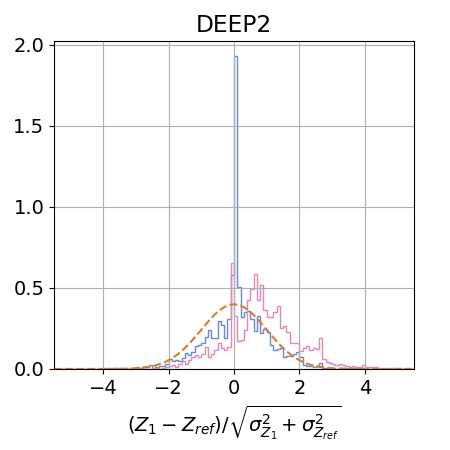}
\hspace*{-0.5cm}
\includegraphics[height=5.6cm]{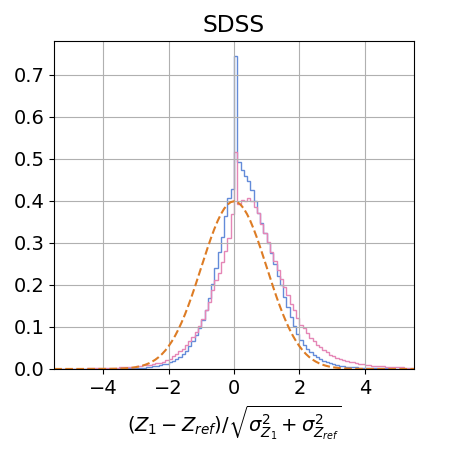}
\hspace*{-0.5cm}
\includegraphics[height=5.6cm]{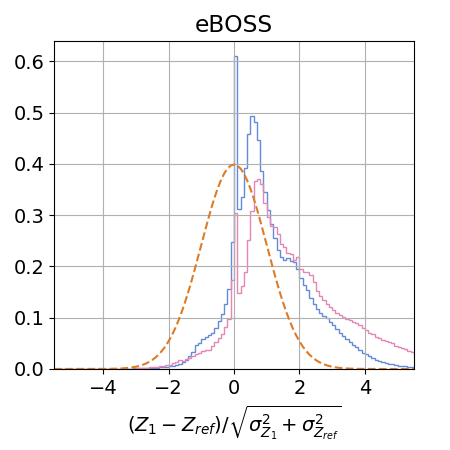}
\end{center}
\end{figure}

At fixed IMF, the metallicities derived with models based on different stellar libraries are not always consistent, but the result also depends on the galaxy sample. 
Fig. \ref{fig:distributions:MwZ} shows the distribution of the difference of metallicity normed by the uncertainty on the parameter when varying the libraries: ELODIE, MILES, STELIB (for a Chabrier IMF).
In case of the SDSS sample, metallicities are nearly consistent. The distribution shows a small systematic shift, but the overall width is comparable to that of a Gaussian. 
For the DEEP2 sample, errors are very large, hence the distribution seems consistent, but the parameter metallicity is quite unconstrained.
For the eBOSS sample, on the other hand, there are systematic deviations when comparing either MILES and ELODIE or MILES and STELIB. The distribution is broader than the Gaussian meaning there is a disagreement between the parameter estimation. This is the case also for the MILES vs ELODIE comparison, even if the two stellar libraries offer a similar metallicity coverage. Hence, the disagreement is not just simply due to the different coverage in input parameters.
We thus warn the future user to be cautious when using the metallicity parameters fitted on the eBOSS data. 
Absolute values of metallicity for the sample galaxies are presented in Section \ref{sec:results}. 

\subsection{Dust and stellar masses}
\label{subsec:res:dust:mass}
\begin{figure}
\begin{center}
\caption{\label{fig:distributions:DM} 
Dust - stellar mass 2d-histograms for the SDSS (top row) eBOSS (middle row) and DEEP2 (bottom row) data in the 3 library setups: columns ELODIE, MILES, STELIB for the Chabrier IMF.} 
\includegraphics[height=5.5cm]{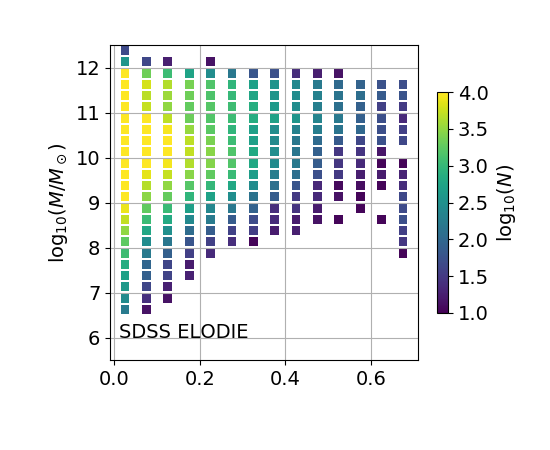}
\hspace*{-1.7cm}
\includegraphics[height=5.5cm]{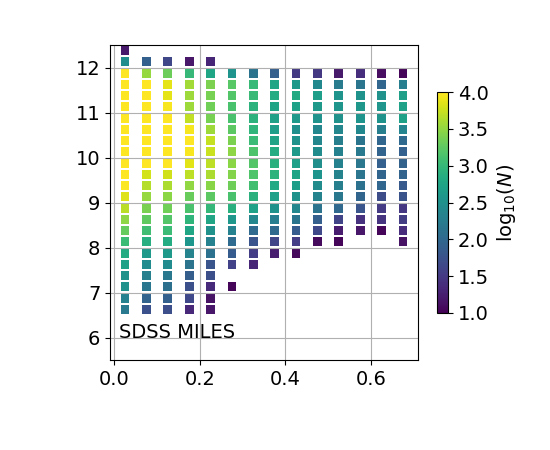}
\hspace*{-1.7cm}
\includegraphics[height=5.5cm]{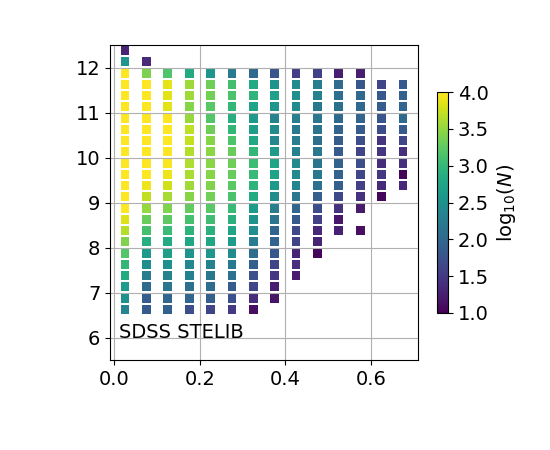} \\
\vspace*{-1.1cm}
\includegraphics[height=5.5cm]{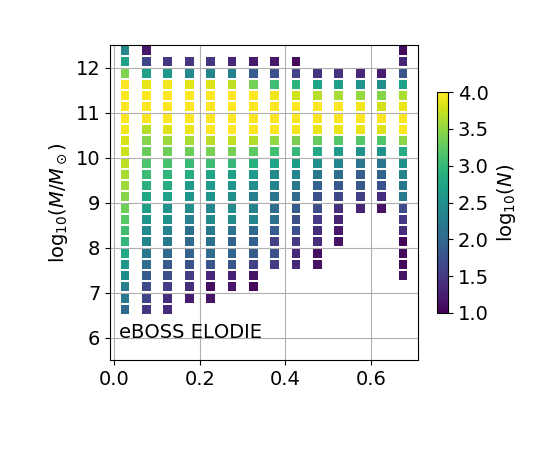}
\hspace*{-1.7cm}
\includegraphics[height=5.5cm]{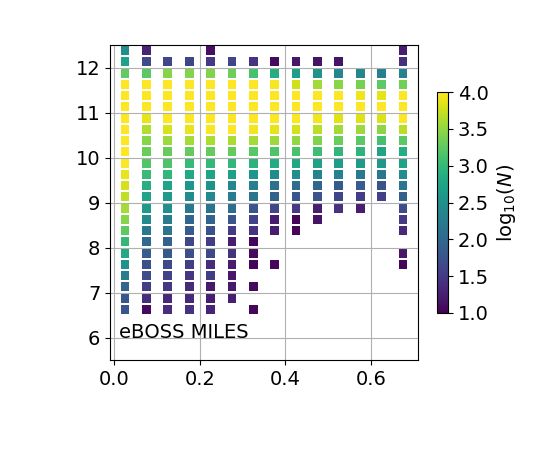}
\hspace*{-1.7cm}
\includegraphics[height=5.5cm]{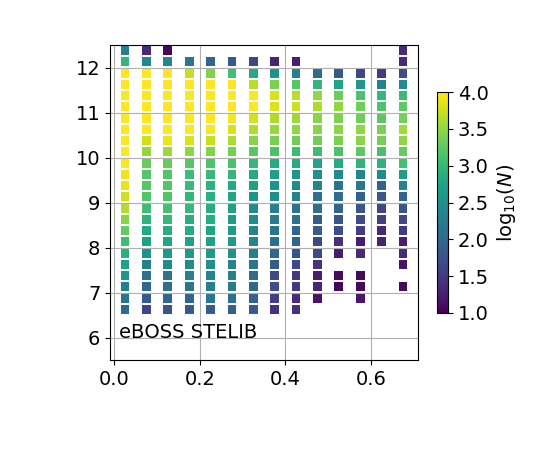}\\
\vspace*{-1.1cm}
\includegraphics[height=5.5cm]{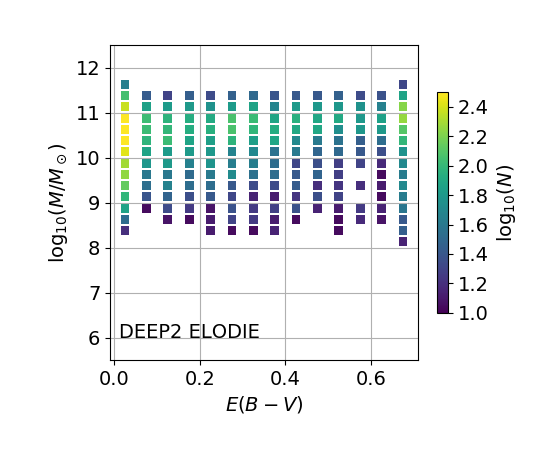}
\hspace*{-1.7cm}
\includegraphics[height=5.5cm]{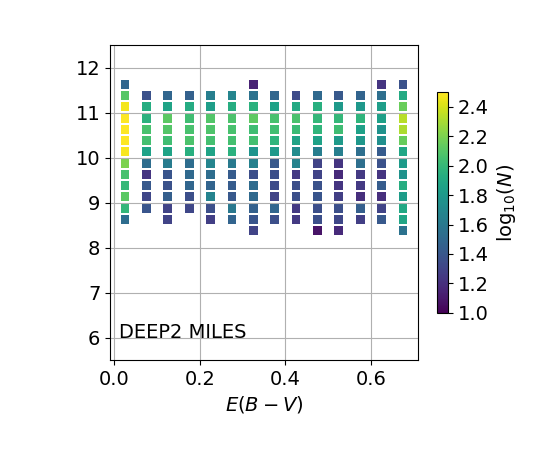}
\hspace*{-1.7cm}
\includegraphics[height=5.5cm]{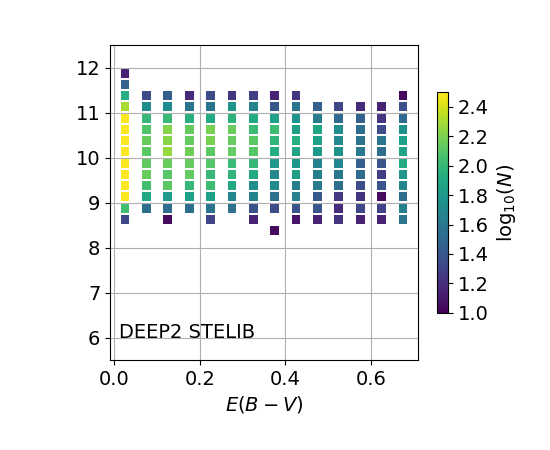}
\end{center}
\end{figure}

Similarly to the age-metallicity degeneracy there exists a dust-age degeneracy which creates a dust-mass degeneracy. 
Both a young and very dusty model or an old dust-free model can provide a good description of the same spectrum \citep{2006ARAA..44..141R}. 
Thus for an old and dust free galaxy, its stellar mass could be significantly underestimated \citep{2012MNRAS.422.3285P}. 
In specific cases, like \citet{Maraston2013}, where they analyzed the sample of massive galaxies in SDSS-DR12, they assumed dust-free models to avoid this bias. 
Indeed, these galaxies should not be dominated by a young stellar population. 
In our analysis, because the galaxy populations considered cover a large redshift range and a variety of different galaxy types, we choose to fit for dust in the analysis. 
In Fig. \ref{fig:distributions:DM} we show how each dataset occupies the input parameter space of dust and stellar mass. 
The distributions obtained with ELODIE and MILES are similar, with a tendency for MILES-based models to find a larger fraction of dusty galaxies at low stellar masses ($M<10^9~M_{\odot}$). 
Fits done with the M11-STELIB models favour a lower dust attenuation for the eBOSS sample due to the lack of low metallicity coverage in these models. 
DEEP2 galaxies seem to cluster on the low E(B-V) side, although this may be due to the small wavelength coverage of the sample.
Note that due to a larger volume and a preselection to include only massive galaxies, eBOSS seems to contain more massive galaxies than SDSS although the mean redshift is greater. 
We directly compare the stellar mass and E(B-V) values obtained in the next two paragraphs and on Figs. \ref{fig:distributions:EBV} (E(B-V)), \ref{fig:comparison:imfs:MS} (stellar mass), \ref{fig:distributions:MwM} (stellar mass).

\subsubsection{Dust}
\label{subsec:res:dust}
\begin{figure}
\begin{center}
\caption{\label{fig:distributions:EBV} 
Distribution of the difference of E(B-V) when varying the libraries: ELODIE, MILES, STELIB (fixed IMF: Chabrier).
Due to different wavelength coverage in each library, the E(B-V) values measured differ by up to 0.2 dex.
The difference between E(B-V) obtained with the MILES and the ELODIE is shown with a blue solid line. 
The difference between E(B-V) obtained with the STELIB and the ELODIE is shown with a pink solid line. 
}  
\includegraphics[width=6cm]{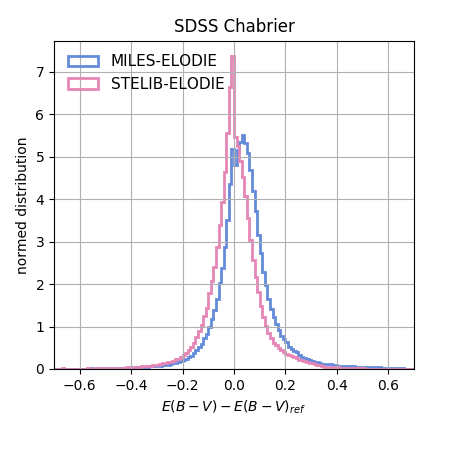}
\hspace*{-0.6cm}
\includegraphics[width=6cm]{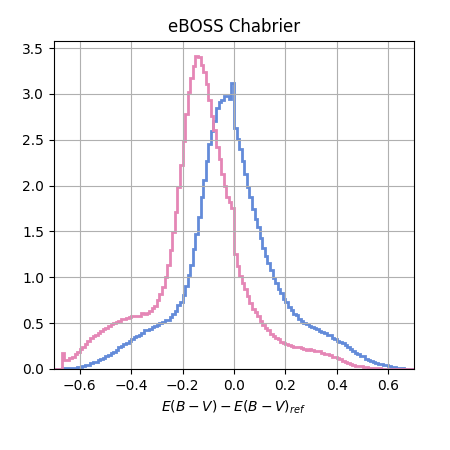}
\hspace*{-0.6cm}
\includegraphics[width=6cm]{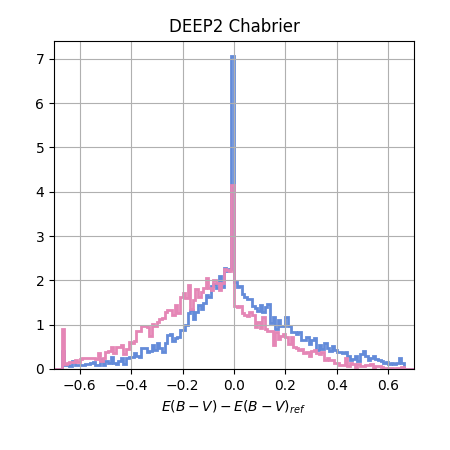}
\end{center}
\end{figure}

% The dust is fitted first on the data and then assumed when fitting for the combination of SSPs. 
The assumed library induces a noticeable difference in the retrieved E(B-V). 
Fits done with M11-STELIB (largest wavelength coverage) provide on average lower values of attenuation than those based on M11-MILES and M11-ELODIE (smallest wavelength coverage), see Fig. \ref{fig:distributions:EBV}. 
SDSS and eBOSS have a larger wavelength coverage than the models so that the models are the limiting factor. 
DEEP2 has a limited wavelength coverage so this directly impacts the ability of the software to constrain E(B-V) and in most cases it is 0.
As expected, the IMF has no impact on the derived $E(B-V)$.%, see Fig. \ref{fig:distributions:EBV2}. 

\subsubsection{Stellar mass}
\label{subsec:mass}
\begin{figure}
\begin{center}
\caption{\label{fig:comparison:imfs:MS} 
Comparison of stellar masses obtained at the fixed M11-MILES model and varying the IMF, for the three samples: SDSS (left); eBOSS (middle); DEEP2, (right). 
Stellar masses are systematically different, as expected. 
Kroupa is systematically less massive than Chabrier, and both Kroupa and Chabrier are systematically less massive than Salpeter.
The difference between stellar masses obtained with the Kroupa and the Chabrier is shown with a blue solid line. 
The difference between stellar masses obtained with the Salpeter and the Chabrier is shown with a pink solid line. 
}
\includegraphics[height=5.6cm]{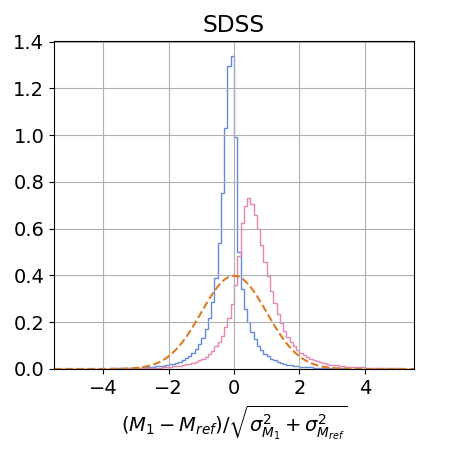}
\includegraphics[height=5.6cm]{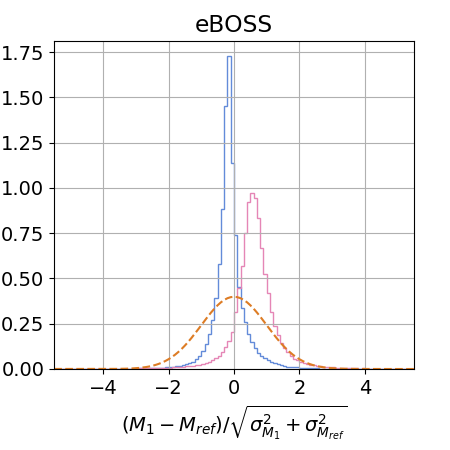}
\includegraphics[height=5.6cm]{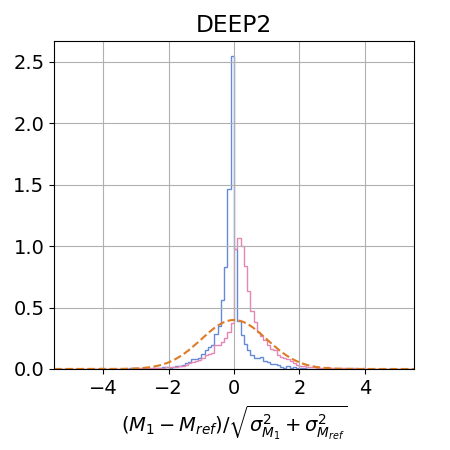}
\end{center}
\end{figure}

\begin{figure}
\begin{center}
\caption{\label{fig:distributions:MwM} 
Global comparison of stellar masses obtained with varying the assumed model (at a fixed Chabrier IMF). 
If the obtained distribution is narrower than a normal distribution (depicted with the line 'N(0,1)') then measurements agree within errors. 
The difference between stellar masses obtained with the MILES and the ELODIE is shown with a blue solid line. 
The difference between stellar masses obtained with the STELIB and the ELODIE is shown with a pink solid line. 
The difference between stellar masses obtained with the STELIB and the MILES is shown with a orange solid line. 
}  
\includegraphics[height=5.6cm]{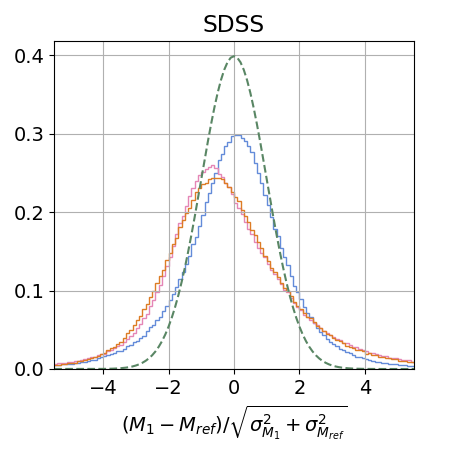}
\hspace*{-0.5cm}
\includegraphics[height=5.6cm]{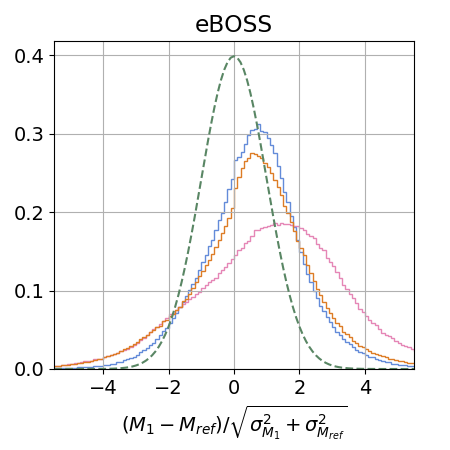}
\includegraphics[height=5.6cm]{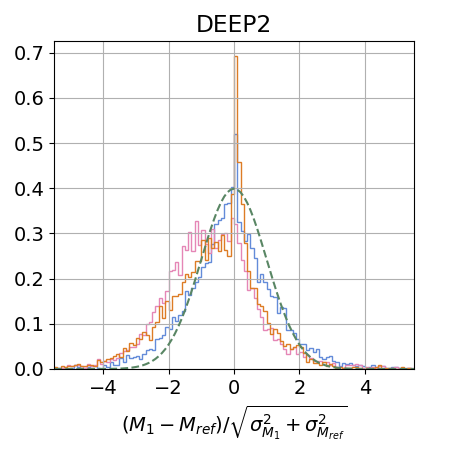}
\hspace*{-0.5cm}
\end{center}
\end{figure}

Stellar masses obtained with the Kroupa IMF are systematically smaller than those obtained with either a Chabrier or a Salpeter IMF.  
Fig. \ref{fig:comparison:imfs:MS} shows the difference in stellar masses obtained when initial mass functions are varied, for the fixed M11-MILES models. The systematic difference in the inferred masses due to the assumed IMF is larger than the statistical uncertainty on stellar mass. 
Fig. \ref{fig:distributions:MwM} shows the difference in the stellar masses obtained when the stellar library is varied. 
It shows that changing library also induces systematic changes in the stellar mass that are larger than the statistical uncertainty on the stellar mass. 
The masses obtained with the Salpeter (Kroupa) IMF are on average 0.08 to 0.1dex (-0.02dex) more (less) massive than with the Chabrier IMF.

Fig. 10 summarises our analysis by assessing the consistency of all derived parameters simultaneously (at fixed IMF) when varying input library. We find that models based on either MILES and ELODIE give statistically consistent sets of galaxy parameters in case of SDSS and DEEP2 while for the eBOSS sample, the parameters derived with the various models are not consistent. Likewise, parameters derived using STELIB and ELODIE models are not consistent independently of the galaxy sample.

\subsection{Star Formation history}
\label{subsec:history}
In cases of high signal to noise ($>20$), the star formation history can be reconstructed. Fig. \ref{fig:firefly:output:sdss} shows a low redshift galaxy observed at high signal to noise. 
The third panel on the third row shows the mass weight of each SSP. 
It indicates that this galaxy spectrum could be explained by the combination of two 10 Gyr old components with a solar metallicity contributing to more than 90\% of the mass and a $\sim1.5$ Gyr old component with a sub solar metallicity contributing to the remaining 10\%.
We show how SSPs components are distributed in parameter space in Sec. \ref{sec:results}. 
%the distribution of SSP weights for all fits on Fig. \ref{fig:SHF:all}.

\clearpage
\section{Results}
\label{sec:results}

\begin{figure*}
\begin{center}
\caption{\label{fig:nz}Galaxy redshift distributions, where galaxies with stellar mass constrained within $\pm0.2$dex are shown as solid lines, and within $\pm0.4$dex as dashed lines. 
Stellar masses used here are obtained using M11-ELODIE and refer to the Chabrier IMF. }
\includegraphics[width=16cm]{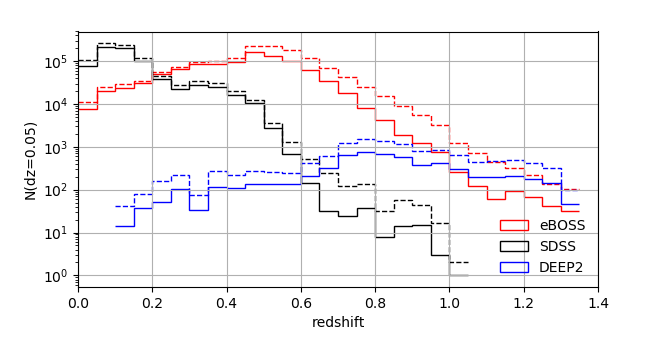}
\end{center}
\end{figure*}

\begin{figure*}
\begin{center}
\caption{\label{fig:MZ}Stellar mass vs redshift distributions for the three considered galaxy samples distinguishing between 
those whose stellar mass is constrained within $\pm0.2$dex (upper panels) and $\pm0.4$dex (lower panels). Stellar masses shown here are obtained using M11-ELODIE and refer to the Chabrier IMF. } 
\includegraphics[height=7.5cm]{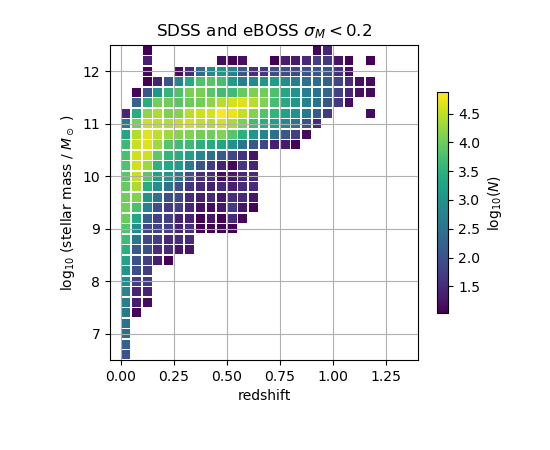}
\hspace*{-0.5cm}
\includegraphics[height=7.5cm]{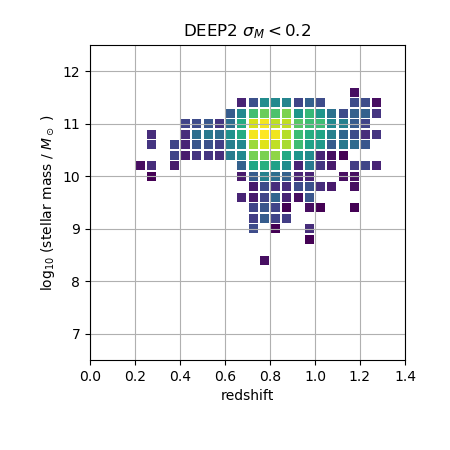} \\
\hspace*{-0.5cm}
\includegraphics[height=7.5cm]{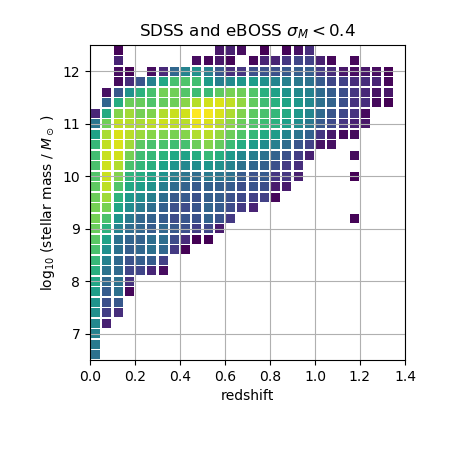}
\hspace*{-0.5cm}
\includegraphics[height=7.5cm]{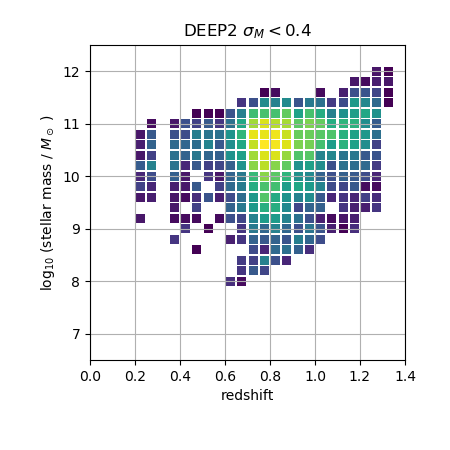}
\end{center}
\end{figure*}

Our results span a large range in redshift ($0<z<1.3$) and stellar mass ($7<\log M^{*}/M_{\odot}<12$) and can be used for a variety of galaxy evolution studies. 
Fig. \ref{fig:nz} shows the redshift distribution of the galaxies with determined stellar properties.
Fig. \ref{fig:MZ} shows their distribution in the stellar mass vs redshift plane. Detailed figures for the plotted quantities are given in Tables~1 and 2.

In the following, we first describe the overall results, e.g. number of galaxy with fitted stellar properties, redshift distribution and parameter - age, metallicity and stellar mass - distributions. We then discuss how the uncertainty of the fitted parameters is related to observational parameters.

%\localtableofcontents

\subsection{SDSS and eBOSS}

The SDSS+eBOSS specObjAll file contains 1,843,200 + 3,008,000 spectra, respectively, among which 950,705 and 1,759,362 are classified as galaxies. Among the latter, 948,259 (99.7\%) and 1,759,362 (100\%) could be run through our model spectral fitting, 
% why DEEP2 here?? this is the section on SDSS
%and 34,851 (DEEP2) spectra, 
see Table \ref{table:single:spectra}. 
The small fraction of missing data in SDSS is due to the fact that we only use the data from the observational run dedicated to galaxy targets (the run is named `26'). 
Indeed in the observational runs dedicated to stellar targets (runs named `103' and `104') some galaxies might have been mistaken for stars i.e. this sample is contaminated by galaxies. 
We will include the runs 10 and 104 in future releases. 
More than 95\% of SDSS and eBOSS galaxies have their output stellar mass constrained i.e. $0<M^*-\sigma_M<M^*+\sigma_M<10^{14} M_\odot$. The remaining 5\% have returned fitted parameters but they are consistent with a stellar mass of 0. We call these fits `unconstrained'. 
%For DEEP2 the fraction of constrained fit is lower due to the lower signal to noise ratio of the data (see 5.1.2).

For each observed galaxy spectrum we provide models of the continuum in up to 9 combinations of stellar population model and IMF. 

This is the first full spectral fitting release of the BOSS+eBOSS high-redshift extension of SDSS which provide stellar population parameters. 
Previous work performing spectral fitting was based on a PCA approach \citep{2012MNRAS.421..314C} aimed at optimizing the stellar mass determination, and for which stellar parameters were not provided.

\begin{table*}
\caption{\label{table:single:spectra} Summary table of observed spectra and fit results for SDSS and BOSS. 
The Table is divided in four sectors.
The first line in each subset gives the total number of spectra available in the survey and how many of them are considered as galaxies. 
The assumed fitting setup (model and IMF) is given in the first 2 columns. 
The third column gives the number of galaxies for which the obtained stellar mass is non zero. 
The last two columns gives the number of galaxies for which the stellar mass parameter is constrained within less than 0.4 dex and 0.2 dex, respectively. 
The number in parenthesis give the percentage relative to the total number of galaxies.}
\begin{center}
\begin{tabular}{ll rrr}
\hline \hline
\multicolumn{5}{c}{eBOSS DR14: $1,759,362$ galaxies} \\
IMF &
Library & 
$M>0$ & 
$\sigma_{\log_{10}M}<0.4$ dex & 
$\sigma_{\log_{10}M}<0.2$ dex \\ \hline
Chabrier & ELODIE & $1758635$ (100.0) & $1427861$ (81.2) & $980810$ (55.7) \\ 
Chabrier & MILES & $1758819$ (100.0) & $1529842$ (87.0) & $1237575$ (70.3) \\ 
Chabrier & STELIB & $1758934$ (100.0) & $1466376$ (83.3) & $1217235$ (69.2) \\ 
Kroupa & ELODIE & $1758635$ (100.0) & $1446616$ (82.2) & $1012728$ (57.6) \\ 
Kroupa & MILES & $1758819$ (100.0) & $1556522$ (88.5) & $1278732$ (72.7) \\ 
Kroupa & STELIB & $1758934$ (100.0) & $1490832$ (84.7) & $1252335$ (71.2) \\ 
Salpeter & ELODIE & $1758635$ (100.0) & $1467855$ (83.4) & $1047431$ (59.5) \\ 
Salpeter & MILES & $1758819$ (100.0) & $1583900$ (90.0) & $1351031$ (76.8) \\ 
Salpeter & STELIB & $1758934$ (100.0) & $1504425$ (85.5) & $1270344$ (72.2) \\ 
\hline 
 \multicolumn{5}{c}{SDSS DR14: $948,259$ galaxies} \\
IMF &
Library & 
$M>0$ & 
$\sigma_{\log_{10}M}<0.4$ dex & 
$\sigma_{\log_{10}M}<0.2$ dex \\ \hline
Chabrier & ELODIE & $938316$ (99.0) & $886081$ (93.4) & $729736$ (77.0) \\ 
Chabrier & MILES & $938317$ (99.0) & $901024$ (95.0) & $765905$ (80.8) \\ 
Chabrier & STELIB & $938317$ (99.0) & $882412$ (93.1) & $722856$ (76.2) \\ 
Kroupa & ELODIE & $938316$ (99.0) & $892330$ (94.1) & $760271$ (80.2) \\ 
Kroupa & MILES & $938317$ (99.0) & $905489$ (95.5) & $798735$ (84.2) \\ 
Kroupa & STELIB & $938317$ (99.0) & $890963$ (94.0) & $752197$ (79.3) \\ 
Salpeter & ELODIE & $938316$ (99.0) & $896355$ (94.5) & $774556$ (81.7) \\ 
Salpeter & MILES & $938317$ (99.0) & $909374$ (95.9) & $814850$ (85.9) \\ 
Salpeter & STELIB & $938317$ (99.0) & $894546$ (94.3) & $768726$ (81.1) \\ 
\hline 
\end{tabular}
\end{center}
\end{table*}

\subsection{DEEP2}
The DEEP2 catalog contains $21,273$ galaxies in the redshift range $0.7<z<1.2$, see Table \ref{table:single:spectra:deep2}. 
The fitting routine converges for all. 
However, due to their generally low SNR, the fraction of spectra for which the stellar mass parameter obtained has an uncertainty smaller than $\sigma_{\log_{10}M}<0.4$ dex, is small, amounting to less than 40\%.
Among the 21,273 spectra, 30\% have a SNR below 1 and only 10\% above 2. Only 200 objects have a SNR above 5. 
This sample will not allow us to explore detailed stellar population properties. 
This distribution of SNRs explains why only roughly 35\% (15\%) have a stellar mass constrained within 0.4 (0.2) dex (Table \ref{table:single:spectra:deep2}). 
From the mock exercises of \citet{firefly2017MNRAS} (e.g. Figures 12-15) we know that stellar masses can be reasonably recovered down to low SNRs (S/N$\sim~5$), but the uncertainty is high and with DEEP2 we are dealing with even lower SNRs. We regard the DEEP2 spectral fitting analysis as an extreme example to explore the boundaries of full spectral fitting. 
  
\begin{table*}
\caption{\label{table:single:spectra:deep2} Same as Table \ref{table:single:spectra} for DEEP2 and DEEP2 \OII galaxies.
The last set shows the subset of the DEEP2 set that have a detection with a signal to noise ratio greater than 5 of the \OII emission line. }
\begin{center}
\begin{tabular}{ll rrr}
\hline \hline
\multicolumn{5}{c}{DEEP2 DR4: $21,273$ galaxies $0.7<z<1.2$} \\                
IMF &                                                                          
Library & 
$M>0$ & 
$\sigma_{\log_{10}M}<0.4$ dex & 
$\sigma_{\log_{10}M}<0.2$ dex \\ \hline
Chabrier & ELODIE& $21268$ (100.0) & $6006$ (28.2) & $2798$ (13.2) \\
Chabrier & MILES&  $21273$ (100.0) & $6934$ (32.6) & $3193$ (15.0) \\ 
Chabrier & STELIB& $21273$ (100.0) & $6823$ (32.1) & $1796$ (8.4) \\ 
Kroupa   & ELODIE& $21268$ (100.0) & $6638$ (31.2) & $3031$ (14.2) \\ 
Kroupa   & MILES&  $21273$ (100.0) & $7274$ (34.2) & $3548$ (16.7) \\ 
Kroupa   & STELIB& $21273$ (100.0) & $7394$ (34.8) & $2163$ (10.2) \\ 
Salpeter & ELODIE& $21268$ (100.0) & $7008$ (32.9) & $3162$ (14.9) \\ 
Salpeter & MILES&  $21273$ (100.0) & $7565$ (35.6) & $3752$ (17.6) \\ 
Salpeter & STELIB& $21273$ (100.0) & $7644$ (35.9) & $2435$ (11.4) \\ 
\hline 
\multicolumn{5}{c}{DEEP2 DR4: $15,498$ \OII galaxies $0.7<z<1.2$} \\
IMF &
Library & 
$M>0$ & 
$\sigma_{\log_{10}M}<0.4$ dex & 
$\sigma_{\log_{10}M}<0.2$ dex \\ \hline
Chabrier & ELODIE & $15493$ (100.0) & $3444$ (22.2) & $1431$ (9.2) \\ 
Chabrier & MILES & $15498$ (100.0) & $4090$ (26.4) & $1822$ (11.8) \\ 
Chabrier & STELIB & $15498$ (100.0) & $3975$ (25.6) & $944$ (6.1) \\ 
Kroupa & ELODIE & $15493$ (100.0) & $3903$ (25.2) & $1529$ (9.9) \\ 
Kroupa & MILES & $15498$ (100.0) & $4267$ (27.5) & $1960$ (12.6) \\ 
Kroupa & STELIB & $15498$ (100.0) & $4390$ (28.3) & $1108$ (7.1) \\ 
Salpeter & ELODIE & $15493$ (100.0) & $4184$ (27.0) & $1619$ (10.4) \\ 
Salpeter & MILES & $15498$ (100.0) & $4545$ (29.3) & $2123$ (13.7) \\ 
Salpeter & STELIB & $15498$ (100.0) & $4559$ (29.4) & $1228$ (7.9) \\ 
\hline
\end{tabular}
\end{center}
\end{table*}
\subsection{General findings as a function of parameters.}
\label{subsec:about:convergence}
Depending on fitting setup, between 81 and 96\% of the fits give constrained parameters for the SDSS+eBOSS data and between 22 and 36\% for the DEEP2 data. For example, in the case M11-MILES and a Chabrier IMF, 80\% (70\%) of the SDSS (eBOSS) data stellar masses are constrained to the 0.2 dex level, where by `constrained' we mean that the statistical errors are smaller than 0.2 dex on the stellar mass. 

For the DEEP2 data, due to the intrinsic faintness of the spectra and the much lower SNR of the continuum, we could tightly constrain stellar parameters for only about 10\% of the sample.
Exact numbers are available in Tables \ref{table:single:spectra}, \ref{table:single:spectra:deep2}. 
%The accuracy in constraining the fitted parameters is directly related to the signal to noise ratio measured in the spectrum, as mentioned in Sec. \ref{subsec:firefly:performances} and further discussed in Sec. \ref{subsec:stat:err}. 

In SDSS+eBOSS, un-constrained fits are dominated by spectra with low signal to noise ratio ($<1-2$) and by a fraction of QSOs that were mis-classified as galaxies by the automated pipeline. 
In the DEEP2 data, the un-constrained fits are split into two components: the restricted wavelength coverage of the spectra; very low signal to noise ratio in the continuum ($\sim$0.1-1) (Redshift determined with emission lines only). 

\begin{figure*}
\begin{center}
\caption{\label{fig:sky_sdss:eboss} 
Distribution on the sky of galaxies in the SDSS (top left) and eBOSS (top right) catalogs. 
All panels have the same dot size and transparency level applied. 
The second line shows the distribution of galaxies for which the stellar mass parameters was determined to better than 0.2 dex. 
In the SDSS bottom left panel, we see small gaps that are not visible in the top left panel e.g. around R.A.=350$^\circ$.
In the eBOSS bottom right panel, we see that the stripe in $40^\circ<Dec.<60^\circ$ is less dense than in the top right panel.
}  
\includegraphics[height=4cm]{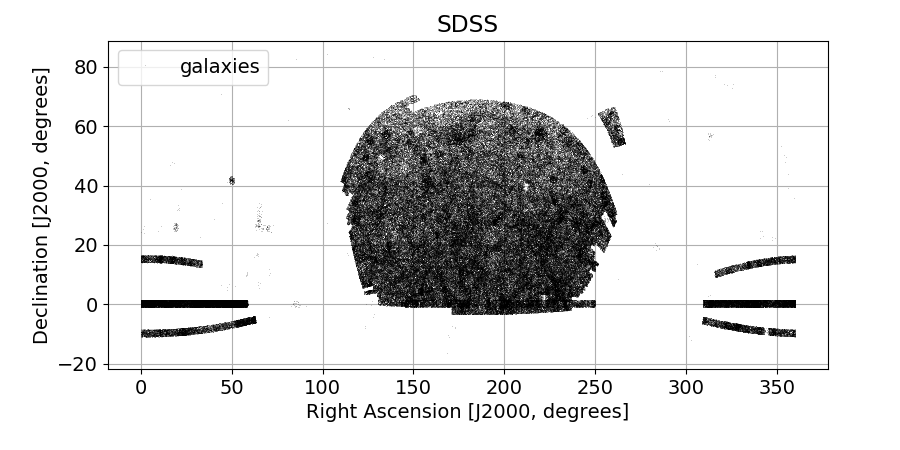}
\includegraphics[height=4cm]{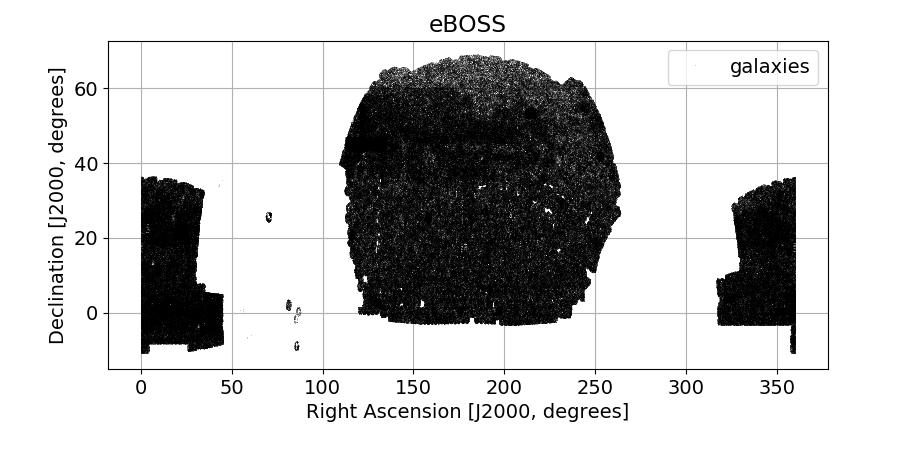}
\includegraphics[height=4cm]{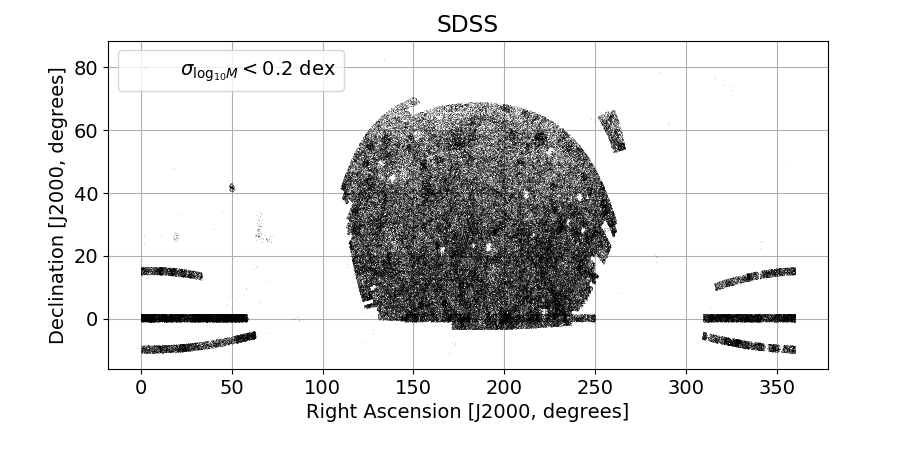}
\includegraphics[height=4cm]{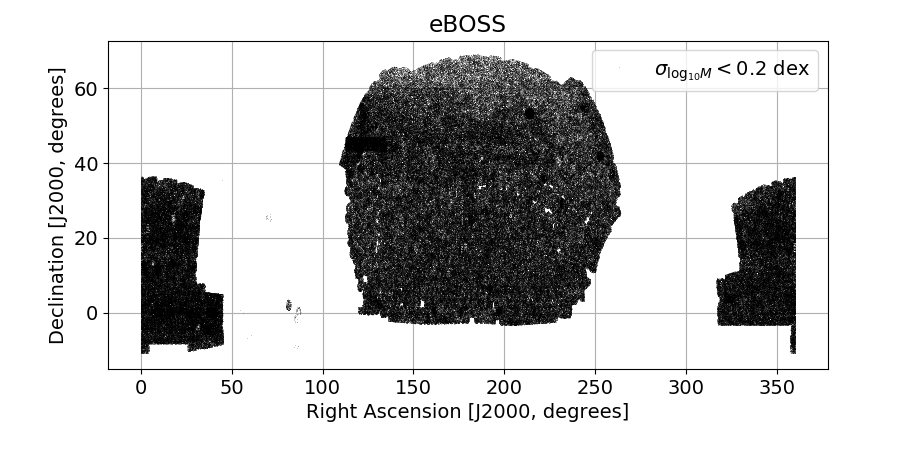}
\end{center}
\end{figure*}

\begin{figure*}
\begin{center}
\caption{\label{fig:rmag:hist:ratio:sdss:eboss} 
Ratio between the number of galaxies for which the stellar mass parameter was determined to better than 0.2 (0.4) dex and the complete galaxy catalog in bins of SDSS $r$-band model magnitude (when available)  for the SDSS catalog (left) and the eBOSS catalog (right). 
As expected, at the faint ends the fraction of galaxies with accurate stellar mass determinations drops.}  
\includegraphics[height=5.5cm]{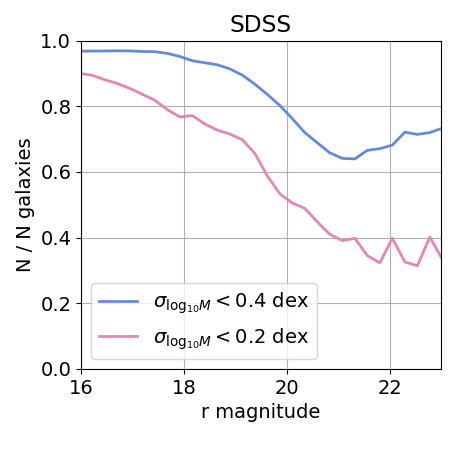}
\includegraphics[height=5.5cm]{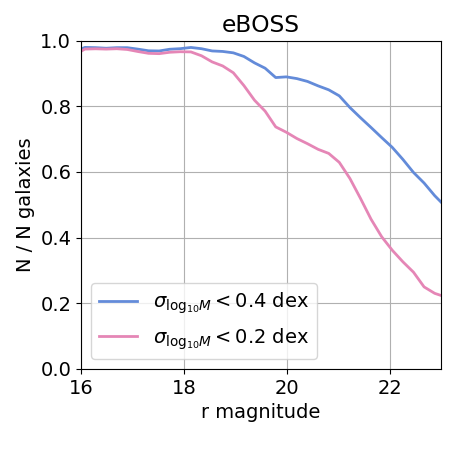}
\end{center}
\end{figure*}

In any case, the sample with fitted properties is not a clean subset of the parent catalogs, rather it is a biased subsample of the parent catalogs. 
It is biased as a function of position on the sky and magnitude. 
Fig. \ref{fig:sky_sdss:eboss} shows the distribution on the sky of galaxies in the SDSS and eBOSS catalogs and the distributions of galaxies for which the stellar mass parameters is determined to better than 0.2 dex. 
We note differences in density of points. 
The gaps correlate with specific plates that have overall a smaller signal to noise. 
Furthermore, the distribution in magnitude of the galaxies with stellar mass parameter determined to better than 0.2 dex is not a fair sub sample of the complete galaxy population, in particular towards the faint end, see Fig. \ref{fig:rmag:hist:ratio:sdss:eboss} that shows the fraction of constrained fits as a function of the SDSS $r$-band magnitude. 

\subsubsection{Statistical error on derived parameters}
\label{subsec:stat:err}
\begin{figure*}
\begin{center}
\caption{\label{fig:uncertainty:HIST} 
Distributions of the 1 $sigma$ errors on the stellar parameters: age (left), metallicity (middle), mass (right).
}  
\includegraphics[height=5.5cm]{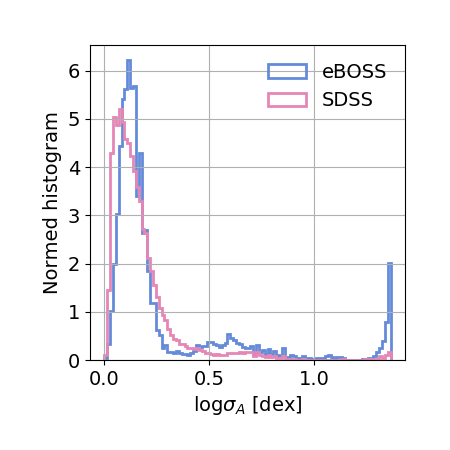}
\includegraphics[height=5.5cm]{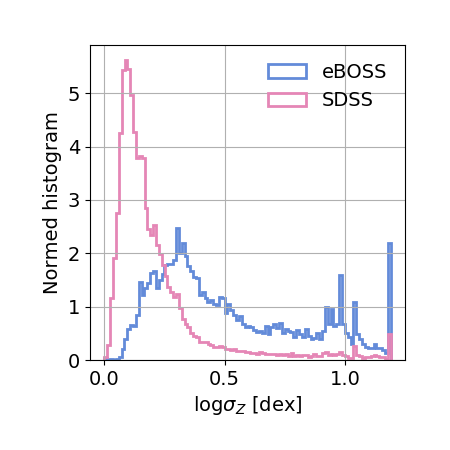}
\includegraphics[height=5.5cm]{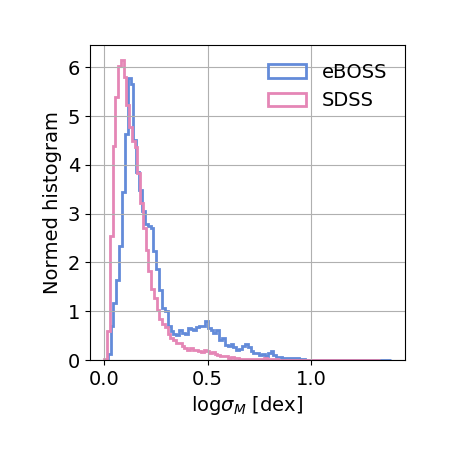}
\end{center}
\end{figure*}

The statistical uncertainty on the stellar age, stellar metallicity and stellar mass is estimated using the full probability distribution function of the parameters derived during the fit. 
We provide the 1 and 2 $\sigma$ uncertainties. 
Fig. \ref{fig:uncertainty:HIST} shows the distribution of the $1\sigma$ uncertainties obtained. 
Although uncertainties on age and metallicity can be quite high, the uncertainty on the stellar mass remains contained. 

\begin{figure*}
\begin{center}
\caption{\label{fig:uncertainty:vs:SNR} 
Individual median signal to noise ratio in the SDSS+eBOSS spectra around the 4000 Angstrom break v.s. uncertainty on the stellar mass (top row), stellar age (middle row), stellar metallicity (bottom row). 
Each column shows a different redshift range. 
The red dashed show the mean and dispersion of the data binned along the x-axis.
}  
\includegraphics[height=5.5cm]{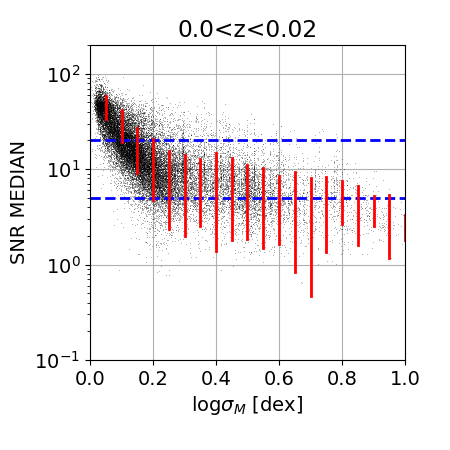}
\includegraphics[height=5.5cm]{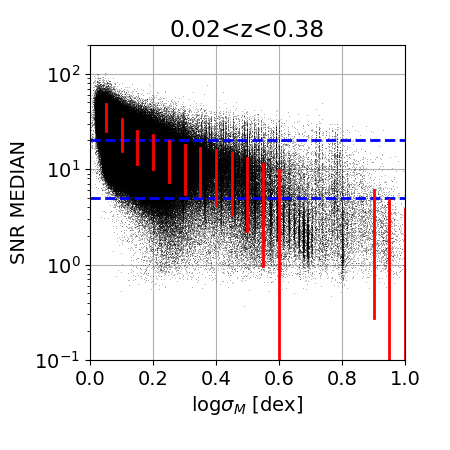}
\includegraphics[height=5.5cm]{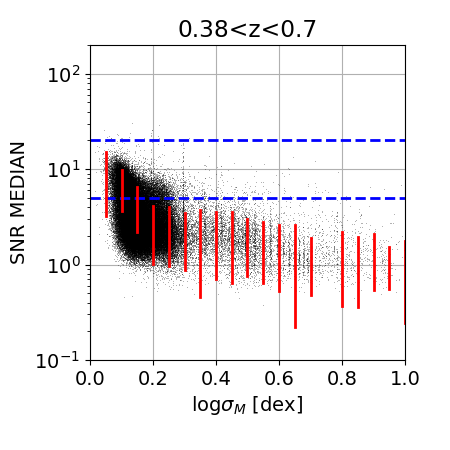}

\includegraphics[height=5.5cm]{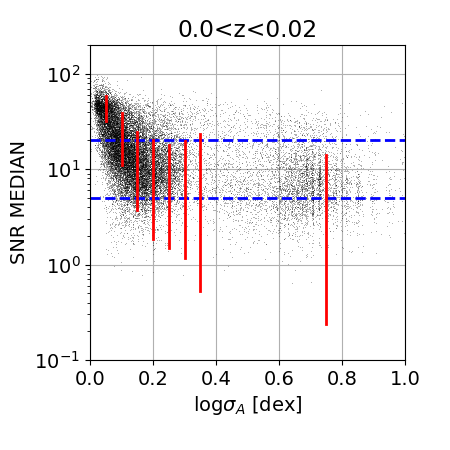}
\includegraphics[height=5.5cm]{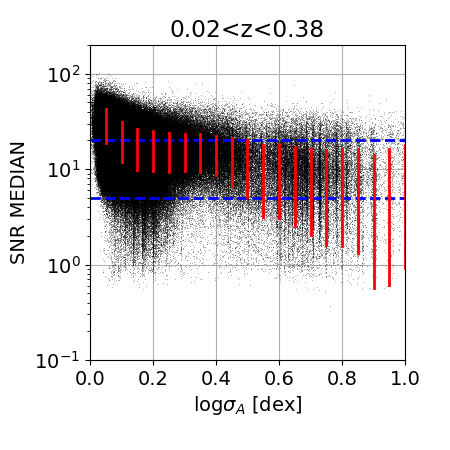}
\includegraphics[height=5.5cm]{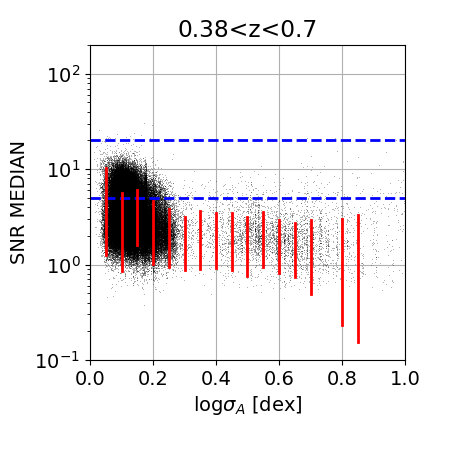}

\includegraphics[height=5.5cm]{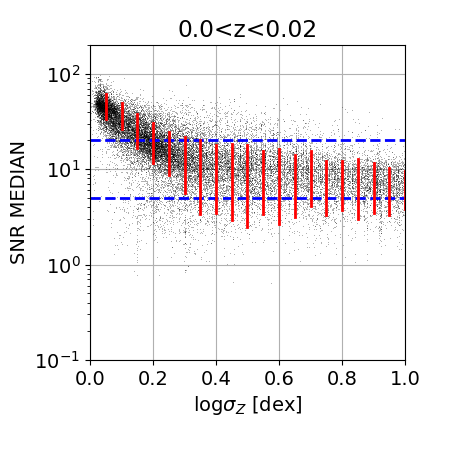}
\includegraphics[height=5.5cm]{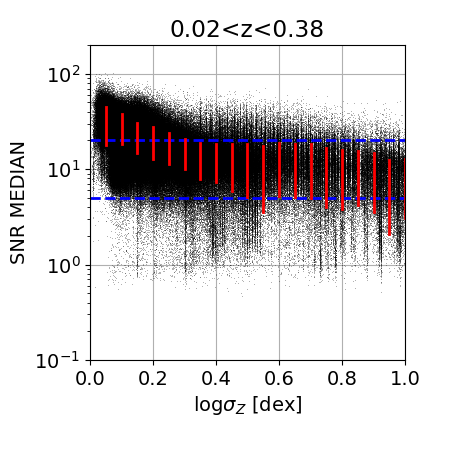}
\includegraphics[height=5.5cm]{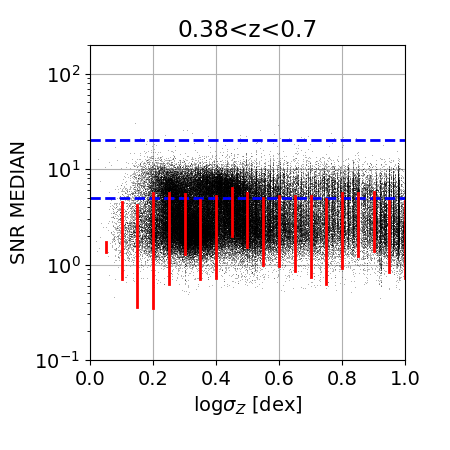}

\end{center}
\end{figure*}

As in \citet{firefly2017MNRAS}, we find that the uncertaintes are mainly driven by the average signal-to-noise per pixel in the spectra in the band where information is localized (roughly 3800-5500\text{\AA}). Fig. \ref{fig:uncertainty:vs:SNR} shows the median signal to noise ratio in the spectrum around the 4000 Angstrom break v.s. uncertainty on the stellar mass, stellar age and stellar metallicity. 

For spectra with large error on the stellar mass, one should be cautious, and combine this measurement with stellar masses (and other parameters in general) based on broad band magnitudes SED fitting \citep[e.g.][for the same galaxies]{Maraston2013}. 
In some sense, our new catalogs provide stellar masses with a better constrains for a subset of the complete sample SDSS+eBOSS.

\subsubsection{Systematical biases and errors}

It is well known that stellar population properties derived from data depend on the assumed models and modelling techniques, as routinely discussed in the literature. Here we focus on comparing stellar masses from this paper to previous releases, as this is the ouput that was previously provided.
\citet{2016MNRAS.455.4122B} discuss in detail systematic uncertainties on stellar mass and stellar mass function, concluding that the dominant one is the assumed stellar population models.  They showed that for the same IMF, different assumptions on the adopted stellar population models cause up to 0.3 dex systematic differences. Different dust models lead to  0.2 dex systematic differences. On top of these systematics, the assumed IMF gives an offset, which is e.g.. of the order 0.2 dex passing from a Chabrier or Kroupa IMF to a Salpeter one  \citep[e.g.][]{2012MNRAS.422.3285P}. As discussed in several papers \citep[][]{2010MNRAS.407..830M,2012MNRAS.422.3285P}, even when the same stellar population model is assumed, different assumption in the star formation history templates reflect themselves in stellar mass offsets. All effects settle around a 0.3 dex systematics in most cases. Sometimes the choice is to average stellar mass determinations from different modelling approaches and use the average for mass function studies \citep[as in][]{2017ApJ...851...34B}.

\begin{figure*}
\begin{center}
\caption{\label{fig:comparison:literature} 
Comparison of the stellar masses obtained with its preceding catalog the 'starforming' flavour of the calculations by the Portsmouth group for the data release 12 \citep{Maraston2013}.
}  
\includegraphics[height=5.5cm]{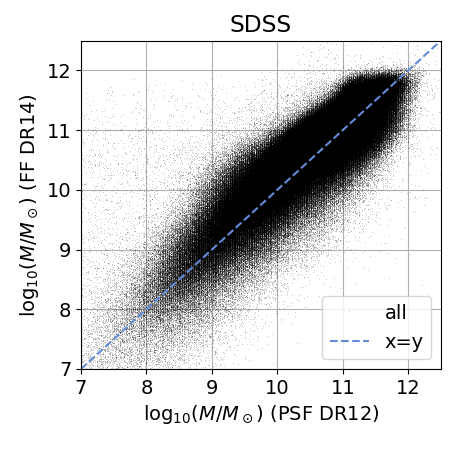}
\includegraphics[height=5.5cm]{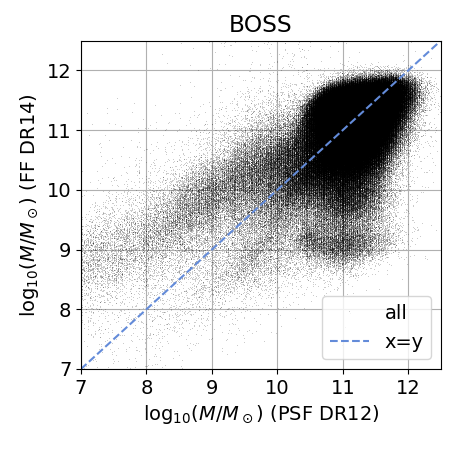}
\end{center}
\end{figure*}

Here we compare the catalogs presented here to the 'starforming' flavour of the calculations by the Portsmouth group for the data release 12 \citep{Maraston2013} and available at \url{http://www.sdss.org/dr14/spectro/galaxy_portsmouth}, see Fig. \ref{fig:comparison:literature}. We choose this flavour as it includes fitting templates with low ages as in our current calculations \footnote{See \citealt{Maraston2013} for the 'passive' flavour and how it compares to the star forming set of calculations.}. 
The agreement between the SDSS stellar masses is excellent. 
A larger scatter is found in the comparison of BOSS galaxies' stellar masses, which is captured by the uncertainties. 
We found the distribution normed by area of the quantity $|M_1-M_2|/\sqrt{\sigma^2_{M1}+\sigma^2_{M2}}$ where $M_1$ and $M_2$ are the DR12 and the DR14 versions of the stellar mass measurement to be very close to a normal distribution if using $2\sigma$ errors. 
If using $1\sigma$ errors, there is a some level of tension between the catalogs between the BOSS catalogs.  
However, when we consider the subset where galaxies have the same redshift (within 0.001) and the same $E(B-V)$~ (within 0.02), then the tension at $1\sigma$ disappears. 
Overall, we conclude that the agreement with previous calculations - considering the different fitting methods and assumed templates (within the same stellar evolution framework) - is good. The question on whether stellar masses calculated via spectroscopy or spectrophotometry compare is an ongoing question (partly addressed in \citealt{Maraston2013},Appendix), which we should.

Furthermore we find that the average uncertainties on stellar mass, age and metallicity are systematically larger when comparing the M11-ELODIE results to the M11-STELIB ones. We disproved that this may be due to the different wavelength coverage. Given the complexity of stellar library input, we are not able to pin down the exact reason and we leave this for future investigations. 

\subsubsection*{Performance on the data}
We follow the \citet{2013AJ....145...69L} procedure to mask pixels affected by the sky and estimate the median SNR. 
The median SNR in all good pixels in the $i$-band of the spectrum is anticorrelated to the uncertainty. 
The higher the SNR, the smaller the uncertainty, see Fig. \ref{fig:uncertainty:vs:SNR}.\footnote{\citet{2012MNRAS.421..314C} achieved a similar conclusion by comparing the stellar mass obtained via their PCA-based full spectral fitting and the one from broad-band SED fitting. The two estimates converge at a $SNR$~around 25. See also discussion in \citet{Maraston2013}, Appendix.} 
This correlation holds up to redshift $\sim0.4$. 
Then the band of importance between $3500$-$5500$\text{\AA} break starts to enter the $z$-band where the estimation of the SNR are much noisier. 
So, at low redshift ($z<0.3$), the selection of the best fits can be based on either the SNR or on the uncertainty on the stellar mass. 
At higher redshift ($z>0.3$), the median SNR measure provided in the SDSS specObj summary files is not a reliable estimate of the actual SNR. 

\subsubsection{Trends in SDSS data: source type, signal to noise ratio, redshift}

SDSS, BOSS and eBOSS have large numbers of programs (or sub components) that we split according to the `SOURCETYPE' flag (there exist of order of $\sim$200 different programs). 
Among these programs only eight contain more than 10,000 spectra classified as galaxies: GALAXY, NONLEGACY, QSO, QA in SDSS and LRG, QSO, SEQUELS\_TARGET, SPIDERS\_RASS\_CLUS in eBOSS. 
Then in eBOSS of order of 40 programs have between 1,000 and 10,000 galaxy spectra. 
We give the number of galaxies split by SDSS (eBOSS) program in Table 
\ref{ref:table:sdss:src:firefly}
(\ref{ref:table:boss:src:firefly}). 
(Note that the SOURCETYPE corresponds to the type it was assigned the first time it was targeted. To construct full sample, one needs to use target bits.) 
We give the number of galaxies considered for the fit and the fraction where the stellar mass parameter was constrained at better than 0.4 and 0.2 dex. 
In SDSS, the target type 'GALAXY' has a constrained stellar mass for more than 95\% of its galaxy spectra which have a positive median signal to noise ratio. 
More than 80\% have an uncertainty on the stellar mass parameter lower than $\pm$0.2 dex. 
In eBOSS, the dominant galaxy type is LRG and 60\% have an uncertainty on the stellar mass parameter lower than $\pm$0.2 dex.  

The fraction of objects with well-constrained parameters have three main dependences: redshift, signal to noise ratio and fraction of the light observed in the fiber. 
For each program with more than 100 galaxy spectra we break down the statistics as a function of these parameters. 
The dependence on signal to noise ratio and redshift is shown in Table 
\ref{ref:table:sdss:src:SNR} 
(\ref{ref:table:boss:src:SNR}). 
It gives for five redshift bins (with boundaries $0$, $0.025$, $0.375$, $0.7$, $0.85$, $1.6$) the percentile of galaxies that corresponds to a median SNR of 5 or 20. 
For redshifts $z<0.375$ more than half of the SDSS GALAXY (total $\sim800,000$) have SNR greater than 20. 
For redshifts $z>0.375$ not a single SDSS GALAXY (total $\sim40,000$) has a SNR greater than 5. 
% The left panel on Fig. \ref{fig:mass:redshift} shows how the median signal to noise in good pixels are distributed with redshift. 
Low redshift galaxies ($z<0.4$) were observed on average at higher signal to noise ratio than higher redshift galaxies.
The panels of Fig. \ref{fig:uncertainty:vs:SNR} show how the median signal to noise in good pixels in a band around the $4000$\text{\AA} break correlates with the uncertainty on the stellar mass (for 3 redshift bins, $0<z<0.02$, $0.02<z<0.38$ and $0.38<z<0.7$).  

The dependence on the fraction of light observed and redshift is given in Table 
\ref{ref:table:sdss:src:fibermag} 
(\ref{ref:table:boss:src:fibermag}) 
It gives, in three redshift bins (with boundaries $0$, $0.17$,  $0.55$, $1.6$), the number of galaxies where the fraction of the light in the fiber compared to the total light is greater than 50\% (10\%) i.e. the fiber magnitude smaller than the total magnitude by no more than 0.75 mag (2.5 mag). 
At redshift $z<0.17$, most SDSS galaxies have between 10 and 50\% of the light in the fiber, indeed their extension is larger than the fiber size.  
At high redshift $z>0.55$, more than a third of the eBOSS LRG galaxies have more than 50\% of their light in the fiber.

\subsubsection{Common objects between SDSS, eBOSS and DEEP2}
There are galaxies which were observed by both SDSS and DEEP2, precisely 64 (493) galaxies were observed by both DEEP2 and SDSS (eBOSS). 
Among these, 31 (261) have redshift values that agree within $|z_{DEEP2}-z_{SDSS (eBOSS)}|<0.0005$. 
3 (31) galaxies have a stellar mass constrained within $\pm0.2$ dex. 
For these 3+31 galaxies, the estimated stellar masses agree within errors. 

\subsubsection{Comparison with previous calculations}
Compared to previous stellar population model catalogs, we roughly double the number of galaxies for which parameters were measured \citep[DR12,][]{Maraston2013,Thomas2013a}. 
In particular the number of well-constrained stellar masses (i.e. constrained to better than 0.2 dex at given IMF) has also doubled. Compared to previous studies of the SDSS data based on spectrophotometry \citep[e.g.][]{Maraston2013}, the sample size of galaxies with parameters constrained to this level is increased by a factor of 2. 
%Therefore we are gaining significant insight on the stellar population parameters. 
Parameters based on broad-band SEDs on the other hand could be calculated for all data as they are less dependent on the SNR of the spectroscopic data. 
Hence, such a gain in precision is enabled by fitting every pixel of the spectra rather than only using the broad-band magnitudes (photometry) at the spectroscopic redshift. 
However, this is at the cost of having a less homogeneous sample. 
Indeed, the uncertainty on the stellar population parameters derived from full spectral fitting depends on the signal to noise ratio obtained in individual spectra, which varies according to observing conditions, position in the spectrograph and survey strategy. 
Hence, the tighter constrain on stellar parameters is gained at the price of completeness. 
\subsection{Ages, metallicities and star formation histories of galaxies}
In this section we use our parameter estimates to construct a statistical view over the key population parameters, age, chemical composition and star formation history across our galaxy samples and across redshift. A more quantitative use of these results for a variety of galaxy evolution studies will be pursued in future investigations.

\subsubsection{Galaxy ages}
\begin{figure*}
\begin{center}
\caption{\label{fig:galaxy:age} 
Cumulative normed distribution of mass-weighted ages.}  
\includegraphics[height=3cm]{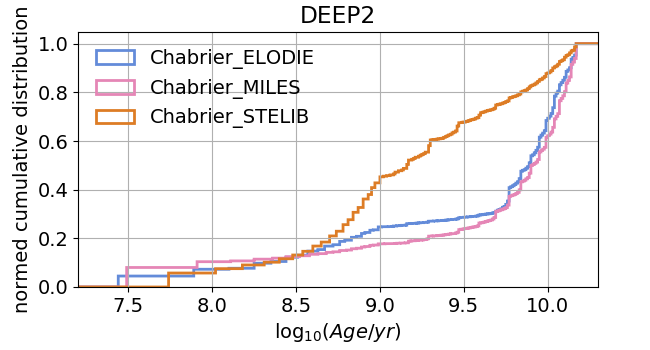}
\includegraphics[height=3cm]{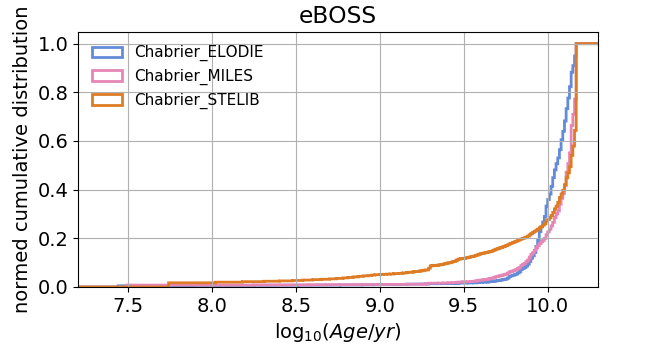}
\includegraphics[height=3cm]{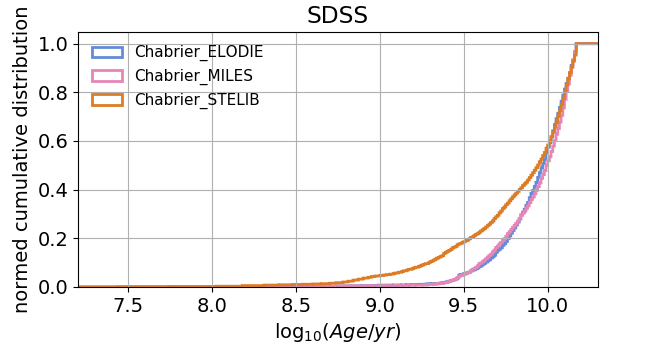}
\end{center}
\end{figure*}
Fig. \ref{fig:galaxy:age} shows the cumulative normed distribution of mass-weighted ages obtained for our three galaxy samples when using the three M11 model types (based on the ELODIE, MILES and STELIB), referred to a Chabrier IMF. We plot results for one IMF only as this parameter has little influence on the derived galaxy physical parameters \citep[e.g.][]{2012MNRAS.422.3285P}, when slope and exponents are not too different among the options.  

Fitted mass-weighted ages obtained with M11-Miles and Elodie are remarkably consistent for the three galaxy samples, pointing to values larger than 3 Gyr, with over 60 \% of the samples around 10 Gyr. Ages obtained with M11-Stelib models imply a $\sim~20\%$~of slightly younger galaxies due to their narrower coverage in metallicity and the age-metallicity degeneracy (by which the fit with high metallicity models releases younger ages, e.g. Worthey~1994). Note also how nicely the age distribution for the DEEP2 sample containing higher-redshift galaxies selected to be star forming shifts to younger ages consistently for all models. Again here the effect at obtaining generally lower ages with M11-Stelib is even more pronounced, and 70\%~of the galaxies are younger than 3 Gyr.

\subsubsection{Galaxy metallcity}
\begin{figure*}
\begin{center}
\caption{\label{fig:galaxy:metal} 
Cumulative normed distribution of mass-weighted metallicities.}  
\includegraphics[height=3cm]{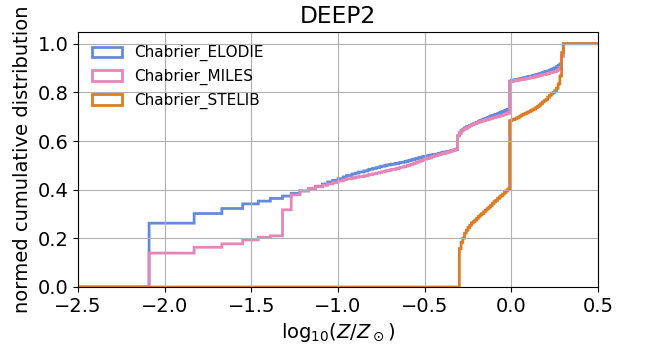}
\includegraphics[height=3cm]{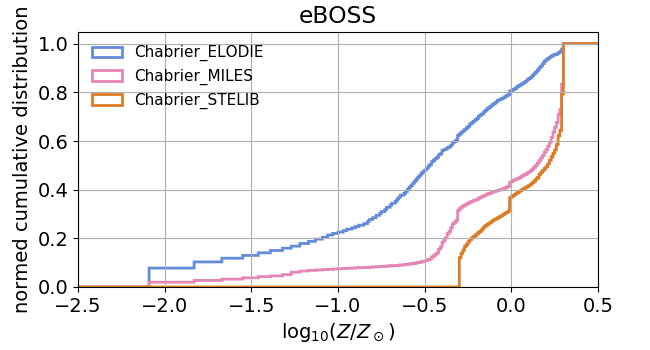}
\includegraphics[height=3cm]{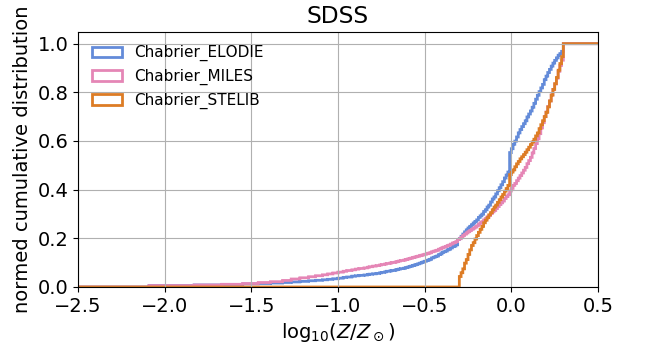}
\end{center}
\end{figure*}

Fig. \ref{fig:galaxy:metal} shows the cumulative normed distribution of mass-weighted metallicities fitted. There are some interesting trends to note. First of all, for the SDSS sample all models agree in pointing to high metallicities, with 80\% of the sample having half-solar and above a chemical composition. The residual 20\%~is found to have lower metallicities by M11-MILES and M11-ELODIE as these contain low-metallicity models whereas M11-STELIB only span the range half-solar to twice solar (cfr. M11, Figure~3). Also for the DEEP2 sample (left-hand panel) M11-MILES and M11-ELODIE models give a very similar answer, with now 60\%~of the sample lying at subsolar metallicities. The M11-STELIB gives higher metallicities by construction, as already explained.

The eBOSS results are those revealing a larger scatter among the models, with M11-ELODIE giving a large fraction of metal-poor galaxies which is not found when fitting the other two models, which instead give the very similar result of high-metallicity also in the eBOSS galaxies.

\subsubsection{Galaxy star formation histories}

\begin{figure*}
\begin{center}
\caption{\label{fig:SHF:all} 
Distribution of the SSP components in the age-metallicity plane color coded with the total number of components in a bin times the SSP mass weight (Chabrier IMF).}  

\includegraphics[height=4.5cm]{age_metallicity_Chabrier_ELODIE_deep2_04.png}
\includegraphics[height=4.5cm]{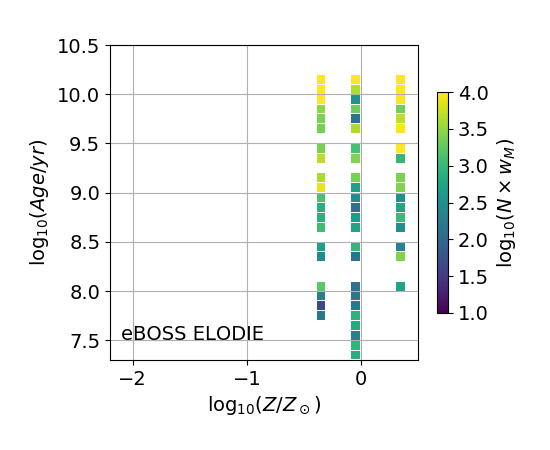}
\includegraphics[height=4.5cm]{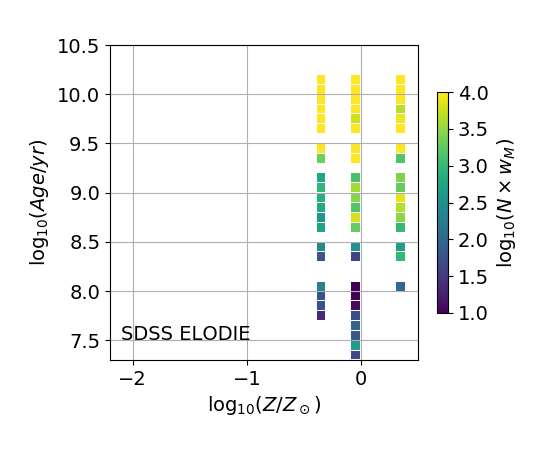}
\includegraphics[height=4.5cm]{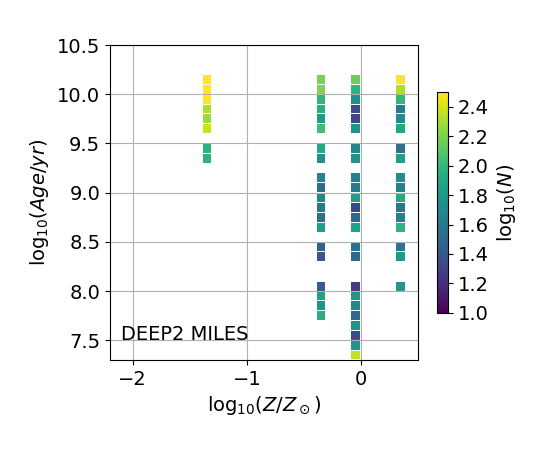}
\includegraphics[height=4.5cm]{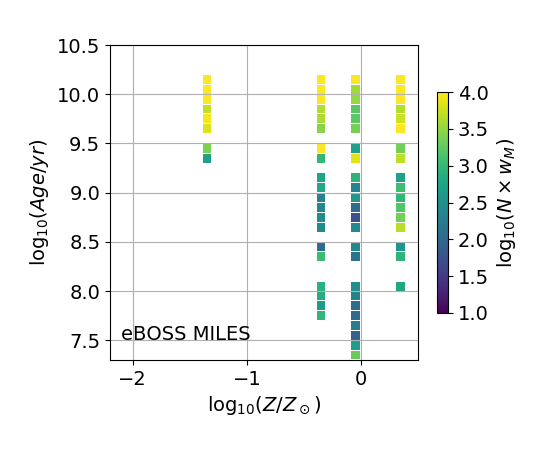}
\includegraphics[height=4.5cm]{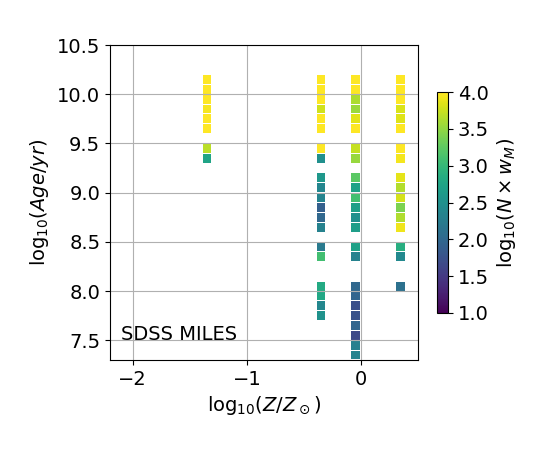}
\includegraphics[height=4.5cm]{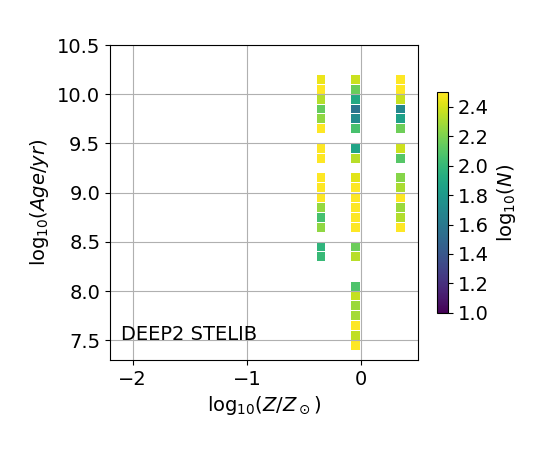}
\includegraphics[height=4.5cm]{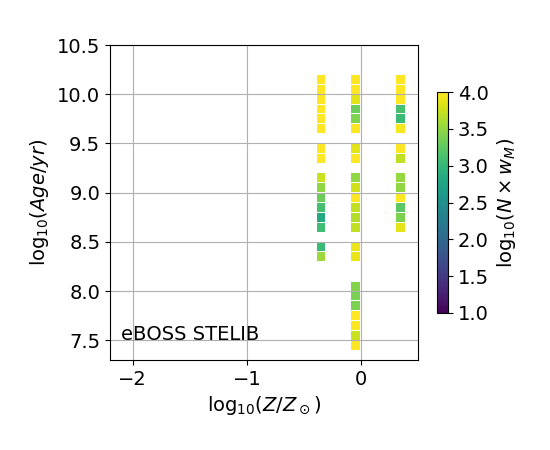}
\includegraphics[height=4.5cm]{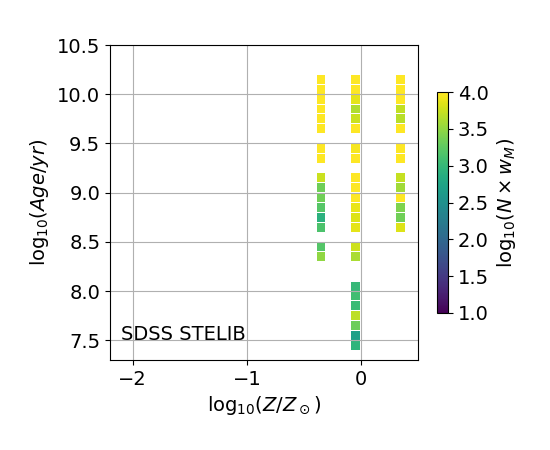}

\end{center}                 
\end{figure*}

Fig. \ref{fig:SHF:all} show the distribution of SSP components of the total best-fit to a galaxy, in the age-metallicity plane, colour-coded with the total number of components in a bin times the SSP weight. Those histograms have been obtained by adding the mass weights of all components in a small bin of age and metallicity. 
As in previous sections, plots refer to the M11-MILES, STELIB, ELODIE fitting models for a Chabrier IMF. 

First of all, the fits with M11-MILES find a metal-poor old component in all galaxy samples. 
This could be interpreted as a metal-poor halo, which is known to exist in most galaxies. The weight of such a component is $\sim15\%$ for SDSS and eBOSS and $\sim45\%$ for DEEP2. 
The rest of the components for SDSS and eBOSS lie at higher metallicity, with the ages of the most metal-rich component spanning towards lower values, as expected by a gradual galaxy build-up where metal-enrichment proceeds with time. Obviously the details of individual objects will vary, for example massive ellipticals within the sample may show coeval, high metallicity populations, while massive spirals will be consistent with an age/metallicity relation of the kind just described. The same model on the DEEP2 sample gives less power to the high-metallicity components, and also a more homogenous age spread. This is reflecting the larger stellar age ranges found in this sample.
M11-ELODIE give similar results for the high-metallicity components - a gradual decrease in age towards the highest metallicity - but does not reveal any low-metallicity part. Finally, M11-STELIB find all solutions at high-metallicity (larger than half solar by construction), but, differently from the other two models, present the largest fractions of lower ages (see yellow parts in the histograms), a fact that was already noted in M11 and \citet{firefly2017MNRAS}.

These examples show how much the details of galaxy evolution are incapsulated in the interpretative model one assumes, although the macro characteristics of the galaxy components are - reassurely - quite independent of the assumed model.

\clearpage
\section{Applications}
\label{sec:application}

\subsection{Stellar mass function of galaxies selected with their \OII emission line luminosity in DEEP2}
\label{sec:application:elg}
\begin{figure}
\begin{center}
\includegraphics[type=png,ext=.png,read=.png,height=8cm]{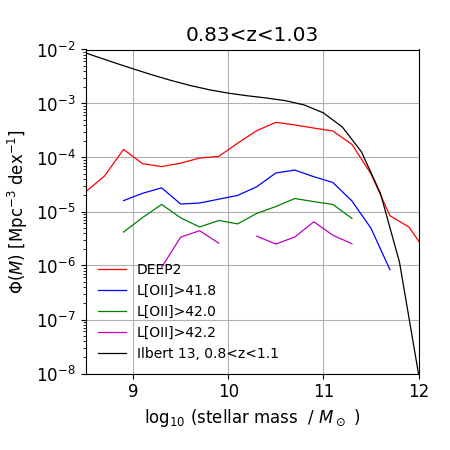}
\label{fig:LF:sampling}
\caption{Stellar mass function as measured in the DEEP2 \OII emitter sample in the redshift range $0.83<z<1.03$ (considering all combinations of IMF and libraries). 
We show the stellar mass function is sampled by the \OII emitters for different thresholds in $\log_{10}($L\OII$)$ luminosity 41.8, 42 and 42.2.}  
\end{center}
\end{figure}

How emission line galaxies populate the cosmic web is a hot topic in cosmology nowadays \citep{favole_2015_elg,2017MNRAS.472..550F,raichoor_2017MNRAS.471.3955R,2018MNRAS.474.4024G,2018arXiv181005318G}. 
To characterize how emission line galaxies are related to the overall galaxy population, we project the DEEP2 observed stellar mass function in the redshift range $0.83<z<1.03$ for three \OII luminosity threshold, $\log_{10}($L[\OII]$)>41.8$, 42 and 42.2, see Fig.  \ref{fig:LF:sampling}. 
It is known that there is scatter in SFR at fixed mass and that the \OII -- SFR relation also has scatter. 
Therefore we do not expect to find that only a narrow range of masses is populated by the strongest \OII emitters.
Indeed, the distributions we find are quite broad, covering the stellar mass range $10^9<M/M_{\odot}<10^{11.5}$. 
More interestingly, these distributions are quite flat and their shape do not seem to depend on the luminosity threshold. 
Similar distributions are found in \citet{raichoor_2017MNRAS.471.3955R,2018arXiv181005318G}. 
Given that the DEEP2 sample is complete for the \OII luminosity limits applied, we conclude that up to $z=1.5$ there is no preferred host galaxy mass (in the range $10^9<M/M_{\odot}<10^{11.5}$) to find a strong \OII emission.

Recently, a broad range of properties of ELGs was predicted by the model presented in \citet{2018MNRAS.474.4024G}. 
In our data, it seems that massive galaxies are selected as \OII emitters, where in the model, some of them are missing. 
This discrepancy is due to the treatment of the dust and to how emission lines are implemented in the model. A further detailed analysis is needed to understand better this very interesting puzzle.

The broad distribution in stellar mass underlying emission-line galaxies implies that stacked spectra of galaxies selected by emission line luminosity thresholds is unlikely to capture the variety of galaxies constituting the emission line galaxy population. The fact that any galaxy can be detected via emission-lines happens also because the light of emission-line selected galaxies is dominated by their latest generation of stars, which overshines the underlying structure of older stellar populations, much independently of the actual galaxy mass. This is \citep[known as the 'iceberg effect'][]{2010MNRAS.407..830M}. A larger redshift range is needed to probe any dependence on mass of the 'emission-line' activity in galaxies. 

\subsection{Stellar mass function probed by SDSS, eBOSS and DEEP2}
\label{subsec:sdss:stellarmassdensity}
The galaxy stellar mass function and its evolution with redshift is a crucial probe of galaxy formation and evolution in a cosmological context \citep{2006ApJ...651..120B,2010AA...523A..13P,Maraston2013,Ilbert2013SMF,2016MNRAS.455.4122B,2017AA...605A..70D,2017MNRAS.466..228E}. 
The study of this function requires large numbers of galaxies. 
For this reason usually the calculation of $\sim M^{*}$ is performed via broad-band photometry. 
It is therefore interesting to see what we obtain for the galaxy mass function when using our results based on spectral fitting rather than broad-band photometry, and for three different datasets in two redshift bins.

% We consider stellar masses constrained to better than 0.2 dex based on the Chabrier IMF for the three models M11-ELODIE, M11-STELIB and M11-MILES. 
For eBOSS galaxies, we consider the area to be 10,000 deg2, for SDSS 7,900 deg2 and for DEEP2 0.5 deg2 (low redshift) or 2.78 deg2 (high redshift).
For SDSS and eBOSS each galaxy represents itself only, no weighting other than the area is applied. 
It should be noted that \citet{Maraston2013} concluded that BOSS is complete down to $M\sim10^{11}M_{\odot}$~(for a Kroupa IMF) up to $z\sim0.55$ \citep[see also][that reached a similar conclusion]{2016MNRAS.457.4021L}. 
The completeness of eBOSS has not yet been studied in depth. 
For DEEP2, we use the statistical weights from \citet{Comparat2016LFs}. These weights correct from the target selection algorithm used in DEEP2 and allow the recovery of the correct galaxy density as a function of redshift and magnitude. 

For each catalog (3 surveys x 3 libraries x 3 IMF), we estimate the observed stellar mass function (OSMF) and its Poisson error (cosmic variance is not taken into account). 
We use only galaxies for which the stellar masses is constrained to better than 0.2 dex. 
Then, per each survey we compute the median of the nine measurements (i.e. over the three libraries x three IMF) and the minimum and maximum of the nine measurements (with only Poisson errors considered). 
The OSMF obtained constitutes a robust a lower limit to the stellar mass function. 
Indeed because we use only stellar masses that are tightly constrained, we are certain that at least this density of stellar mass exists in galaxies. 

\begin{figure}
\begin{center}
\includegraphics[type=png,ext=.png,read=.png,height=8.5cm]{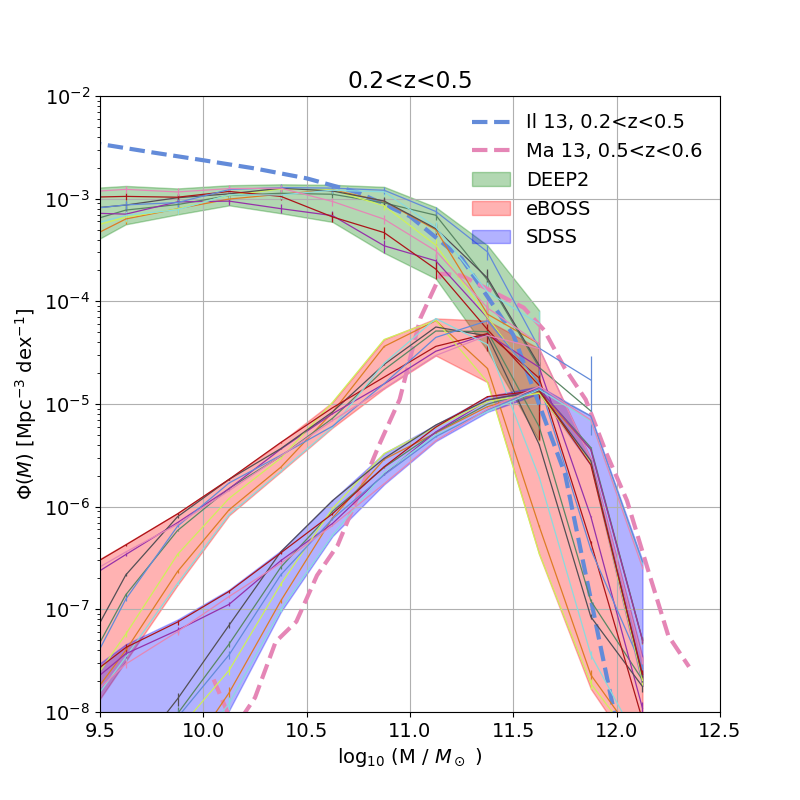}
\includegraphics[type=png,ext=.png,read=.png,height=8.5cm]{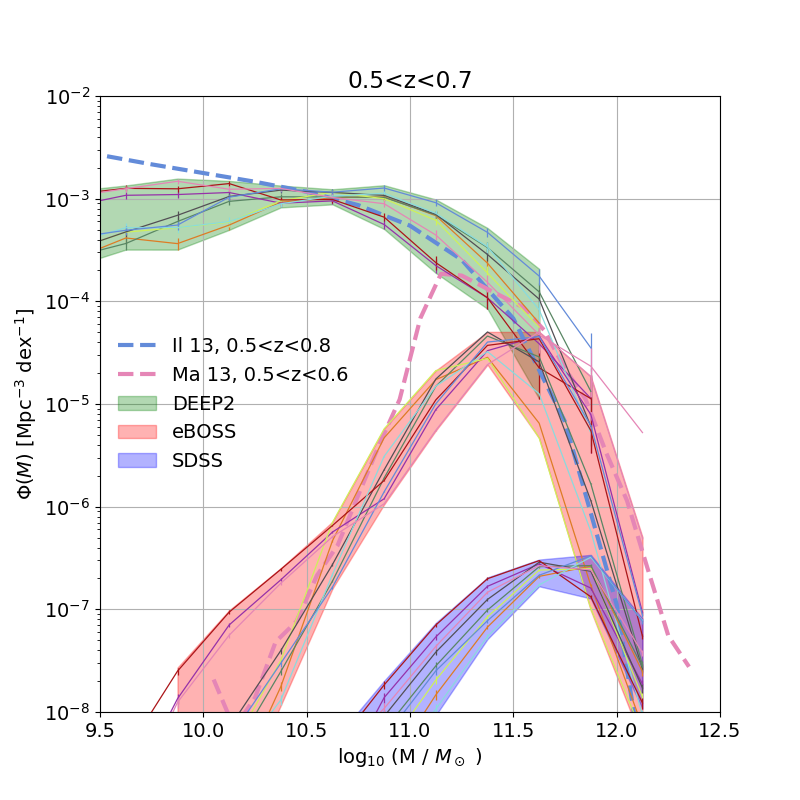}
\caption{\label{fig:smf:all}Observed stellar mass function measured on SDSS, eBOSS and DEEP2 samples containing only well constrained stellar masses. 
Two redshift bins $0.2<z<0.5$, $0.5<z<0.7$ show how each survey spans the stellar mass function. 
For comparison, we added the \citet{Ilbert2013SMF} model (green, they have the exact same redshift bins) and the BOSS SMF from \citet{Maraston2013}, their mean measurement is at redshift 0.55 (magenta line).}
\end{center}
\end{figure}

We compare our results with the stellar mass functions obtained in COSMOS \citet{Ilbert2013SMF} and in BOSS \citep{Maraston2013}. 
We show the results in two redshift bins: $0.2<z<0.5$, $0.5<z<0.8$, see Fig. \ref{fig:smf:all}. 
We see how each sample (DEEP2, SDSS, eBOSS), considering only the tightly constrained stellar masses, is related to the bulk of the galaxy population depicted by the COSMOS and BOSS stellar mass functions. 
This COSMOS stellar mass function is based on a $K$-band selected sample that is known to be biased at low redshift as it misses a fraction of the massive star-forming galaxy population i.e. at the high mass end our measurements are expected to be above that of COSMOS. This is what we find.
The comparison with \citet{Maraston2013} (purple line on the Figure) shows the level of incompleteness we have due to our selection on the error on the stellar mass. 
Indeed we are incomplete as a function of magnitude and position on the sky. 
A detailed stellar mass function accounting for all selection function biases is left for future studies. 

\clearpage
\section{Summary and future releases}

We provide the stellar population properties as obtained by full spectral fitting of models to the observed spectra for SDSS DR14 galaxies, including their high-z extensions (eBOSS), and for the DEEP2 survey mostly containing emission-line galaxies. We adopted the newly released \textsc{Firefly} fitting code coupled with the stellar population  models by \citet{Maraston_2011}. 
This combination has improved the precision of derived parameters. 
In particular, the stellar mass for SDSS galaxies is obtained with a precision of about 0.2 dex for a given IMF, when SNR$>20$. 
Thanks to the high performance computing environment "SCIAMA" at the University of Portsmouth, we could create models of the continuum for nine configurations of IMF and stellar libraries for about 2.7 million galaxies. 
This catalog is the continuation of the Portsmouth SDSS galaxy property catalogs, which were using the spectroscopic redshift combined with the broad-band photometry.
Compared to previous releases, our work doubles the number of galaxies with a tightly constrained stellar mass parameter.

We explore for the first time stellar population modelling of DEEP2 emission line galaxies and find that these galaxies span a variety of properties, which is in broad agreement with predictions of the state-of-the-art semi-analytical models. In particular, DEEP2 galaxies selected by their [OII] luminosity in the redshift range 0.83<z<1.03, have stellar masses with a constant number density in the range $10^9<M(M_{\odot})<10^{11.5}$.

Ongoing \textsc{firefly} developments expected for the next SDSS public release (DR16 onwards) are:
\begin{itemize}
% \item A plugin to the SDSS webapp\footnote{\url{https://dr14.sdss.org/optical/spectrum/view}} to access interactively the continuum models.
\item Emission line measurements on the residuals.
\item An AGN mode added to \textsc{firefly} to allow for fitting all the pixels of AGN spectra \citep[e.g.][]{2017MNRAS.472.4051C}.
\end{itemize}

A variety of science cases will be explored in upcoming companion papers. 
 
\section{Acknowledgements}
JC thanks the \textsc{firefly} team for the team work and this milestone ! 
We all thank Gary Burton for his superb help and technical support with the SCIAMA machine, and Anne-Marie Weijmans for her guidance through the VAC emission.

VGP acknowledges support from the University of Portsmouth through the Dennis Sciama Fellowship award. 
Numerical computations were done on the SCIAMA High Performance Compute cluster which is supported by the ICG, SEPNet and the University of Portsmouth. 
This research used resources of the National Energy Research Scientific Computing Center, a DOE Office of Science User Facility supported by the Office of Science of the U.S. 
Department of Energy under Contract No. DE-AC02-05CH11231. 
This work has benefited from the publicly available programming language {\sc python}.

Funding for the Sloan Digital Sky Survey IV has been provided by the Alfred P. Sloan Foundation, the U.S. Department of Energy Office of Science, and the Participating Institutions. SDSS acknowledges support and resources from the Center for High-Performance Computing at the University of Utah. The SDSS web site is www.sdss.org.

SDSS is managed by the Astrophysical Research Consortium for the Participating Institutions of the SDSS Collaboration including the Brazilian Participation Group, the Carnegie Institution for Science, Carnegie Mellon University, the Chilean Participation Group, the French Participation Group, Harvard-Smithsonian Center for Astrophysics, Instituto de Astrof\'{i}sica de Canarias, The Johns Hopkins University, Kavli Institute for the Physics and Mathematics of the Universe (IPMU) / University of Tokyo, Lawrence Berkeley National Laboratory, Leibniz Institut für Astrophysik Potsdam (AIP), Max-Planck-Institut für Astronomie (MPIA Heidelberg), Max-Planck-Institut für Astrophysik (MPA Garching), Max-Planck-Institut für Extraterrestrische Physik (MPE), National Astronomical Observatories of China, New Mexico State University, New York University, University of Notre Dame, Observatório Nacional / MCTI, The Ohio State University, Pennsylvania State University, Shanghai Astronomical Observatory, United Kingdom Participation Group, Universidad Nacional Autónoma de México, University of Arizona, University of Colorado Boulder, University of Oxford, University of Portsmouth, University of Utah, University of Virginia, University of Washington, University of Wisconsin, Vanderbilt University, and Yale University.

\bibliographystyle{aa}
\bibliography{biblio}

\clearpage
\appendix
\section{Description of the \textsc{firefly} data}

\subsection{Processing}
\label{subsec:processing}

The processing was done on  SCIAMA\footnote{\url{http://www.sciama.icg.port.ac.uk/}}, a high performance computing facility belonging to the University of Portsmouth (United Kingdom). 
A fit for a single model takes about a minute cpu so that the whole run required about 350,000 cpu hours. 
The total data volume is about 3.2T, as in Table \ref{table:processing}.
\begin{table}
\caption{\label{table:processing} Data volume generated in this study.}
\begin{center}
\begin{tabular}{ccccc}
\hline \hline
Survey &
catalog &
spectra & 
models & total \\ \hline
SDSS
& 23G
& 278G
& 860G 
& 1.1T \\
eBOSS
& 49G
& 0.5T 
& 1.5T 
& 2.1T \\
DEEP2 
& 1.4G
& 6.2G
& 17G
& 24.6G \\
\hline 
\end{tabular}
\end{center}
\end{table}

\subsection{Public \textsc{firefly} catalogs and data}
\label{subsec:public:data}

For SDSS, all products are available here 
\url{https://firefly.mpe.mpg.de/v1_1_0/26}. 
For eBOSS, all products are available here 
\url{https://firefly.mpe.mpg.de/v1_1_0/v5_10_0}. 
A previous version of the SDSS/eBOSS firefly catalogs is available here 
\url{https://data.sdss.org/sas/dr14/eboss/spectro/firefly/v1_0_4}.
For DEEP2, they are available here 
\url{https://firefly.mpe.mpg.de/v1_1_0/DEEP2} (both versions).
The data model is available at \url{http://www.sdss.org/dr14/spectro/eboss-firefly-value-added-catalog/}.

\subsubsection{Model spectrum}
The lowest level data product is the model file. 
There is one such file per galaxy spectrum considered. 
It is available for both DEEP2 and SDSS data sets. 
It is a fits file with a header and 9 data units. 
The primary header contains all input parameters used during the fit.
The following nine data units each contain 
\begin{itemize}
\item Header. The best fit parameters for each SSP entering the best model. 
\item Data extension. The best-fitting model spectrum: wavelength (Unit Angstrom) and model flux ($f_\lambda$ convention, unit $10^{-17}$ erg cm$^{-2}$ s$^{-1}$ A$^{-1}$). 
\end{itemize} 

The nine data units contain the results obtained for the different combinations of stellar libraries and initial mass functions. 
\begin{itemize}
\item HDU1 M11-MILES x Chabrier IMF
\item HDU2 M11-MILES x Salpeter IMF
\item HDU3 M11-MILES x Kroupa IMF
\item HDU4 M11-ELODIE x Chabrier IMF
\item HDU5 M11-ELODIE x Salpeter IMF
\item HDU6 M11-ELODIE x Kroupa IMF
\item HDU7 M11-STELIB x Chabrier IMF
\item HDU8 M11-STELIB x Salpeter IMF
\item HDU9 M11-STELIB x Kroupa IMF
\end{itemize}

The data model for this product, named 'spFly', is available here \url{https://data.sdss.org/datamodel/files/EBOSS_FIREFLY/FIREFLY_VER/RUN2D/SPMODELS_VER/PLATE/spFly.html} and the corresponding files here \url{https://firefly.mpe.mpg.de/v1_1_0/26/stellarpop} and here \url{https://firefly.mpe.mpg.de/v1_1_0/v5_10_0/stellarpop}. 
The DEEP2 result model spectra are provided here \url{http://www.mpe.mpg.de/~comparat/DEEP2/stellarpop/} and follow the naming convention 'spFly-deep2-MASK-OBJNUM.fits'. 
An automated summary plot is provided for each model file. 
It illustrates the spectrum, the models, and the fitted parameters, see Fig. \ref{fig:firefly:output:sdss}, \ref{fig:firefly:output:boss}, \ref{fig:firefly:output:deep2}. 

\subsubsection{Plate-level summary catalogs (SDSS only)}
Due to the size of the SDSS data set and its structure, we create summary catalogs for each plate, named 'spFlyPlate'.
They contain all output parameters from the fitting procedure. 
The data model for these catalogs is given here 
\url{https://data.sdss.org/datamodel/files/EBOSS_FIREFLY/FIREFLY_VER/RUN2D/SPMODELS_VER/PLATE/spFlyPlate.html}. 

\subsubsection{Summary catalogs}
Top-level summary catalogs contain all of the fitted parameters. They are available here 
\begin{itemize}
\item SDSS: \url{https://firefly.mpe.mpg.de/v1_1_0/26/sdss_firefly-26.fits}
\item eBOSS: \url{https://firefly.mpe.mpg.de/v1_1_0/v5_10_0/eboss_firefly-v5_10_0.fits}
\item DEEP2: \url{https://firefly.mpe.mpg.de/v1_1_0/DEEP2/catalogs/zcat.deep2.dr4.v4.LFcatalogTC.Planck13.spm.v2.fits.gz}
\end{itemize}
They can be download as follows
\begin{verbatim}
wget --no-check-certificate https://firefly.mpe.mpg.de/v1_1_0/26/sdss_firefly-26.fits .
\end{verbatim}

The data model for the summary file for the eBOSS data is available at: 
\url{https://data.sdss.org/datamodel/files/EBOSS_FIREFLY/FIREFLY_VER/RUN2D/eboss_firefly.html} while the one for the SDSS data \url{https://data.sdss.org/datamodel/files/EBOSS_FIREFLY/FIREFLY_VER/RUN2D/sdss_firefly.html}.
They differ by the redshift column: 'Z' for SDSS and 'Z\_NOQSO' for eBOSS.  

Assumptions in the IMF or templates induce systematic differences in the constrained parameters. 
Therefore, we do not provide a mean stellar mass based on these nine runs.

The SDSS and eBOSS summary files contain the columns described in Table \ref{table:data:model}. The DEEP2 summary file columns are described in the readme on the website.
\section{Warning about fits with unconstrained parameters}

We would like to warn the future users about fits that led to unconstrained age, mass or metallicity parameters. 
Indeed for every spectrum, firefly provides an answer and the full pdf of each parameter. But sometimes, the pdf is so large that the uncertainties are as large as the parameters space itself. 
In such cases the value output should not be trusted. 

\section{Large tables}

\begin{table*}
\begin{center}
\caption{\label{ref:table:sdss:src:firefly}SDSS data sorted per source type. Ordered by decreasing number of galaxies (third column). 
The second column, 'N', gives the number of spectra labeled under this sourcetype. 
The third column, 'N galaxies', gives the number of spectra considered as galaxies and their fraction relatively to the second column. 
The fourth column, 'SNR ALL$>0$', gives the number of spectra considered as galaxies with a positive median SNR over its spectrum. We consider this column as the reference sample to be fitted by \textsc{firefly}, hence the `100\%'. 
The fifth column, 'firefly constrained', gives the fraction of constrained fits and their fraction relative to the reference sample (column 4). 
For the `GALAXY' sourcetype (first line), it is very high at $810540/829464 \sim 97.7\%$. 
The last two columns, '$\sigma_{\log_M}<0.4$' ('$\sigma_{\log_M}<0.2$'), give the fraction of fits that have a stellar mass constrained within 0.4 (0.2) dex. 
It is around 80\% for the GALAXY and QA sourcetype. 
}
\begin{tabular}{c rrr rrr rrrrrrrrrrrrrr}
\hline \hline
sourcetype & N & \multicolumn{2}{c}{N galaxies} & \multicolumn{2}{c}{SNR ALL$>0$} & \multicolumn{2}{c}{firefly constrained} & \multicolumn{2}{c}{$\sigma_{\log_M}<0.4$} & \multicolumn{2}{c}{$\sigma_{\log_M}<0.2$} \\ 
 & N & N & \% & N & \% & N & \% & N & \% & N & \%  \\ 
\hline 
GALAXY & 858244 & 829464 & 96 & 829464 & 100 & 810540 & 97 & 794149 & 95 & 673690 & 81 \\ 
NONLEGACY & 294745 & 75154 & 25 & 74899 & 100 & 68906 & 92 & 57586 & 76 & 33134 & 44 \\ 
QSO & 168563 & 22724 & 13 & 22724 & 100 & 19803 & 87 & 15252 & 67 & 8167 & 35 \\ 
QA & 20408 & 14608 & 71 & 14608 & 100 & 14286 & 97 & 13919 & 95 & 11630 & 79 \\ 
SERENDIPITY\_FIRST & 7526 & 4163 & 55 & 4163 & 100 & 3953 & 95 & 3576 & 85 & 2249 & 54 \\ 
ROSAT\_D & 7057 & 1704 & 24 & 1704 & 100 & 1642 & 96 & 1350 & 79 & 739 & 43 \\ 
SERENDIPITY\_DISTANT & 4911 & 251 & 5 & 251 & 100 & 221 & 88 & 146 & 58 & 68 & 27 \\ 
SERENDIPITY\_BLUE & 22871 & 70 & 0 & 70 & 100 & 54 & 77 & 31 & 44 & 12 & 17 \\ 
STAR\_WHITE\_DWARF & 2290 & 32 & 1 & 32 & 100 & 26 & 81 & 16 & 50 & 10 & 31 \\ 
SERENDIPITY\_MANUAL & 68 & 27 & 39 & 27 & 100 & 25 & 92 & 20 & 74 & 14 & 51 \\ 
STAR\_CARBON & 3458 & 22 & 0 & 22 & 100 & 20 & 90 & 15 & 68 & 9 & 40 \\ 
REDDEN\_STD & 15363 & 13 & 0 & 13 & 100 & 11 & 84 & 10 & 76 & 8 & 61 \\ 
STAR\_BHB & 14230 & 12 & 0 & 12 & 100 & 9 & 75 & 7 & 58 & 4 & 33 \\ 
STAR\_PN & 14 & 8 & 57 & 8 & 100 & 6 & 75 & 0 & 0 & 0 & 0 \\ 
SERENDIPITY\_RED & 2771 & 2 & 0 & 2 & 100 & 1 & 50 & 0 & 0 & 0 & 0 \\ 
HOT\_STD & 3413 & 2 & 0 & 2 & 100 & 2 & 100 & 2 & 100 & 1 & 50 \\ 
SPECTROPHOTO\_STD & 15366 & 1 & 0 & 1 & 100 & 1 & 100 & 1 & 100 & 1 & 100 \\ 
STAR\_BROWN\_DWARF & 560 & 1 & 0 & 1 & 100 & 0 & 0 & 0 & 0 & 0 & 0 \\ 
STAR\_CATY\_VAR & 6853 & 1 & 0 & 1 & 100 & 1 & 100 & 1 & 100 & 0 & 0 \\ 
SKY & 61967 & 0 & 0 & 0 & 100 & 0 & 0 & 0 & 0 & 0 & 0 \\ 
STAR\_SUB\_DWARF & 1226 & 0 & 0 & 0 & 100 & 0 & 0 & 0 & 0 & 0 & 0 \\ 
STAR\_RED\_DWARF & 14496 & 0 & 0 & 0 & 100 & 0 & 0 & 0 & 0 & 0 & 0 \\ 
\hline
\end{tabular}
\end{center}
\end{table*}

\begin{center}
\begin{longtable}{c rrr rrr rrrrrrrrrrrrrr}
\caption{\label{ref:table:boss:src:firefly} Same as Table \ref{ref:table:sdss:src:firefly} for eBOSS.}\\
\hline \hline
source type & N & \multicolumn{2}{c}{N galaxies} & \multicolumn{2}{c}{SNR ALL$>0$} & \multicolumn{2}{c}{firefly constrained} & \multicolumn{2}{c}{$\sigma_{\log_M}<0.4$} & \multicolumn{2}{c}{$\sigma_{\log_M}<0.2$} \\ 
 & N & N & \% & N & \% & N & \% & N & \% & N & \%  \\ 
\hline 
\endfirsthead

\caption{continued.}\\
\hline \hline
source type & N & \multicolumn{2}{c}{N galaxies} & \multicolumn{2}{c}{SNR ALL$>0$} & \multicolumn{2}{c}{firefly constrained} & \multicolumn{2}{c}{$\sigma_{\log_M}<0.4$} & \multicolumn{2}{c}{$\sigma_{\log_M}<0.2$} \\ 
 & N & N & \% & N & \% & N & \% & N & \% & N & \%  \\ 
\hline 
\endhead

\hline \multicolumn{3}{|r|}{{Continued on next page}} \\ \hline
\endfoot

\hline \hline
\endlastfoot
LRG & 1579529 & 1479496 & 93 & 906577 & 100 & 887235 & 97 & 769382 & 84 & 546392 & 60 \\ 
QSO & 645501 & 116243 & 18 & 61892 & 100 & 43481 & 70 & 30035 & 48 & 12669 & 20 \\ 
SEQUELS\_TARGET & 100810 & 38871 & 38 & 38871 & 100 & 36251 & 93 & 26190 & 67 & 11253 & 28 \\ 
SPIDERS\_RASS\_CLUS & 10623 & 10433 & 98 & 2573 & 100 & 2555 & 99 & 2530 & 98 & 2279 & 88 \\ 
WISE\_COMPLETE & 10019 & 9113 & 91 & 0 & 100 & 0 & 0 & 0 & 0 & 0 & 0 \\ 
HIZ\_LRG & 9541 & 7463 & 78 & 0 & 100 & 0 & 0 & 0 & 0 & 0 & 0 \\ 
RM\_TILE1 & 11270 & 7342 & 65 & 7342 & 100 & 6351 & 86 & 5579 & 76 & 3075 & 41 \\ 
WISE\_BOSS\_QSO & 15851 & 6664 & 42 & 4512 & 100 & 3732 & 82 & 3182 & 70 & 1957 & 43 \\ 
RM\_TILE2 & 30331 & 6475 & 21 & 6475 & 100 & 4586 & 70 & 3391 & 52 & 1847 & 28 \\ 
BRIGHTGAL & 5565 & 5361 & 96 & 3541 & 100 & 3403 & 96 & 3370 & 95 & 3292 & 93 \\ 
ELG\_TEST1 & 7200 & 4525 & 62 & 0 & 100 & 0 & 0 & 0 & 0 & 0 & 0 \\ 
ELG1\_EBOSS & 4741 & 3616 & 76 & 0 & 100 & 0 & 0 & 0 & 0 & 0 & 0 \\ 
SN\_GAL1 & 4042 & 3502 & 86 & 3496 & 100 & 3345 & 95 & 2797 & 80 & 1900 & 54 \\ 
ELG\_DECALS\_TEST1 & 5356 & 3423 & 63 & 0 & 100 & 0 & 0 & 0 & 0 & 0 & 0 \\ 
SEQUELS\_ELG & 4217 & 3059 & 72 & 3059 & 100 & 2763 & 90 & 1290 & 42 & 175 & 5 \\ 
BLUE\_RADIO & 2920 & 2874 & 98 & 1791 & 100 & 1744 & 97 & 1687 & 94 & 1390 & 77 \\ 
QSO\_VAR\_LF & 9069 & 2507 & 27 & 2313 & 100 & 1888 & 81 & 1164 & 50 & 456 & 19 \\ 
QSO\_WISE\_FULL\_SKY & 8271 & 2482 & 30 & 1247 & 100 & 1037 & 83 & 856 & 68 & 408 & 32 \\ 
QSO\_VAR\_SDSS & 22752 & 2297 & 10 & 1527 & 100 & 1205 & 78 & 855 & 56 & 339 & 22 \\ 
SEQUELS\_ELG\_LOWP & 2982 & 2269 & 76 & 2269 & 100 & 2025 & 89 & 922 & 40 & 153 & 6 \\ 
FAINT\_ELG & 2699 & 2235 & 82 & 2235 & 100 & 1942 & 86 & 855 & 38 & 85 & 3 \\ 
QSO1\_REOBS & 16021 & 2153 & 13 & 443 & 100 & 242 & 54 & 142 & 32 & 41 & 9 \\ 
CLUSTER\_MEMBER & 2072 & 2056 & 99 & 577 & 100 & 568 & 98 & 565 & 97 & 521 & 90 \\ 
LRG\_ROUND3 & 2486 & 2032 & 81 & 0 & 100 & 0 & 0 & 0 & 0 & 0 & 0 \\ 
GAL\_NEAR\_QSO & 2037 & 1954 & 95 & 1140 & 100 & 1111 & 97 & 990 & 86 & 670 & 58 \\ 
TDSS\_B & 18880 & 1734 & 9 & 489 & 100 & 410 & 83 & 358 & 73 & 225 & 46 \\ 
ELG & 3479 & 1650 & 47 & 409 & 100 & 348 & 85 & 153 & 37 & 25 & 6 \\ 
S82X\_TILE1 & 2775 & 1608 & 57 & 0 & 100 & 0 & 0 & 0 & 0 & 0 & 0 \\ 
QSO\_SUPPZ & 4839 & 1512 & 31 & 1213 & 100 & 825 & 68 & 749 & 61 & 578 & 47 \\ 
STRIPE82BCG & 1407 & 1388 & 98 & 1380 & 100 & 1359 & 98 & 1302 & 94 & 1122 & 81 \\ 
QSO\_WISE\_SUPP & 3992 & 1065 & 26 & 1065 & 100 & 945 & 88 & 771 & 72 & 366 & 34 \\ 
SPIDERS\_RASS\_AGN & 1378 & 1064 & 77 & 275 & 100 & 236 & 85 & 224 & 81 & 181 & 65 \\ 
XMM\_PRIME & 2444 & 1052 & 43 & 1052 & 100 & 911 & 86 & 761 & 72 & 447 & 42 \\ 
QSO\_EBOSS\_W3\_ADM & 4383 & 1044 & 23 & 1044 & 100 & 915 & 87 & 536 & 51 & 149 & 14 \\ 
TAU\_STAR & 1468 & 784 & 53 & 784 & 100 & 123 & 15 & 28 & 3 & 5 & 0 \\ 
ELAIS\_N1\_GMRT\_TAYLOR & 1032 & 766 & 74 & 766 & 100 & 731 & 95 & 550 & 71 & 313 & 40 \\ 
QSO\_GRI & 1963 & 696 & 35 & 503 & 100 & 406 & 80 & 303 & 60 & 148 & 29 \\ 
QSO\_DEEP & 3358 & 661 & 19 & 661 & 100 & 576 & 87 & 257 & 38 & 29 & 4 \\ 
QSO1\_VAR\_S82 & 5499 & 624 & 11 & 0 & 100 & 0 & 0 & 0 & 0 & 0 & 0 \\ 
S82X\_TILE2 & 2621 & 624 & 23 & 0 & 100 & 0 & 0 & 0 & 0 & 0 & 0 \\ 
VARBAL & 1077 & 616 & 57 & 399 & 100 & 283 & 70 & 270 & 67 & 223 & 55 \\ 
CHANDRAv1 & 1179 & 601 & 51 & 383 & 100 & 349 & 91 & 298 & 77 & 176 & 46 \\ 
XMMHR & 731 & 575 & 78 & 360 & 100 & 346 & 96 & 304 & 84 & 217 & 60 \\ 
SPIDERS\_XCLASS\_CLUS & 568 & 559 & 98 & 85 & 100 & 85 & 100 & 84 & 98 & 70 & 82 \\ 
FAINT\_HIZ\_LRG & 685 & 543 & 79 & 543 & 100 & 525 & 96 & 369 & 68 & 120 & 22 \\ 
ELG\_UGRIZW\_TEST1 & 892 & 535 & 60 & 0 & 100 & 0 & 0 & 0 & 0 & 0 & 0 \\ 
ELG1\_EXTENDED & 659 & 500 & 75 & 0 & 100 & 0 & 0 & 0 & 0 & 0 & 0 \\ 
TEMPLATE\_GAL\_PHOTO & 604 & 485 & 80 & 485 & 100 & 460 & 94 & 395 & 81 & 271 & 55 \\ 
XMMBRIGHT & 622 & 461 & 74 & 317 & 100 & 292 & 92 & 254 & 80 & 187 & 59 \\ 
XMMRED & 524 & 434 & 82 & 319 & 100 & 304 & 95 & 298 & 93 & 262 & 82 \\ 
QSO\_AALs & 680 & 379 & 55 & 249 & 100 & 162 & 65 & 156 & 62 & 132 & 53 \\ 
SN\_LOC & 416 & 373 & 89 & 372 & 100 & 355 & 95 & 295 & 79 & 213 & 57 \\ 
TDSS\_FES\_VARBAL & 769 & 365 & 47 & 119 & 100 & 81 & 68 & 67 & 56 & 56 & 47 \\ 
25ORI\_WISE\_W3 & 484 & 349 & 72 & 349 & 100 & 325 & 93 & 317 & 90 & 264 & 75 \\ 
XMM\_SECOND & 669 & 332 & 49 & 332 & 100 & 305 & 91 & 254 & 76 & 163 & 49 \\ 
QSO\_AAL & 507 & 327 & 64 & 208 & 100 & 125 & 60 & 114 & 54 & 99 & 47 \\ 
CXORED & 337 & 311 & 92 & 193 & 100 & 192 & 99 & 188 & 97 & 166 & 86 \\ 
TDSS\_CP & 1054 & 297 & 28 & 116 & 100 & 90 & 77 & 71 & 61 & 29 & 25 \\ 
BLAZXR & 355 & 287 & 80 & 186 & 100 & 160 & 86 & 152 & 81 & 131 & 70 \\ 
ELAIS\_N1\_LOFAR & 413 & 285 & 69 & 285 & 100 & 273 & 95 & 227 & 79 & 162 & 56 \\ 
KOEKAPbSTAR & 542 & 282 & 52 & 282 & 100 & 1 & 0 & 1 & 0 & 0 & 0 \\ 
ELAIS\_N1\_GMRT\_GARN & 367 & 280 & 76 & 280 & 100 & 266 & 95 & 223 & 79 & 154 & 55 \\ 
QSO\_VAR & 1043 & 273 & 26 & 273 & 100 & 226 & 82 & 197 & 72 & 140 & 51 \\ 
QSO\_RIZ & 1471 & 262 & 17 & 206 & 100 & 183 & 88 & 130 & 63 & 63 & 30 \\ 
ELAIS\_N1\_FIRST & 324 & 256 & 79 & 256 & 100 & 241 & 94 & 211 & 82 & 151 & 59 \\ 
RADIO\_2LOBE\_QSO & 1084 & 244 & 22 & 148 & 100 & 130 & 87 & 104 & 70 & 52 & 35 \\ 
TEMPLATE\_QSO\_SDSS1 & 496 & 233 & 47 & 233 & 100 & 170 & 73 & 133 & 57 & 71 & 30 \\ 
ELG\_DES\_TEST1 & 450 & 230 & 51 & 0 & 100 & 0 & 0 & 0 & 0 & 0 & 0 \\ 
PTF\_GAL & 178 & 174 & 97 & 67 & 100 & 64 & 95 & 64 & 95 & 55 & 82 \\ 
ELG\_GRIW\_TEST1 & 275 & 172 & 62 & 0 & 100 & 0 & 0 & 0 & 0 & 0 & 0 \\ 
QSO\_IAL & 292 & 172 & 58 & 117 & 100 & 79 & 67 & 75 & 64 & 62 & 53 \\ 
SPIDERS\_PILOT & 446 & 167 & 37 & 9 & 100 & 8 & 88 & 8 & 88 & 4 & 44 \\ 
TDSS\_FES\_NQHISN & 169 & 158 & 93 & 3 & 100 & 1 & 33 & 1 & 33 & 1 & 33 \\ 
WHITEDWARF\_SDSS & 3898 & 157 & 4 & 112 & 100 & 46 & 41 & 36 & 32 & 27 & 24 \\ 
25ORI\_WISE & 290 & 154 & 53 & 154 & 100 & 148 & 96 & 138 & 89 & 115 & 74 \\ 
QSO\_RADIO & 239 & 148 & 61 & 96 & 100 & 69 & 71 & 68 & 70 & 57 & 59 \\ 
CXOBRIGHT & 169 & 145 & 85 & 104 & 100 & 90 & 86 & 83 & 79 & 63 & 60 \\ 
TDSS\_FES\_DE & 146 & 143 & 97 & 2 & 100 & 2 & 100 & 1 & 50 & 1 & 50 \\ 
ELG\_UGRIZWbright\_TEST1 & 183 & 142 & 77 & 0 & 100 & 0 & 0 & 0 & 0 & 0 & 0 \\ 
TDSS\_FES\_HYPQSO & 352 & 138 & 39 & 4 & 100 & 3 & 75 & 1 & 25 & 0 & 0 \\ 
SPIDERS\_XMMSL\_AGN & 170 & 138 & 81 & 51 & 100 & 40 & 78 & 39 & 76 & 29 & 56 \\ 
KOE2068\_STAR & 276 & 129 & 46 & 129 & 100 & 46 & 35 & 21 & 16 & 9 & 7 \\ 
TDSS\_PILOT & 998 & 120 & 12 & 96 & 100 & 78 & 81 & 47 & 49 & 18 & 18 \\ 
QSO\_VAR\_FPG & 614 & 117 & 19 & 115 & 100 & 94 & 81 & 63 & 54 & 19 & 16 \\ 
WHITEDWARF\_NEW & 4989 & 114 & 2 & 70 & 100 & 36 & 51 & 31 & 44 & 22 & 31 \\ 
QSO\_XD\_KDE\_PAIR & 631 & 112 & 17 & 43 & 100 & 33 & 76 & 27 & 62 & 13 & 30 \\ 
TEMPLATE\_STAR\_PHOTO & 462 & 111 & 24 & 111 & 100 & 86 & 77 & 75 & 67 & 47 & 42 \\ 
QSO\_RADIO\_AAL & 150 & 110 & 73 & 78 & 100 & 53 & 67 & 50 & 64 & 44 & 56 \\ 
DISKEMITTER\_REPEAT & 98 & 98 & 100 & 62 & 100 & 34 & 54 & 30 & 48 & 25 & 40 \\ 
SN\_GAL2 & 107 & 92 & 86 & 92 & 100 & 87 & 94 & 77 & 83 & 57 & 62 \\ 
KOEKAP\_STAR & 302 & 80 & 26 & 80 & 100 & 12 & 15 & 5 & 6 & 3 & 3 \\ 
BLAZGXR & 136 & 79 & 58 & 50 & 100 & 44 & 88 & 40 & 80 & 35 & 70 \\ 
KOE2023\_STAR & 234 & 64 & 27 & 64 & 100 & 23 & 35 & 9 & 14 & 5 & 7 \\ 
VARS & 134 & 52 & 38 & 52 & 100 & 39 & 75 & 34 & 65 & 21 & 40 \\ 
QSO\_RADIO\_IAL & 77 & 45 & 58 & 34 & 100 & 26 & 76 & 25 & 73 & 22 & 64 \\ 
LBG & 242 & 45 & 18 & 45 & 100 & 39 & 86 & 14 & 31 & 1 & 2 \\ 
IAMASERS & 48 & 43 & 89 & 11 & 100 & 11 & 100 & 11 & 100 & 11 & 100 \\ 
RQSS\_SF & 103 & 35 & 34 & 35 & 100 & 33 & 94 & 17 & 48 & 2 & 5 \\ 
BLAZGXQSO & 51 & 35 & 68 & 19 & 100 & 11 & 57 & 11 & 57 & 9 & 47 \\ 
ELAIS\_N1\_JVLA & 58 & 34 & 58 & 34 & 100 & 31 & 91 & 22 & 64 & 13 & 38 \\ 
TDSS\_SPIDERS\_PILOT & 113 & 34 & 30 & 3 & 100 & 3 & 100 & 2 & 66 & 2 & 66 \\ 
QSO\_noAALs & 66 & 33 & 50 & 16 & 100 & 12 & 75 & 12 & 75 & 10 & 62 \\ 
XMMGRIZ & 98 & 33 & 33 & 24 & 100 & 23 & 95 & 12 & 50 & 4 & 16 \\ 
KQSO\_BOSS & 136 & 32 & 23 & 32 & 100 & 28 & 87 & 23 & 71 & 19 & 59 \\ 
BLAZGRQSO & 75 & 29 & 38 & 16 & 100 & 13 & 81 & 13 & 81 & 7 & 43 \\ 
BLAZGRFLAT & 71 & 27 & 38 & 21 & 100 & 19 & 90 & 17 & 81 & 17 & 81 \\ 
STD & 62522 & 25 & 0 & 19 & 100 & 10 & 52 & 10 & 52 & 7 & 36 \\ 
SN\_GAL3 & 25 & 23 & 92 & 23 & 100 & 22 & 95 & 21 & 91 & 16 & 69 \\ 
KOE2023bSTAR & 563 & 19 & 3 & 19 & 100 & 4 & 21 & 3 & 15 & 1 & 5 \\ 
KOE2068bSTAR & 602 & 17 & 2 & 17 & 100 & 4 & 23 & 4 & 23 & 1 & 5 \\ 
CXOGRIZ & 32 & 14 & 43 & 7 & 100 & 6 & 85 & 4 & 57 & 3 & 42 \\ 
ODDBAL & 21 & 13 & 61 & 8 & 100 & 6 & 75 & 6 & 75 & 2 & 25 \\ 
TDSS\_FES\_MGII & 24 & 12 & 50 & 0 & 100 & 0 & 0 & 0 & 0 & 0 & 0 \\ 
OTBAL & 12 & 11 & 91 & 8 & 100 & 7 & 87 & 7 & 87 & 4 & 50 \\ 
XMMSDSS & 15 & 10 & 66 & 0 & 100 & 0 & 0 & 0 & 0 & 0 & 0 \\ 
RQSS\_SFC & 15 & 10 & 66 & 10 & 100 & 9 & 90 & 9 & 90 & 2 & 20 \\ 
SEGUE1 & 2805 & 9 & 0 & 9 & 100 & 3 & 33 & 2 & 22 & 2 & 22 \\ 
SDSSFILLER & 2485 & 9 & 0 & 9 & 100 & 5 & 55 & 5 & 55 & 5 & 55 \\ 
TDSS\_PILOT\_SNHOST & 8 & 7 & 87 & 7 & 100 & 7 & 100 & 6 & 85 & 5 & 71 \\ 
BLAZGX & 18 & 7 & 38 & 6 & 100 & 5 & 83 & 5 & 83 & 4 & 66 \\ 
SEGUE2 & 1953 & 7 & 0 & 7 & 100 & 3 & 42 & 1 & 14 & 0 & 0 \\ 
FBQSBAL & 8 & 6 & 75 & 4 & 100 & 0 & 0 & 0 & 0 & 0 & 0 \\ 
PREVBAL & 25 & 5 & 20 & 5 & 100 & 3 & 60 & 2 & 40 & 0 & 0 \\ 
BLAZR & 6 & 4 & 66 & 4 & 100 & 4 & 100 & 4 & 100 & 3 & 75 \\ 
fainterM & 2806 & 4 & 0 & 4 & 100 & 3 & 75 & 1 & 25 & 0 & 0 \\ 
QSO\_HIZ & 520 & 4 & 0 & 4 & 100 & 4 & 100 & 4 & 100 & 1 & 25 \\ 
STD\_WD & 543 & 3 & 0 & 0 & 100 & 0 & 0 & 0 & 0 & 0 & 0 \\ 
BLAZXRSAM & 4 & 3 & 75 & 3 & 100 & 2 & 66 & 2 & 66 & 2 & 66 \\ 
AMC & 21 & 2 & 9 & 2 & 100 & 0 & 0 & 0 & 0 & 0 & 0 \\ 
RED\_KG & 10457 & 2 & 0 & 2 & 100 & 1 & 50 & 1 & 50 & 1 & 50 \\ 
SPEC\_SN & 2 & 2 & 100 & 2 & 100 & 2 & 100 & 2 & 100 & 2 & 100 \\ 
HIZQSO82 & 64 & 2 & 3 & 2 & 100 & 2 & 100 & 1 & 50 & 0 & 0 \\ 
LBQSBAL & 6 & 1 & 16 & 1 & 100 & 0 & 0 & 0 & 0 & 0 & 0 \\ 
LOW\_MET & 53 & 1 & 1 & 1 & 100 & 1 & 100 & 1 & 100 & 1 & 100 \\ 
SPOKE & 1229 & 1 & 0 & 1 & 100 & 0 & 0 & 0 & 0 & 0 & 0 \\ 
TDSS\_FES\_HYPSTAR & 208 & 1 & 0 & 0 & 100 & 0 & 0 & 0 & 0 & 0 & 0 \\ 
S82X\_TILE3 & 4 & 1 & 25 & 0 & 100 & 0 & 0 & 0 & 0 & 0 & 0 \\ 
BLAZGVAR & 2 & 1 & 50 & 1 & 100 & 1 & 100 & 0 & 0 & 0 & 0 \\ 
TDSS\_PILOT\_PM & 132 & 1 & 0 & 1 & 100 & 1 & 100 & 0 & 0 & 0 & 0 \\ 
HIZQSOIR & 121 & 1 & 0 & 0 & 100 & 0 & 0 & 0 & 0 & 0 & 0 \\ 
HPM & 75 & 1 & 1 & 1 & 100 & 0 & 0 & 0 & 0 & 0 & 0 \\ 
fainterL & 1276 & 1 & 0 & 0 & 100 & 0 & 0 & 0 & 0 & 0 & 0 \\ 
brighterM & 1787 & 0 & 0 & 0 & 100 & 0 & 0 & 0 & 0 & 0 & 0 \\ 
TEMPLATE\_STAR\_SPECTRO & 160 & 0 & 0 & 0 & 100 & 0 & 0 & 0 & 0 & 0 & 0 \\ 
TDSS\_FES\_WDDM & 54 & 0 & 0 & 0 & 100 & 0 & 0 & 0 & 0 & 0 & 0 \\ 
TDSS\_FES\_DWARFC & 96 & 0 & 0 & 0 & 100 & 0 & 0 & 0 & 0 & 0 & 0 \\ 
brighterL & 422 & 0 & 0 & 0 & 100 & 0 & 0 & 0 & 0 & 0 & 0 \\ 
TDSS\_FES\_ACTSTAR & 154 & 0 & 0 & 0 & 100 & 0 & 0 & 0 & 0 & 0 & 0 \\ 
FLARE1 & 30 & 0 & 0 & 0 & 100 & 0 & 0 & 0 & 0 & 0 & 0 \\ 
BLAZXRVAR & 1 & 0 & 0 & 0 & 100 & 0 & 0 & 0 & 0 & 0 & 0 \\ 
SPOKE2 & 95 & 0 & 0 & 0 & 100 & 0 & 0 & 0 & 0 & 0 & 0 \\ 
CALSTAR & 40 & 0 & 0 & 0 & 100 & 0 & 0 & 0 & 0 & 0 & 0 \\ 
FLARE2 & 59 & 0 & 0 & 0 & 100 & 0 & 0 & 0 & 0 & 0 & 0 \\ 
RVTEST & 84 & 0 & 0 & 0 & 100 & 0 & 0 & 0 & 0 & 0 & 0 \\ 
RQSS\_STMC & 2 & 0 & 0 & 0 & 100 & 0 & 0 & 0 & 0 & 0 & 0 \\ 
RQSS\_STM & 6 & 0 & 0 & 0 & 100 & 0 & 0 & 0 & 0 & 0 & 0 \\ 
QSO\_STD & 1776 & 0 & 0 & 0 & 100 & 0 & 0 & 0 & 0 & 0 & 0 \\ 
NA & 280513 & 0 & 0 & 0 & 100 & 0 & 0 & 0 & 0 & 0 & 0 \\ 
MTEMP & 75 & 0 & 0 & 0 & 100 & 0 & 0 & 0 & 0 & 0 & 0 \\ 
GES & 263 & 0 & 0 & 0 & 100 & 0 & 0 & 0 & 0 & 0 & 0 \\ 
\hline
\end{longtable}
\end{center}

\clearpage

\begin{table*}
\begin{center}
\caption{\label{ref:table:sdss:src:SNR}SDSS sourcetypes containing more than 100 galaxies (column 3 of the previous table). 
We divide the data in five redshift using the boundaries $0$, $0.025$, $0.375$, $0.7$, $0.85$ and $1.6$. 
In each bin, we show the percentiles corresponding to SNR 5 and 20 around $(1+z)4000$ \text{\AA}. 
The percentages shown correspond to the fraction of the data in the redshift bin with a SNR lower than 5 (20). 
The lower the percentile the higher the fraction of higher SNR data. 
More than half of the spectra, with a sourcetype='GALAXY' and a redshift lower than 0.375, has a signal to noise ratio above 20.
All the data at redshift higher than 0.375 has a median SNR lower than 5. 
'-1' means that data is not available. 
}
\begin{tabular}{c r rrrr rrrr rrrrr rrrrr rrrrr}
\hline \hline
programname & N galaxies &  \multicolumn{3}{c}{$0<z<0.025$} &  \multicolumn{3}{c}{$0.025<z<0.375$} \\
       		&   & N & \%$_{5}$ & \%$_{20}$ & N & \%$_{5}$ & \%$_{20}$ \\
\hline
GALAXY              & 829464 & 24203 & 8   & 67  & 764226 & 4  & 48  \\
NONLEGACY           & 74899  & 1826  & 35  & 72  & 62961  & 12 & 90  \\
QSO                 & 22724  & 1119  & 2   & 59  & 20524  & 3  & 85  \\
QA                  & 14608  & 358   & 7   & 67  & 13425  & 2  & 50  \\
SERENDIPITY\_FIRST   & 4163   & 1     & 100 & 100 & 1667   & 13 & 95  \\
ROSAT\_D             & 1704   & 17    & 19  & 78  & 1360   & 32 & 95  \\
SERENDIPITY\_DISTANT & 251    & 1     & 100 & 100 & 247    & 96 & 100 \\
\hline
programname & N galaxies &   \multicolumn{3}{c}{$0.375<z<0.7$} &  \multicolumn{3}{c}{$0.7<z<0.85$} \\
       		&   & N & \%$_{5}$ & \%$_{20}$ & N & \%$_{5}$ & \%$_{20}$  \\
\hline
GALAXY              & 829464 & 40916 & 100 & 100 & 54  & 100 & 100 \\ 
NONLEGACY           & 74899  & 9775  & 100 & 100 & 255 & 100 & 100 \\ 
QSO                 & 22724  & 886   & 100 & 100 & 135 & 100 & 100 \\ 
QA                  & 14608  & 814   & 100 & 100 & 6   & 100 & 100 \\ 
SERENDIPITY\_FIRST   & 4163   & 2377  & 100 & 100 & 88  & 100 & 100 \\ 
ROSAT\_D             & 1704   & 319   & 100 & 100 & 6   & 100 & 100 \\ 
SERENDIPITY\_DISTANT & 251    & 2     & 100 & 100 & 0   & -1  & -1  \\
\hline
programname & N galaxies &   \multicolumn{3}{c}{$0.85<z<1.6$} \\
       		&   & N & \%$_{5}$ & \%$_{20}$ \\
\hline 
GALAXY              & 829464 & 65 & 100 & 100 \\ 
NONLEGACY           & 74899  & 82 & 100 & 100 \\
QSO                 & 22724  & 60 & 100 & 100 \\
QA                  & 14608  & 5  & 100 & 100 \\
SERENDIPITY\_FIRST   & 4163   & 30 & 100 & 100 \\
ROSAT\_D             & 1704   & 2  & 100 & 100 \\
SERENDIPITY\_DISTANT & 251    & 1  & 100 & 100 \\
\hline
\end{tabular}
\end{center}
\end{table*}

\begin{center}
\begin{longtable}{c r rrrr rrrr rrrrr rrrrr rrrrr}
\caption{\label{ref:table:boss:src:SNR}Same as Table \ref{ref:table:sdss:src:SNR} for eBOSS spectra.}\\
\hline \hline
\endfirsthead

\caption{continued.}\\
\hline\hline
\endhead

\hline
\endfoot

programname        & N galaxies        &  \multicolumn{3}{c}{$0<z<0.025$}        &  \multicolumn{3}{c}{$0.025<z<0.375$} \\ 
       		&          & N        & \%$_{5}$        & \%$_{20}$        & N        & \%$_{5}$        & \%$_{20}$       \\ 
       		
\hline
LRG                      & 906577 & 117  & 31  & 73  & 168026 & 4 & 71    \\\ 
QSO                      & 61892  & 1652 & 35  & 95  & 35238  & 57 & 95   \\\ 
SEQUELS\_TARGET          & 38871  & 533  & 44  & 95  & 9819   & 43 & 94   \\\ 
RM\_TILE1                & 7342   & 3    & 81  & 100 & 2398   & 8 & 65    \\\ 
RM\_TILE2                & 6475   & 410  & 13  & 89  & 4357   & 34 & 92   \\
WISE\_BOSS\_QSO          & 4512   & 157  & 4   & 96  & 2257   & 4  & 95   \\ 
BRIGHTGAL                & 3541   & 543  & 3   & 43  & 2996   & 0  & 4    \\
SN\_GAL1                 & 3496   & 1    & 100 & 100 & 2474   & 28 & 87   \\
SEQUELS\_ELG             & 3059   & 2    & 100 & 100 & 624    & 90 & 100  \\ 
SPIDERS\_RASS\_CLUS      & 2573   & 2    & 100 & 100 & 2230   & 4 & 65    \\
QSO\_VAR\_LF             & 2313   & 98   & 27  & 82  & 1328   & 64 & 95   \\ 
SEQUELS\_ELG\_LOWP       & 2269   & 3    & 100 & 100 & 556    & 93 & 99   \\ 
FAINT\_ELG               & 2235   & 1    & 100 & 100 & 14     & 59 & 100  \\ 
BLUE\_RADIO              & 1791   & 1    & 100 & 100 & 830    & 4 & 95    \\
QSO\_VAR\_SDSS           & 1527   & 45   & 35  & 100 & 706    & 53 & 95   \\ 
STRIPE82BCG              & 1380   & 0    & -1  & -1  & 895    & 3 & 73    \\
QSO\_WISE\_FULL\_SKY     & 1247   & 25   & 24  & 96  & 688    & 35 & 96   \\ 
QSO\_SUPPZ               & 1213   & 9    & 100 & 72  & 646    & 3 & 77    \\
GAL\_NEAR\_QSO           & 1140   & 0    & -1  & -1  & 479    & 18 & 99   \\
QSO\_WISE\_SUPP          & 1065   & 28   & 41  & 100 & 376    & 24 & 95   \\
XMM\_PRIME               & 1052   & 18   & 48  & 100 & 424    & 23 & 78   \\
QSO\_EBOSS\_W3\_ADM      & 1044   & 19   & 33  & 96  & 510    & 66 & 95   \\
TAU\_STAR                & 784    & 11   & 44  & 86  & 608    & 41 & 99   \\ 
ELAIS\_N1\_GMRT\_TAYLOR  & 766    & 0    & -1  & -1  & 372    & 20 & 90   \\
QSO\_DEEP                & 661    & 2    & 100 & 100 & 288    & 97 & 100  \\
CLUSTER\_MEMBER          & 577    & 0    & -1  & -1  & 499    & 2 & 77    \\
FAINT\_HIZ\_LRG          & 543    & 0    & -1  & -1  & 1      & 100 & 100 \\
QSO\_GRI                 & 503    & 49   & 92  & 100 & 140    & 52 & 97   \\
TDSS\_B                  & 489    & 5    & 75  & 94  & 281    & 23 & 90   \\
TEMPLATE\_GAL\_PHOTO     & 485    & 5    & 25  & 100 & 307    & 14 & 77   \\
QSO1\_REOBS              & 443    & 3    & 24  & 100 & 300    & 82 & 99   \\
ELG                      & 409    & 16   & 75  & 100 & 128    & 94 & 100  \\
VARBAL                   & 399    & 3    & 100 & 100 & 270    & 0 & 58    \\
CHANDRAv1                & 383    & 21   & 10  & 57  & 154    & 10 & 90   \\
SN\_LOC                  & 372    & 22   & 4   & 47  & 334    & 27 & 75   \\
XMMHR                    & 360    & 0    & -1  & -1  & 186    & 4 & 91    \\
25ORI\_WISE\_W3          & 349    & 0    & -1  & -1  & 255    & 3 & 86    \\
XMM\_SECOND              & 332    & 2    & 100 & 93  & 161    & 22 & 75   \\
XMMRED                   & 319    & 0    & -1  & -1  & 237    & 2 & 89    \\
XMMBRIGHT                & 317    & 3    & 100 & 100 & 179    & 4 & 90    \\
ELAIS\_N1\_LOFAR         & 285    & 4    & 100 & 38  & 139    & 20 & 60   \\
KOEKAPbSTAR              & 282    & 0    & -1  & -1  & 0      & -1  & -1  \\
ELAIS\_N1\_GMRT\_GARN    & 280    & 0    & -1  & -1  & 160    & 15 & 75   \\
SPIDERS\_RASS\_AGN       & 275    & 1    & 100 & 100 & 192    & 0 & 47    \\
QSO\_VAR                 & 273    & 12   & 100 & 70  & 164    & 5 & 90    \\
ELAIS\_N1\_FIRST         & 256    & 0    & -1  & -1  & 135    & 7 & 55    \\
QSO\_AALs                & 249    & 6    & 100 & 100 & 130    & 100 & 4   \\
TEMPLATE\_QSO\_SDSS1     & 233    & 3    & 100 & 100 & 126    & 14 & 83   \\
QSO\_AAL                 & 208    & 1    & 100 & 100 & 112    & 100 & 100 \\
QSO\_RIZ                 & 206    & 5    & 8   & 44  & 64     & 66 & 99   \\
CXORED                   & 193    & 1    & 100 & 100 & 153    & 100 & 93  \\
BLAZXR                   & 186    & 1    & 100 & 100 & 119    & 3 & 70    \\
25ORI\_WISE              & 154    & 0    & -1  & -1  & 98     & 4 & 90    \\
RADIO\_2LOBE\_QSO        & 148    & 2    & 100 & 100 & 54     & 38 & 100  \\
KOE2068\_STAR            & 129    & 1    & 100 & 100 & 112    & 5 & 88    \\
TDSS\_FES\_VARBAL        & 119    & 1    & 100 & 100 & 92     & 4 & 90    \\
QSO\_IAL                 & 117    & 1    & 100 & 100 & 66     & 100 & 1   \\
TDSS\_CP                 & 116    & 0    & -1  & -1  & 51     & 27 & 100  \\
QSO\_VAR\_FPG            & 115    & 1    & 100 & 100 & 65     & 52 & 100  \\
WHITEDWARF\_SDSS         & 112    & 10   & 100 & 33  & 40     & 100 & 54  \\
TEMPLATE\_STAR\_PHOTO    & 111    & 6    & 100 & 57  & 40     & 100 & 59  \\
CXOBRIGHT                & 104    & 1    & 100 & 100 & 53     & 100 & 82  \\

\hline \hline
programname        & N galaxies   &   \multicolumn{3}{c}{$0.375<z<0.7$}        &  \multicolumn{3}{c}{$0.7<z<0.85$}        &  \multicolumn{3}{c}{$0.85<z<1.6$} \\
       		&          & N        & \%$_{5}$        & \%$_{20}$        & N        & \%$_{5}$        & \%$_{20}$         & N        & \%$_{5}$        & \%$_{20}$       \\
\hline
LRG                      & 906577 & 699627 & 95  & 100 & 36243 & 100 & 100  & 2564 & 100 & 100 \\ 
QSO                      & 61892  & 12247  & 95  & 100 & 8020  & 100 & 100  & 4735 & 100 & 100 \\ 
SEQUELS\_TARGET          & 38871  & 14835  & 95  & 98  & 9114  & 100 & 100  & 4570 & 100 & 100 \\ 
RM\_TILE1                & 7342   & 2144   & 93  & 100 & 1576  & 100 & 100  & 1221 & 100 & 100 \\ 
RM\_TILE2                & 6475   & 1359   & 95  & 99  & 271   & 100 & 100  & 78 & 100 & 100 \\ 
WISE\_BOSS\_QSO          & 4512   & 1335   & 94  & 99  & 360   & 100 & 100  & 403 & 100 & 100 \\ 
BRIGHTGAL                & 3541   & 2      & 100 & 100 & 0     & -1  & -1   & 0 & -1  & -1 \\ 
SN\_GAL1                 & 3496   & 900    & 91  & 100 & 90    & 100 & 100  & 31 & 100 & 100 \\ 
SEQUELS\_ELG             & 3059   & 1594   & 100 & 100 & 444   & 100 & 100  & 395 & 100 & 100 \\ 
SPIDERS\_RASS\_CLUS      & 2573   & 340    & 91  & 100 & 1     & 100 & 100  & 0 & -1  & -1 \\ 
QSO\_VAR\_LF             & 2313   & 446    & 95  & 100 & 260   & 100 & 100  & 181 & 100 & 100 \\ 
SEQUELS\_ELG\_LOWP       & 2269   & 1151   & 100 & 100 & 291   & 100 & 100  & 268 & 100 & 100 \\ 
FAINT\_ELG               & 2235   & 460    & 100 & 100 & 868   & 100 & 100  & 892 & 100 & 100 \\ 
BLUE\_RADIO              & 1791   & 942    & 80  & 100 & 14    & 100 & 100  & 4 & 100 & 100 \\ 
QSO\_VAR\_SDSS           & 1527   & 371    & 95  & 100 & 241   & 100 & 100  & 164 & 100 & 100 \\ 
STRIPE82BCG              & 1380   & 478    & 81  & 100 & 7     & 100 & 100  & 0 & -1  & -1 \\ 
QSO\_WISE\_FULL\_SKY     & 1247   & 220    & 96  & 100 & 193   & 100 & 100  & 121 & 100 & 100 \\ 
QSO\_SUPPZ               & 1213   & 413    & 82  & 98  & 132   & 100 & 100  & 13 & 100 & 100 \\ 
GAL\_NEAR\_QSO           & 1140   & 645    & 92  & 100 & 15    & 100 & 100  & 1  & 100 & 100 \\ 
QSO\_WISE\_SUPP          & 1065   & 432    & 95  & 100 & 147   & 100 & 100  & 82 & 100 & 100 \\ 
XMM\_PRIME               & 1052   & 390    & 95  & 99  & 143   & 100 & 100  & 77 & 100 & 100 \\ 
QSO\_EBOSS\_W3\_ADM      & 1044   & 341    & 96  & 100 & 107   & 100 & 100  & 67 & 100 & 100 \\ 
TAU\_STAR                & 784    & 44     & 100 & 100 & 11    & 100 & 100  & 110 & 100 & 100 \\ 
ELAIS\_N1\_GMRT\_TAYLOR  & 766    & 306    & 96  & 100 & 49    & 100 & 100  & 39 & 100 & 100 \\ 
QSO\_DEEP                & 661    & 162    & 100 & 100 & 123   & 100 & 100  & 86 & 100 & 100 \\ 
CLUSTER\_MEMBER          & 577    & 78     & 89  & 100 & 0     & -1  & -1   & 0 & -1  & -1 \\ 
FAINT\_HIZ\_LRG          & 543    & 370    & 100 & 100 & 153   & 100 & 100  & 19 & 100 & 100 \\ 
QSO\_GRI                 & 503    & 305    & 96  & 100 & 8     & 100 & 100  & 1  & 100 & 100 \\ 
TDSS\_B                  & 489    & 107    & 90  & 98  & 34    & 100 & 100  & 62 & 100 & 100 \\ 
TEMPLATE\_GAL\_PHOTO     & 485    & 166    & 96  & 100 & 5     & 100 & 100  & 2 & 100 & 100 \\ 
QSO1\_REOBS              & 443    & 90     & 100 & 100 & 4     & 100 & 100  & 46 & 100 & 100 \\ 
ELG                      & 409    & 96     & 100 & 100 & 80    & 100 & 100  & 89 & 100 & 100 \\ 
VARBAL                   & 399    & 91     & 77  & 100 & 28    & 100 & 100  & 7 & 100 & 100 \\ 
CHANDRAv1                & 383    & 158    & 96  & 100 & 39    & 100 & 100  & 11 & 100 & 100 \\ 
SN\_LOC                  & 372    & 13     & 97  & 100 & 3     & 100 & 100  & 0 & -1  & -1 \\ 
XMMHR                    & 360    & 151    & 93  & 100 & 16    & 100 & 100  & 7 & 100 & 100 \\ 
25ORI\_WISE\_W3          & 349    & 88     & 99  & 100 & 2     & 100 & 100  & 4 & 100 & 100 \\ 
XMM\_SECOND              & 332    & 111    & 96  & 100 & 41    & 100 & 100  & 17 & 100 & 100 \\ 
XMMRED                   & 319    & 82     & 83  & 100 & 0     & -1  & -1   & 0 & -1  & -1 \\ 
XMMBRIGHT                & 317    & 110    & 87  & 100 & 10    & 100 & 100  & 15 & 100 & 100 \\ 
ELAIS\_N1\_LOFAR         & 285    & 96     & 93  & 100 & 30    & 100 & 100  & 16 & 100 & 100 \\ 
KOEKAPbSTAR              & 282    & 274    & 100 & 100 & 8     & 100 & 100  & 0 & -1  & -1 \\ 
ELAIS\_N1\_GMRT\_GARN    & 280    & 88     & 94  & 100 & 15    & 100 & 100  & 17 & 100 & 100 \\ 
SPIDERS\_RASS\_AGN       & 275    & 40     & 75  & 100 & 15    & 100 & 100  & 27 & 100 & 100 \\ 
QSO\_VAR                 & 273    & 52     & 87  & 100 & 17    & 100 & 100  & 28 & 100 & 100 \\ 
ELAIS\_N1\_FIRST         & 256    & 89     & 90  & 100 & 19    & 100 & 100  & 13 & 100 & 100 \\ 
QSO\_AALs                & 249    & 83     & 79  & 98  & 19    & 100 & 100  & 11 & 100 & 100 \\ 
TEMPLATE\_QSO\_SDSS1     & 233    & 74     & 93  & 100 & 18    & 100 & 100  & 12 & 100 & 100 \\ 
QSO\_AAL                 & 208    & 52     & 74  & 99  & 19    & 100 & 100  & 24 & 100 & 100 \\ 
QSO\_RIZ                 & 206    & 78     & 97  & 100 & 55    & 100 & 100  & 4 & 100 & 100 \\ 
CXORED                   & 193    & 39     & 85  & 100 & 0     & -1  & -1   & 0 & -1  & -1 \\ 
BLAZXR                   & 186    & 44     & 80  & 100 & 15    & 100 & 100  & 7 & 100 & 100 \\ 
25ORI\_WISE              & 154    & 55     & 99  & 100 & 0     & -1  & -1   & 1  & 100 & 100 \\ 
RADIO\_2LOBE\_QSO        & 148    & 45     & 95  & 100 & 29    & 100 & 100  & 18 & 100 & 100 \\ 
KOE2068\_STAR            & 129    & 5      & 46  & 100 & 4     & 100 & 100  & 7 & 100 & 100 \\ 
TDSS\_FES\_VARBAL        & 119    & 14     & 94  & 100 & 4     & 100 & 100  & 8 & 100 & 100 \\ 
QSO\_IAL                 & 117    & 35     & 51  & 96  & 4     & 100 & 100  & 11 & 100 & 100 \\ 
TDSS\_CP                 & 116    & 25     & 96  & 100 & 14    & 100 & 100  & 26 & 100 & 100 \\ 
QSO\_VAR\_FPG            & 115    & 20     & 100 & 100 & 11    & 100 & 100  & 18 & 100 & 100 \\ 
WHITEDWARF\_SDSS         & 112    & 45     & 67  & 96  & 11    & 100 & 100  & 6 & 100 & 100 \\ 
TEMPLATE\_STAR\_PHOTO    & 111    & 32     & 90  & 100 & 22    & 100 & 100  & 11 & 100 & 100 \\ 
CXOBRIGHT                & 104    & 36     & 88  & 100 & 4     & 100 & 100  & 10 & 100 & 100 \\ 
\hline
\end{longtable}
\end{center}

\begin{landscape}

\begin{table*}
\begin{center}
\caption{\label{ref:table:sdss:src:fibermag}SDSS source types containing more than 100 galaxies divided in three redshift to match the 4000\text{\AA} break region to a SDSS broad band change to $g,r,i$ : $0<z<0.17$, $0.17<z<0.55$, $0.55<z<1.6$. 
This table is created only for objects with SDSS photometry. 
In each bin, we count the fraction of objects where the difference between fiber magnitude and model magnitude is smaller than 0.75 (2.5) i.e. where 50\% (10\%) or more of the galaxy light went in the fiber. The SDSS galaxies at low redshift are extended objects, so that most of them only have a small fraction of their light going in the fiber.}
\begin{tabular}{c r rrrr rrrr rrrrr rrrrr rrrrr}
\hline \hline
programname  & N 
&  \multicolumn{4}{c}{$0<z<0.17$, g} 
&  \multicolumn{4}{c}{$0.17<z<0.55$, r} 
&  \multicolumn{4}{c}{$0.55<z<1.6$, i} \\
  &                              
  & z                            & mag                            & $>50\%$                            & $>10\%$
  & z                            & mag                            & $>50\%$                            & $>10\%$
  & z                            & mag                            & $>50\%$                            & $>10\%$ \\
\hline
GALAXY & 829464 & 634828 & 627598 & 12823 & 585371 & 194198 & 192141 & 4646 & 191720 & 438 & 407 & 30 & 395 \\ 
NONLEGACY & 74899 & 35110 & 33981 & 6553 & 33257 & 38003 & 37445 & 10991 & 37405 & 1786 & 1723 & 918 & 1718 \\ 
QSO & 22724 & 15321 & 15028 & 4518 & 14753 & 7042 & 6891 & 4053 & 6887 & 361 & 351 & 347 & 350 \\ 
QA & 14608 & 10717 & 10584 & 285 & 9876 & 3869 & 3832 & 171 & 3822 & 22 & 21 & 11 & 20 \\ 
SERENDIPITY\_FIRST & 4163 & 154 & 152 & 32 & 152 & 3240 & 3215 & 791 & 3215 & 769 & 759 & 333 & 759 \\ 
ROSAT\_D & 1704 & 458 & 385 & 168 & 378 & 1188 & 1132 & 626 & 1130 & 58 & 54 & 27 & 54 \\ 
SERENDIPITY\_DISTANT & 251 & 12 & 11 & 11 & 11 & 236 & 232 & 232 & 232 & 3 & 2 & 2 & 2 \\ 
\hline
\end{tabular}
\end{center}
\end{table*}

\begin{center}
\begin{longtable}{c r rrrr rrrr rrrrr rrrrr rrrrr}
\caption{\label{ref:table:boss:src:fibermag} Same as Table \ref{ref:table:sdss:src:fibermag} for eBOSS sourcetypes.} \\
\hline \hline
\endfirsthead

\caption{continued.}\\
\hline\hline
\endhead

\hline \multicolumn{14}{|c|}{{Continued on next page}} \\ \hline
\endfoot

\hline \hline
\endlastfoot

% 0, 0.17, 0.55,  1.6
programname  & N 
&  \multicolumn{4}{c}{$0<z<0.17$, g} 
&  \multicolumn{4}{c}{$0.17<z<0.55$, r} 
&  \multicolumn{4}{c}{$0.55<z<1.6$, i} \\
  &                              
  & z                            & mag                            & $>50\%$                            & $>10\%$
  & z                            & mag                            & $>50\%$                            & $>10\%$
  & z                            & mag                            & $>50\%$                            & $>10\%$ \\
\hline

LRG 					 & 906577 & 21079 & 20411 & 1168 & 20147 & 557046 & 554121 & 146236 & 553802 & 328452 & 327943 & 118872 & 327548 \\ 
QSO 					 & 61892 & 15656 & 13105 & 13032 & 13105 & 27561 & 26871 & 26795 & 26868 & 18675 & 18365 & 18321 & 18364 \\ 
SEQUELS\_TARGET 		 & 38871 & 3885 & 3091 & 2599 & 3087 & 12122 & 11553 & 9487 & 11552 & 22864 & 22678 & 13736 & 22677 \\ 
RM\_TILE1 				 & 7342 & 445 & 445 & 398 & 445 & 3279 & 3226 & 3035 & 3226 & 3618 & 3553 & 3460 & 3553 \\ 
RM\_TILE2 				 & 6475 & 2553 & 2530 & 2530 & 2530 & 3031 & 2995 & 2993 & 2995 & 891 & 885 & 880 & 885 \\ 
WISE\_BOSS\_QSO 		 & 4512 & 1154 & 1151 & 1150 & 1151 & 2096 & 2085 & 2085 & 2085 & 1262 & 1261 & 1261 & 1261 \\ 
BRIGHTGAL 				 & 3541 & 3534 & 3301 & 23 & 2388 & 7 & 3 & 0 & 2 & 0 & 0 & 0 & 0 \\ 
SN\_GAL1 				 & 3496 & 532 & 520 & 165 & 510 & 2661 & 2599 & 1611 & 2599 & 303 & 287 & 256 & 287 \\ 
SEQUELS\_ELG 			 & 3059 & 150 & 130 & 105 & 130 & 1358 & 1282 & 1096 & 1282 & 1551 & 1530 & 1271 & 1530 \\ 
SPIDERS\_RASS\_CLUS		 & 2573 & 1076 & 1075 & 239 & 1063 & 1490 & 1484 & 936 & 1483 & 7 & 7 & 6 & 7 \\ 
QSO\_VAR\_LF 			 & 2313 & 549 & 501 & 489 & 497 & 1146 & 1116 & 1105 & 1116 & 618 & 598 & 586 & 598 \\ 
SEQUELS\_ELG\_LOWP 		 & 2269 & 120 & 111 & 83 & 111 & 1087 & 1050 & 899 & 1050 & 1062 & 1034 & 852 & 1033 \\ 
FAINT\_ELG 				 & 2235 & 6 & 5 & 5 & 5 & 94 & 81 & 62 & 81 & 2135 & 1841 & 1405 & 1835 \\ 
BLUE\_RADIO 			 & 1791 & 6 & 5 & 3 & 5 & 1713 & 1709 & 677 & 1709 & 72 & 72 & 41 & 71 \\ 
QSO\_VAR\_SDSS 			 & 1527 & 359 & 314 & 313 & 314 & 554 & 529 & 526 & 529 & 614 & 597 & 592 & 597 \\ 
STRIPE82BCG 			 & 1380 & 164 & 161 & 35 & 158 & 1139 & 1116 & 497 & 1115 & 77 & 75 & 47 & 75 \\ 
QSO\_WISE\_FULL\_SKY	 & 1247 & 292 & 287 & 287 & 287 & 549 & 547 & 545 & 547 & 406 & 406 & 406 & 406 \\ 
QSO\_SUPPZ & 1213 		 & 365 & 365 & 365 & 365 & 450 & 450 & 450 & 450 & 398 & 398 & 398 & 398 \\ 
GAL\_NEAR\_QSO 			 & 1140 & 2 & 2 & 1 & 2 & 1045 & 1027 & 517 & 1027 & 93 & 87 & 22 & 87 \\ 
QSO\_WISE\_SUPP 		 & 1065 & 169 & 158 & 158 & 158 & 455 & 414 & 414 & 414 & 441 & 435 & 435 & 435 \\ 
XMM\_PRIME & 1052 		 & 148 & 141 & 83 & 139 & 518 & 513 & 380 & 512 & 386 & 381 & 320 & 380 \\ 
QSO\_EBOSS\_W3\_ADM	 	 & 1044 & 193 & 140 & 138 & 140 & 551 & 522 & 520 & 522 & 300 & 288 & 288 & 288 \\ 
TAU\_STAR 			 	 & 784 & 338 & 334 & 78 & 331 & 291 & 282 & 79 & 280 & 155 & 154 & 33 & 154 \\ 
ELAIS\_N1\_GMRT\_TAYLOR  & 766 & 84 & 82 & 18 & 82 & 504 & 495 & 264 & 495 & 178 & 171 & 127 & 171 \\ 
QSO\_DEEP 				 & 661 & 47 & 40 & 37 & 40 & 334 & 287 & 271 & 285 & 280 & 248 & 222 & 248 \\ 
CLUSTER\_MEMBER 		 & 577 & 170 & 170 & 32 & 170 & 405 & 405 & 213 & 405 & 2 & 2 & 2 & 2 \\ 
FAINT\_HIZ\_LRG 		 & 543 & 1 & 0 & 0 & 0 & 50 & 50 & 44 & 50 & 492 & 489 & 346 & 489 \\ 
QSO\_GRI & 503 & 80 & 45 & 45 & 45 & 311 & 249 & 240 & 244 & 112 & 110 & 110 & 110 \\ 
TDSS\_B & 489 & 58 & 55 & 53 & 55 & 305 & 289 & 242 & 289 & 126 & 126 & 125 & 126 \\ 
TEMPLATE\_GAL\_PHOTO & 485 & 201 & 201 & 57 & 200 & 225 & 224 & 127 & 224 & 59 & 58 & 29 & 58 \\ 
QSO1\_REOBS & 443 & 122 & 121 & 121 & 121 & 220 & 220 & 220 & 220 & 101 & 101 & 101 & 101 \\ 
ELG & 409 & 94 & 65 & 32 & 63 & 106 & 86 & 62 & 83 & 209 & 180 & 141 & 180 \\ 
VARBAL & 399 & 165 & 165 & 165 & 165 & 148 & 145 & 145 & 145 & 86 & 86 & 86 & 86 \\ 
CHANDRAv1 & 383 & 62 & 46 & 34 & 44 & 235 & 214 & 175 & 213 & 86 & 74 & 66 & 74 \\ 
SN\_LOC & 372 & 278 & 132 & 17 & 129 & 89 & 61 & 20 & 61 & 5 & 3 & 3 & 3 \\ 
XMMHR & 360 & 34 & 33 & 14 & 33 & 259 & 253 & 138 & 252 & 67 & 62 & 44 & 62 \\ 
25ORI\_WISE\_W3 & 349 & 102 & 75 & 6 & 74 & 173 & 110 & 38 & 110 & 74 & 37 & 14 & 36 \\ 
XMM\_SECOND & 332 & 44 & 44 & 8 & 42 & 180 & 175 & 97 & 175 & 108 & 107 & 77 & 107 \\ 
XMMRED & 319 & 34 & 33 & 10 & 33 & 274 & 266 & 99 & 266 & 11 & 8 & 5 & 8 \\ 
XMMBRIGHT & 317 & 32 & 31 & 23 & 31 & 227 & 217 & 153 & 217 & 58 & 57 & 53 & 57 \\ 
ELAIS\_N1\_LOFAR & 285 & 62 & 55 & 11 & 48 & 136 & 130 & 56 & 130 & 87 & 87 & 57 & 87 \\ 
KOEKAPbSTAR & 282 & 0 & 0 & 0 & 0 & 3 & 0 & 0 & 0 & 279 & 0 & 0 & 0 \\ 
ELAIS\_N1\_GMRT\_GARN & 280 & 53 & 52 & 6 & 50 & 164 & 162 & 78 & 162 & 63 & 60 & 38 & 60 \\ 
SPIDERS\_RASS\_AGN & 275 & 73 & 72 & 23 & 72 & 153 & 152 & 98 & 152 & 49 & 49 & 43 & 49 \\ 
QSO\_VAR & 273 & 110 & 95 & 93 & 95 & 99 & 88 & 88 & 88 & 64 & 58 & 58 & 58 \\ 
ELAIS\_N1\_FIRST & 256 & 49 & 48 & 4 & 48 & 133 & 130 & 40 & 130 & 74 & 74 & 41 & 73 \\ 
QSO\_AALs & 249 & 69 & 69 & 69 & 69 & 103 & 103 & 103 & 103 & 77 & 76 & 76 & 76 \\ 
TEMPLATE\_QSO\_SDSS1 & 233 & 33 & 33 & 33 & 33 & 140 & 140 & 138 & 140 & 60 & 60 & 60 & 60 \\ 
QSO\_AAL & 208 & 36 & 36 & 36 & 36 & 108 & 108 & 108 & 108 & 64 & 64 & 64 & 64 \\ 
QSO\_RIZ & 206 & 36 & 21 & 19 & 19 & 73 & 52 & 44 & 47 & 97 & 79 & 74 & 76 \\ 
CXORED & 193 & 28 & 26 & 11 & 26 & 160 & 157 & 62 & 157 & 5 & 3 & 1 & 3 \\ 
BLAZXR & 186 & 30 & 26 & 10 & 23 & 122 & 113 & 59 & 113 & 34 & 31 & 30 & 31 \\ 
25ORI\_WISE & 154 & 27 & 19 & 1 & 18 & 77 & 52 & 14 & 52 & 50 & 31 & 17 & 31 \\ 
RADIO\_2LOBE\_QSO & 148 & 17 & 15 & 15 & 15 & 68 & 41 & 41 & 41 & 63 & 58 & 58 & 58 \\ 
KOE2068\_STAR & 129 & 89 & 88 & 9 & 86 & 29 & 29 & 14 & 29 & 11 & 9 & 7 & 9 \\ 
TDSS\_FES\_VARBAL & 119 & 20 & 20 & 20 & 20 & 79 & 79 & 78 & 79 & 20 & 20 & 20 & 20 \\ 
QSO\_IAL & 117 & 18 & 18 & 16 & 18 & 79 & 79 & 79 & 79 & 20 & 20 & 20 & 20 \\ 
TDSS\_CP & 116 & 13 & 13 & 11 & 13 & 55 & 54 & 51 & 54 & 48 & 47 & 47 & 47 \\ 
QSO\_VAR\_FPG & 115 & 29 & 29 & 29 & 29 & 45 & 43 & 42 & 43 & 41 & 38 & 38 & 38 \\ 
WHITEDWARF\_SDSS & 112 & 22 & 21 & 21 & 21 & 57 & 57 & 57 & 57 & 33 & 33 & 33 & 33 \\ 
TEMPLATE\_STAR\_PHOTO & 111 & 21 & 21 & 21 & 21 & 38 & 38 & 38 & 38 & 52 & 52 & 52 & 52 \\ 
CXOBRIGHT & 104 & 12 & 11 & 5 & 11 & 69 & 68 & 48 & 68 & 23 & 23 & 22 & 23 \\ 
\hline
\end{longtable}
\end{center}
\end{landscape}

\begin{center}
\begin{longtable}{c r rrrr rrrr rrrrr rrrrr rrrrr}
\caption{Same as Table \ref{ref:table:sdss:src:fibermag} for eBOSS sourcetypes.} \\
\hline \hline
\endfirsthead

\caption{continued.}\\
\hline\hline
\endhead

\hline \multicolumn{3}{|r|}{{Continued on next page}} \\ \hline
\endfoot

\hline \hline
\endlastfoot

programname                            & N                            &  \multicolumn{4}{c}{$0<z<0.17$, g}           \\ 
                               &                              & z                            & mag                            & $>10\%$                            & $>1\%$      \\ 
                               
                  \hline
LRG             	& 906577 & 21079 & 20411 & 464   & 1259 	\\ 
QSO             	& 61892  & 15656 & 13105 & 11238 & 13044 	\\
SEQUELS\_TARGET 	& 38871  & 3885  & 3091  & 1833 & 2648		\\
RM\_TILE1 		& 7342   & 445   & 445   & 282 & 445 		\\
RM\_TILE2 		& 6475   & 2553  & 2530  & 2327 & 2530 		\\
WISE\_BOSS\_QSO 	& 4512   & 1154  & 1151  & 1069 & 1151 		\\
BRIGHTGAL 		& 3541   & 3534  & 3301  & 10 & 28 		\\
SN\_GAL1 		& 3496   & 532   & 520   & 33 & 190 		\\
SEQUELS\_ELG 		& 3059   & 150   & 130   & 55 & 107 		\\
SPIDERS\_RASS\_CLUS 	& 2573   & 1076  & 1075  & 7 & 296 		\\
QSO\_VAR\_LF 		& 2313   & 549   & 501   & 396 & 489 		\\
SEQUELS\_ELG\_LOWP 	& 2269   & 120   & 111   & 36 & 87 		\\
FAINT\_ELG 		& 2235   & 6     & 5     & 3 & 5 		\\
BLUE\_RADIO 		& 1791   & 6     & 5     & 0 & 4 		\\
QSO\_VAR\_SDSS 		& 1527   & 359   & 314   & 274 & 314 		\\
STRIPE82BCG 		& 1380   & 164   & 161   & 5 & 41 		\\
QSO\_WISE\_FULL\_SKY 	& 1247   & 292   & 287   & 253 & 287 		\\
QSO\_SUPPZ 		& 1213   & 365   & 365   & 356 & 365 		\\
GAL\_NEAR\_QSO 		& 1140   & 2     & 2     & 0 & 1 		\\
QSO\_WISE\_SUPP 	& 1065   & 169   & 158   & 148 & 158 		\\
XMM\_PRIME 		& 1052   & 148   & 141   & 71 & 87 		\\
QSO\_EBOSS\_W3\_ADM 	& 1044   & 193   & 140   & 112 & 138 		\\
TAU\_STAR 		& 784    & 338   & 334   & 20 & 89 		\\
ELAIS\_N1\_GMRT\_TAYLOR & 766    & 84    & 82    & 3 & 21 		\\
QSO\_DEEP 		& 661    & 47    & 40    & 22 & 37 		\\
CLUSTER\_MEMBER 	& 577    & 170   & 170   & 1 & 44 		\\
FAINT\_HIZ\_LRG 	& 543    & 1     & 0     & 0 & 0 		\\
QSO\_GRI 		& 503    & 80    & 45    & 30 & 45 		\\
TDSS\_B 		& 489    & 58    & 55    & 33 & 53 		\\
TEMPLATE\_GAL\_PHOTO 	& 485    & 201   & 201   & 3 & 73 		\\
QSO1\_REOBS 		& 443    & 122   & 121   & 99 & 121 		\\
ELG 			& 409    & 94    & 65    & 14 & 39 		\\
VARBAL 			& 399    & 165   & 165   & 161 & 165 		\\
CHANDRAv1 		& 383    & 62    & 46    & 30 & 37 		\\
SN\_LOC 		& 372    & 278   & 132   & 0 & 19 		\\
XMMHR 			& 360    & 34    & 33    & 3 & 18 		\\
25ORI\_WISE\_W3 	& 349    & 102   & 75    & 3 & 9 		\\
XMM\_SECOND 		& 332    & 44    & 44    & 5 & 9 		\\
XMMRED			& 319    & 34    & 33    & 1 & 12 		\\
XMMBRIGHT 		& 317    & 32    & 31    & 16 & 25 		\\
ELAIS\_N1\_LOFAR 	& 285    & 62    & 55    & 3 & 11 		\\
KOEKAPbSTAR 		& 282    & 0     & 0     & 0 & 0 		\\
ELAIS\_N1\_GMRT\_GARN 	& 280    & 53    & 52    & 2 & 7 		\\
SPIDERS\_RASS\_AGN 	& 275    & 73    & 72    & 4 & 24 		\\
QSO\_VAR 		& 273    & 110   & 95    & 72 & 93 		\\
ELAIS\_N1\_FIRST 	& 256    & 49    & 48    & 3 & 5 		\\
QSO\_AALs 		& 249    & 69    & 69    & 67 & 69 		\\
TEMPLATE\_QSO\_SDSS1 	& 233    & 33    & 33    & 33 & 33 		\\
QSO\_AAL 		& 208    & 36    & 36    & 33 & 36 		\\
QSO\_RIZ 		& 206    & 36    & 21    & 14 & 19 		\\
CXORED 			& 193    & 28    & 26    & 2 & 11 		\\
BLAZXR 			& 186    & 30    & 26    & 3 & 11 		\\
25ORI\_WISE 		& 154    & 27    & 19    & 1 & 2 		\\
RADIO\_2LOBE\_QSO 	& 148    & 17    & 15    & 14 & 15 		\\
KOE2068\_STAR 		& 129    & 89    & 88    & 4 & 12 		\\
TDSS\_FES\_VARBAL 	& 119    & 20    & 20    & 19 & 20 		\\
QSO\_IAL 		& 117    & 18    & 18    & 16 & 16 		\\
TDSS\_CP 		& 116    & 13    & 13    & 8 & 11 		\\
QSO\_VAR\_FPG 		& 115    & 29    & 29    & 27 & 29 		\\
WHITEDWARF\_SDSS	& 112    & 22    & 21    & 20 & 21 		\\
TEMPLATE\_STAR\_PHOTO 	& 111    & 21    & 21    & 21 & 21 		\\
CXOBRIGHT 		& 104    & 12    & 11    & 3 & 5 		\\
\hline \hline
programname                            & N                           &  \multicolumn{4}{c}{$0.17<z<0.55$, r}                            &  \multicolumn{4}{c}{$0.55<z<1.6$, i} \\
                               &                               & z                            & mag                            & $>10\%$                            & $>1\%$                            & z                            & mag                            & $>10\%$                            & $>1\%$ \\
\hline
LRG             	& 906577 & 557046 & 554121 & 7783 & 184634 & 328452 & 327943 & 9200 & 141815 \\ 
QSO             	& 61892  & 27561 & 26871 & 22589 & 26814   & 18675 & 18365 & 14854 & 18329 \\ 
SEQUELS\_TARGET 	& 38871  & 12122 & 11553 & 3964 & 9945 	  & 22864 & 22678 & 3771 & 15167 \\ 
RM\_TILE1 		& 7342   & 3279 & 3226 & 1762 & 3035 	  & 3618 & 3553 & 2219 & 3460 \\ 
RM\_TILE2 		& 6475   & 3031 & 2995 & 2628 & 2993 	  & 891 & 885 & 752 & 883 \\ 
WISE\_BOSS\_QSO 	& 4512   & 2096 & 2085 & 1702 & 2085 	  & 1262 & 1261 & 1079 & 1261 \\ 
BRIGHTGAL 		& 3541   & 7 & 3 & 0 & 0 		  & 0 & 0 & 0 & 0 \\ 
SN\_GAL1 		& 3496   & 2661 & 2599 & 233 & 1762 	  & 303 & 287 & 149 & 259 \\ 
SEQUELS\_ELG 		& 3059   & 1358 & 1282 & 357 & 1144 	  & 1551 & 1530 & 391 & 1331 \\ 
SPIDERS\_RASS\_CLUS 	& 2573   & 1490 & 1484 & 69 & 1033 	  & 7 & 7 & 1 & 7 \\ 
QSO\_VAR\_LF 		& 2313   & 1146 & 1116 & 751 & 1108	  & 618 & 598 & 502 & 590 \\ 
SEQUELS\_ELG\_LOWP 	& 2269   & 1087 & 1050 & 272 & 935 	  & 1062 & 1034 & 279 & 888 \\ 
FAINT\_ELG 		& 2235   & 94 & 81 & 16 & 71 		  & 2135 & 1841 & 479 & 1479 \\ 
BLUE\_RADIO 		& 1791   & 1713 & 1709 & 41 & 828 	  & 72 & 72 & 18 & 45 \\ 
QSO\_VAR\_SDSS 		& 1527   & 554 & 529 & 422 & 527 	  & 614 & 597 & 478 & 592 \\ 
STRIPE82BCG 		& 1380   & 1139 & 1116 & 61 & 577 	  & 77 & 75 & 11 & 54 \\ 
QSO\_WISE\_FULL\_SKY 	& 1247   & 549 & 547 & 414 & 545 	  & 406 & 406 & 337 & 406 \\ 
QSO\_SUPPZ 		& 1213   & 450 & 450 & 438 & 450 	  & 398 & 398 & 377 & 398 \\ 
GAL\_NEAR\_QSO 		& 1140   & 1045 & 1027 & 41 & 603 	  & 93 & 87 & 8 & 27 \\ 
QSO\_WISE\_SUPP 	& 1065   & 455 & 414 & 339 & 414 	  & 441 & 435 & 353 & 435 \\ 
XMM\_PRIME 		& 1052   & 518 & 513 & 155 & 398 	  & 386 & 381 & 171 & 328 \\ 
QSO\_EBOSS\_W3\_ADM 	& 1044   & 551 & 522 & 361 & 520 	  & 300 & 288 & 213 & 288 \\ 
TAU\_STAR 		& 784    & 291 & 282 & 10 & 98 		  & 155 & 154 & 8 & 40 \\ 
ELAIS\_N1\_GMRT\_TAYLOR & 766    & 504 & 495 & 35 & 293		  & 178 & 171 & 41 & 138 \\ 
QSO\_DEEP 		& 661    & 334 & 287 & 128 & 275 	  & 280 & 248 & 132 & 225 \\ 
CLUSTER\_MEMBER 	& 577    & 405 & 405 & 6 & 248 		  & 2 & 2 & 0 & 2 \\ 
FAINT\_HIZ\_LRG 	& 543    & 50 & 50 & 5 & 47 		  & 492 & 489 & 45 & 380 \\ 
QSO\_GRI 		& 503    & 311 & 249 & 177 & 241 	  & 112 & 110 & 77 & 110 \\ 
TDSS\_B 		& 489    & 305 & 289 & 111 & 249 	  & 126 & 126 & 100 & 126 \\ 
TEMPLATE\_GAL\_PHOTO 	& 485    & 225 & 224 & 2 & 132 		  & 59 & 58 & 18 & 29 \\ 
QSO1\_REOBS 		& 443    & 220 & 220 & 167 & 220 	  & 101 & 101 & 60 & 101 \\ 
ELG 			& 409    & 106 & 86 & 18 & 67 		  & 209 & 180 & 59 & 143 \\ 
VARBAL 			& 399    & 148 & 145 & 142 & 145 	  & 86 & 86 & 84 & 86 \\ 
CHANDRAv1 		& 383    & 235 & 214 & 58 & 182 		  & 86 & 74 & 44 & 66 \\ 
SN\_LOC 		& 372    & 89 & 61 & 3 & 22 		  & 5 & 3 & 2 & 3 \\ 
XMMHR 			& 360    & 259 & 253 & 20 & 152 		  & 67 & 62 & 16 & 48 \\ 
25ORI\_WISE\_W3 	& 349    & 173 & 110 & 9 & 45 		  & 74 & 37 & 8 & 14 \\ 
XMM\_SECOND 		& 332    & 180 & 175 & 22 & 101 		  & 108 & 107 & 27 & 79 \\ 
XMMRED			& 319    & 274 & 266 & 6 & 123 		  & 11 & 8 & 1 & 5 \\ 
XMMBRIGHT 		& 317    & 227 & 217 & 58 & 161 		  & 58 & 57 & 38 & 53 \\ 
ELAIS\_N1\_LOFAR 	& 285    & 136 & 130 & 13 & 60 		  & 87 & 87 & 22 & 59 \\ 
KOEKAPbSTAR 		& 282    & 3 & 0 & 0 & 0 		  & 279 & 0 & 0 & 0 \\ 
ELAIS\_N1\_GMRT\_GARN 	& 280    & 164 & 162 & 13 & 86 		  & 63 & 60 & 13 & 40 \\ 
SPIDERS\_RASS\_AGN 	& 275    & 153 & 152 & 21 & 103 		  & 49 & 49 & 33 & 44 \\ 
QSO\_VAR 		& 273    & 99 & 88 & 76 & 88 		  & 64 & 58 & 55 & 58 \\ 
ELAIS\_N1\_FIRST 	& 256    & 133 & 130 & 8 & 52 		  & 74 & 74 & 10 & 45 \\ 
QSO\_AALs 		& 249    & 103 & 103 & 100 & 103 	  & 77 & 76 & 69 & 76 \\ 
TEMPLATE\_QSO\_SDSS1 	& 233    & 140 & 140 & 133 & 138 	  & 60 & 60 & 58 & 60 \\ 
QSO\_AAL 		& 208    & 108 & 108 & 105 & 108 	  & 64 & 64 & 62 & 64 \\ 
QSO\_RIZ 		& 206    & 73 & 52 & 31 & 45 		  & 97 & 79 & 42 & 74 \\ 
CXORED 			& 193    & 160 & 157 & 2 & 74 		  & 5 & 3 & 1 & 1 \\ 
BLAZXR 			& 186    & 122 & 113 & 19 & 65 		  & 34 & 31 & 26 & 30 \\ 
25ORI\_WISE 		& 154    & 77 & 52 & 2 & 16 		  & 50 & 31 & 5 & 17 \\ 
RADIO\_2LOBE\_QSO 	& 148    & 68 & 41 & 31 & 41		  & 63 & 58 & 45 & 58 \\ 
KOE2068\_STAR 		& 129    & 29 & 29 & 3 & 14 		  & 11 & 9 & 7 & 7 \\ 
TDSS\_FES\_VARBAL 	& 119    & 79 & 79 & 72 & 78 		  & 20 & 20 & 19 & 20 \\ 
QSO\_IAL 		& 117    & 79 & 79 & 76 & 79 		  & 20 & 20 & 19 & 20 \\ 
TDSS\_CP 		& 116    & 55 & 54 & 38 & 51 		  & 48 & 47 & 41 & 47 \\ 
QSO\_VAR\_FPG 		& 115    & 45 & 43 & 37 & 42 		  & 41 & 38 & 30 & 38 \\ 
WHITEDWARF\_SDSS	& 112    & 57 & 57 & 54 & 57 		  & 33 & 33 & 31 & 33 \\ 
TEMPLATE\_STAR\_PHOTO 	& 111    & 38 & 38 & 38 & 38 		  & 52 & 52 & 52 & 52 \\ 
CXOBRIGHT 		& 104    & 69 & 68 & 10 & 51 		  & 23 & 23 & 17 & 23 \\ 
\hline
\end{longtable}
\end{center}

\clearpage
\begin{table}
\caption{\label{table:data:model} Data model of the summary files. \$\{IMF\} takes values in 'Chabrier', 'Kroupa', 'Salpeter'; \$\{LIBRARY\} takes values in 'MILES', 'STELIB', 'ELODIE' and \$\{i\} takes values in 0 to 9.}
\begin{center}
\begin{tabular}{ L{3.5cm} L{4cm} L{11cm} }
\hline \hline
\multicolumn{2}{c}{columns from Spec Obj All} & \textsc{firefly} columns \\
\hline
SURVEY:  \newline 
INSTRUMENT:  \newline 
CHUNK:  \newline 
PROGRAMNAME:  \newline 
PLATERUN:  \newline 
PLATEQUALITY:  \newline 
PLATESN2:  \newline 
DEREDSN2:  \newline 
LAMBDA\_EFF:  \newline 
BLUEFIBER:  \newline 
ZOFFSET:  \newline 
SNTURNOFF:  \newline 
NTURNOFF:  \newline 
SPECPRIMARY:  \newline 
SPECSDSS:  \newline 
SPECLEGACY:  \newline 
SPECSEGUE:  \newline 
SPECSEGUE1:  \newline 
SPECSEGUE2:  \newline 
SPECBOSS:  \newline 
BOSS\_SPECOBJ\_ID:  \newline 
SPECOBJID:  \newline 
FLUXOBJID:  \newline 
BESTOBJID:  \newline 
TARGETOBJID:  \newline 
PLATEID:  \newline 
NSPECOBS:  \newline 
FIRSTRELEASE:  \newline 
RUN2D:  \newline 
RUN1D:  \newline 
DESIGNID:  \newline 
CX:  
CY:  
CZ:  \newline 
XFOCAL:  \newline 
YFOCAL:  \newline 
SOURCETYPE:  \newline 
TARGETTYPE:  \newline 
THING\_ID\_TARGETING:  \newline 
THING\_ID:  \newline 
PRIMTARGET:  \newline 
SECTARGET:  \newline 
LEGACY\_TARGET1:  \newline 
LEGACY\_TARGET2:  \newline 
SPECIAL\_TARGET1:  \newline 
SPECIAL\_TARGET2:  \newline 
SEGUE1\_TARGET1:  \newline 
SEGUE1\_TARGET2:  \newline 
SEGUE2\_TARGET1:  \newline 
SEGUE2\_TARGET2:  \newline 
MARVELS\_TARGET1:  \newline 
MARVELS\_TARGET2:  \newline 
BOSS\_TARGET1:  \newline 
BOSS\_TARGET2:  \newline 
EBOSS\_TARGET0:  \newline 
EBOSS\_TARGET1:  \newline 
EBOSS\_TARGET2:  \newline 
EBOSS\_TARGET\_ID:  \newline 
ANCILLARY\_TARGET1:  \newline 
ANCILLARY\_TARGET2:  \newline 
SPECTROGRAPHID:  \newline 
PLATE:  \newline 
TILE:   \newline 
MJD:    \newline 
FIBERID:  \newline 
OBJID:  \newline 
 & 
PLUG\_RA:  \newline 
PLUG\_DEC:  \newline 
CLASS:  \newline 
SUBCLASS:  \newline 
Z:  \newline 
Z\_ERR:  \newline 
RCHI2:  \newline 
DOF:  \newline 
RCHI2DIFF:  \newline 
TFILE:  \newline 
TCOLUMN:  \newline 
NPOLY:  \newline 
THETA:  \newline 
VDISP:  \newline 
VDISP\_ERR:  \newline 
VDISPZ:  \newline 
VDISPZ\_ERR:  \newline 
VDISPCHI2:  \newline 
VDISPNPIX:  \newline 
VDISPDOF:  \newline 
WAVEMIN:  \newline 
WAVEMAX:  \newline 
WCOVERAGE:  \newline 
ZWARNING:  \newline 
SN\_MEDIAN\_ALL:  \newline 
SN\_MEDIAN:  \newline 
CHI68P:  \newline 
FRACNSIGMA:  \newline 
FRACNSIGHI:  \newline 
FRACNSIGLO:  \newline 
SPECTROFLUX:  \newline 
SPECTROFLUX\_IVAR:  \newline 
SPECTROSYNFLUX:  \newline 
SPECTROSYNFLUX\_IVAR:  \newline 
SPECTROSKYFLUX:  \newline 
ANYANDMASK:  \newline 
ANYORMASK:  \newline 
SPEC1\_G:  
SPEC1\_R:  
SPEC1\_I:  \newline 
SPEC2\_G:   
SPEC2\_R:   
SPEC2\_I:  \newline 
ELODIE\_FILENAME:  \newline 
ELODIE\_OBJECT:  \newline 
ELODIE\_SPTYPE:  \newline 
ELODIE\_BV:  \newline 
ELODIE\_TEFF:  \newline 
ELODIE\_LOGG:  \newline 
ELODIE\_FEH:  \newline 
ELODIE\_Z:  \newline 
ELODIE\_Z\_ERR:  \newline 
ELODIE\_Z\_MODELERR:  \newline 
ELODIE\_RCHI2:  \newline 
ELODIE\_DOF:  \newline 
Z\_NOQSO:  \newline 
Z\_ERR\_NOQSO:  \newline 
ZWARNING\_NOQSO:  \newline 
CLASS\_NOQSO:  \newline 
SUBCLASS\_NOQSO:  \newline 
RCHI2DIFF\_NOQSO:  \newline 
Z\_PERSON:  \newline 
CLASS\_PERSON:  \newline 
Z\_CONF\_PERSON:  \newline 
COMMENTS\_PERSON:  \newline 
CALIBFLUX:  \newline 
CALIBFLUX\_IVAR:  \newline 
 
& 

\$\{IMF\}\_\$\{LIBRARY\}\_age\_lightW: light weighted age in years \newline 
\$\{IMF\}\_\$\{LIBRARY\}\_age\_lightW\_up\_1sig:  
\$\{IMF\}\_\$\{LIBRARY\}\_age\_lightW\_low\_1sig: 
\$\{IMF\}\_\$\{LIBRARY\}\_age\_lightW\_up\_2sig:  
\$\{IMF\}\_\$\{LIBRARY\}\_age\_lightW\_low\_2sig: errors \newline 
\$\{IMF\}\_\$\{LIBRARY\}\_metallicity\_lightW: light weighted metallicity in solar metallicity \newline 
\$\{IMF\}\_\$\{LIBRARY\}\_metallicity\_lightW\_up\_1sig:  
\$\{IMF\}\_\$\{LIBRARY\}\_metallicity\_lightW\_low\_1sig:  
\$\{IMF\}\_\$\{LIBRARY\}\_metallicity\_lightW\_up\_2sig:  
\$\{IMF\}\_\$\{LIBRARY\}\_metallicity\_lightW\_low\_2sig: errors \newline 
\$\{IMF\}\_\$\{LIBRARY\}\_age\_massW: mass weighted age in years \newline 
\$\{IMF\}\_\$\{LIBRARY\}\_age\_massW\_up\_1sig: 
\$\{IMF\}\_\$\{LIBRARY\}\_age\_massW\_low\_1sig:
\$\{IMF\}\_\$\{LIBRARY\}\_age\_massW\_up\_2sig: 
\$\{IMF\}\_\$\{LIBRARY\}\_age\_massW\_low\_2sig:errors \newline 
\$\{IMF\}\_\$\{LIBRARY\}\_metallicity\_massW: mass weighted metallicity in solar metallicity \newline 
\$\{IMF\}\_\$\{LIBRARY\}\_metallicity\_massW\_up\_1sig: 
\$\{IMF\}\_\$\{LIBRARY\}\_metallicity\_massW\_low\_1sig: 
\$\{IMF\}\_\$\{LIBRARY\}\_metallicity\_massW\_up\_2sig: 
\$\{IMF\}\_\$\{LIBRARY\}\_metallicity\_massW\_low\_2sig:errors \newline 
\$\{IMF\}\_\$\{LIBRARY\}\_stellar\_mass: total stellar mass in living stars + remants + gas in solar mass \newline 
\$\{IMF\}\_\$\{LIBRARY\}\_stellar\_mass\_up\_1sig: 
\$\{IMF\}\_\$\{LIBRARY\}\_stellar\_mass\_low\_1sig:
\$\{IMF\}\_\$\{LIBRARY\}\_stellar\_mass\_up\_2sig: 
\$\{IMF\}\_\$\{LIBRARY\}\_stellar\_mass\_low\_2sig:errors \newline 
\$\{IMF\}\_\$\{LIBRARY\}\_living\_star\_mass: stellar mass in living stars + remnants in solar mass \newline 
\$\{IMF\}\_\$\{LIBRARY\}\_remnant\_mass: stellar mass in remnants \newline 
\$\{IMF\}\_\$\{LIBRARY\}\_remnant\_mass\_in\_whitedwarfs: stellar mass in white dwarfs in solar mass \newline 
\$\{IMF\}\_\$\{LIBRARY\}\_remnant\_mass\_in\_neutronstars: stellar mass in neutron stars in solar mass \newline 
\$\{IMF\}\_\$\{LIBRARY\}\_remnant\_mass\_blackholes: stellar mass in black holes in solar mass \newline 
\$\{IMF\}\_\$\{LIBRARY\}\_mass\_of\_ejecta: stellar mass ejected in solar mass \newline 
\$\{IMF\}\_\$\{LIBRARY\}\_spm\_EBV: reddenning value fitted \newline 
\$\{IMF\}\_\$\{LIBRARY\}\_nComponentsSSP:         number of single stellar population components  \newline 
\$\{IMF\}\_\$\{LIBRARY\}\_stellar\_mass\_ssp\_\$\{i\}:  Stellar mass in the ith SSP  \newline 
\$\{IMF\}\_\$\{LIBRARY\}\_age\_ssp\_\$\{i\}:            Age of the ith SSP          \newline 
\$\{IMF\}\_\$\{LIBRARY\}\_metal\_ssp\_\$\{i\}:          Metallicity of the ith SSP   \newline 
\$\{IMF\}\_\$\{LIBRARY\}\_weightMass\_ssp\_\$\{i\}:     mass weight of the ith SSP in the overall solution \newline 
\$\{IMF\}\_\$\{LIBRARY\}\_weightLight\_ssp\_\$\{i\}:    light weight of the ith SSP in the overall solution \newline 
\$\{IMF\}\_\$\{LIBRARY\}\_chi2: chi square \newline 
\$\{IMF\}\_\$\{LIBRARY\}\_ndof: Number of degree of freedom \newline 
abs\_mag\_u\_spec: 
abs\_mag\_g\_spec: 
abs\_mag\_r\_spec: 
abs\_mag\_i\_spec: Absolute magnitude (u,g,r,i-band) measured on the observed spectrum \newline 
abs\_mag\_u\_noise:  
abs\_mag\_g\_noise:
abs\_mag\_r\_noise:
abs\_mag\_i\_noise: Absolute magnitude (u,g,r,i-band) measured on the observed noise spectrum \newline
SNR\_ALL: Median signal to noise in the spectrum \newline 
SNR\_32\_35: Median signal to noise ratio in the band 3200-3500\text{\AA}; 
SNR\_35\_39: - 3500-3900\text{\AA};  
SNR\_39\_41: - 3900-4100\text{\AA};  
SNR\_41\_55: - 4100-5500\text{\AA};  
SNR\_55\_68: - 5500-6800\text{\AA};  
SNR\_68\_74: - 6800-7400\text{\AA};  
SNR\_74\_93: - 7400-9300\text{\AA}  \newline 
% Separation:  \newline 

\\
\hline
\end{tabular}
\end{center}
\end{table}

\end{document}